\documentclass[prb,showpacs,amsmath,amssymb,floatfix,superscriptaddress]{revtex4}
\usepackage[toc,page]{appendix}
\usepackage{graphicx,pdflscape,epsf,epsfig,amsmath}% Include figure files
\usepackage{alltt,dsfont,amsmath,amssymb,bm}
\usepackage{float}

\usepackage{hyperref}
\usepackage{color}

\newcommand{\G}{{\cal{G}}}

\newcommand{\GH}{{\bf g}}
\newcommand{\GHI}{\GH^{-1}}

\newcommand{\tr}{{\text {tr} \ }}

\newcommand{\X}[2]{X_{{#1}}^{#2}}
\newcommand{\si}{\sigma}
\newcommand{\sib}{\bar{\sigma}}

\newcommand{\tJ}{\ $t$-$J$ \ }

\newcommand{\U}{{\cal U}}

\newcommand{\V}{{\cal V}}

\newcommand{\bb}[1]{{{\mathbf #1}}}
\newcommand{\nn}{\nonumber}
\newcommand{\barray}{\begin{eqnarray}}
\newcommand{\earray}{\end{eqnarray}}
\newcommand{\beq}{\begin{eqnarray}}
\newcommand{\eeq}{\end{eqnarray}}

\newcommand{\disp}[1]{Eq.~(\ref{#1})}
\newcommand{\refdisp}[1]{Ref.~(\onlinecite{#1})}
\newcommand{\figdisp}[1]{Fig.~(\ref{#1})}

\newcommand{\vk}{\vec{k}}

\newcommand{\lab}[1]{\label{#1}}

\newcommand{\Lim}[1]{\raisebox{0.5ex}{\scalebox{0.8}{$\displaystyle \lim_{#1}\;$}}}

\begin{document}
\title{Diagrammatic $\lambda$ series for  extremely correlated Fermi liquids}
\author{Edward Perepelitsky}
\affiliation{ Physics Department, University of California, Santa Cruz, CA 95064, USA}
\affiliation{Centre de Physique Th\'eorique, Ecole Polytechnique, CNRS, 91128 Palaiseau Cedex, France}
\affiliation{Coll\`{e}ge de France, 11 place Marcelin Berthelot, 75005 Paris, France.}
\author{B. Sriram Shastry}
\affiliation{ Physics Department, University of California, Santa Cruz, CA 95064, USA}
\affiliation{Centre de Physique Th\'eorique, Ecole Polytechnique, CNRS, 91128 Palaiseau Cedex, France}
\affiliation{Coll\`{e}ge de France, 11 place Marcelin Berthelot, 75005 Paris, France.}
\date{\today}
\begin{abstract}
The recently developed theory of extremely correlated Fermi liquids (ECFL), applicable to models involving the physics of  Gutzwiller projected electrons,  shows considerable promise in understanding the phenomena displayed by the \tJ model. Its formal equations for the Greens function  are reformulated by  a new procedure that is intuitively close to that used in the usual Feynman-Dyson theory.  We provide a systematic procedure by which one can draw diagrams for the $\lambda$-expansion of the ECFL introduced in \refdisp{ECFL}, where the parameter $\lambda \in (0,1)$ counts the order of the terms. In contrast to the  Schwinger method originally  used for this problem, we are able to write down the $n^{th}$ order diagrams ($O(\lambda^n)$) directly with the appropriate coefficients,  without enumerating {\em all} the previous order terms. This is a considerable advantage since it thereby enables the  possible implementation of Monte Carlo methods to evaluate the $\lambda$ series directly. The new procedure also provides a useful and intuitive alternative to the earlier methods.    

\smallskip
\noindent \textbf{Keywords:} $\lambda$ expansion;    \tJ model;     Hubbard model;         Extremely Correlated Fermi Liquid Model;        Strongly correlated electrons.

 \pacs{71.10.Fd}

   \end{abstract}
\maketitle

%\maketitle

\section{Introduction\lab{sec1} }

\subsection{ Motivation}
The $\tJ$ model is a model of fundamental importance in condensed matter physics, and is supposed  to have the necessary ingredients to explain the physics of the high-temperature cuprates \cite{Anderson}. Its Hamiltonian can be written in terms of the Hubbard $X$ operators as \cite{ECQL}
\barray
 H = -\sum_{ij\sigma}t_{ij}X_i^{\sigma 0 }X_j^{0\sigma} - \mu\sum_{i\sigma}X_i^{\sigma\sigma} +\frac{1}{2}\sum_{ij\sigma}J_{ij}X_i^{\sigma\sigma} + \frac{1}{4}\sum_{ij\sigma_1\sigma_2}J_{ij}\{X_i^{\sigma_1\sigma_2 }X_j^{\sigma_2\sigma_1} -X_i^{\sigma_1\sigma_1 }X_j^{\sigma_2\sigma_2}\}.\label{tJmodel}
\earray
The operator $X_i^{ab} = |a\rangle\langle b|$ takes the electron at  site $i$ from the state $|b\rangle$ to the state $|a\rangle$, where $|a\rangle$ and $|b\rangle$ are one of the two occupied states $|\uparrow\rangle$, $|\downarrow\rangle$, or the unoccupied state $|0\rangle$. {In terms of   electron operators $C, C^\dagger$, and the  Gutzwiller projection operator $P_G$ that eliminates double occupancy, we may explicitly write   $X_{i}^{ \sigma 0} = P_G\, C^\dagger_{i \sigma} \, P_G$, $X_{i}^{ 0 \sigma } = P_G\, C_{i \sigma} \, P_G$ and $X_{i}^{ \sigma \sigma'} =  C^\dagger_{i \sigma}C_{i \sigma'} \, P_G$.} 
The key object of study for this model is the single-particle Green's function, given by the expression
\beq \G_{\sigma_1\sigma_2}(i,f) = -\langle T_\tau X_{i}^{0\sigma_1}(\tau_i) X_{f}^{\sigma_2 0}(\tau_f)\rangle, 
\label{defG}
\eeq
as well as higher order dynamical correlation functions. Several novel approaches for computing these objects have been tried in literature\cite{Wang,Greco,Foussats,Zeyher,Zaitsev,Izyumov}, but it has been found difficult to impose the Luttinger Ward volume theorem in a consistent way, while providing a realistic description of both quasiparticle peaks and  background terms in the spectral function.

The essential difficulties in computing these objects are (I) the non-canonical nature of the $X$ operators, and hence the absence of the standard Wick's theorem, and (II) the lack of a convenient expansion parameter. In {the} recently developed {\em  extremely correlated Fermi liquid theory} (ECFL) \cite{ECFL,Monster, Shastry-AOP}, Shastry proposed a formalism which successfully resolves both difficulties. This formalism is based on Schwinger's approach to field theory, which bypasses Wick's theorem, and is more generally applicable than the Feynman approach that is fundamentally based upon Wick's theorem. Building atop  this powerful formalism, the ECFL theory  consists of the following main ingredients:
\begin{itemize}
\item{(1)} The product ansatz, in which the physical Green's function $\G[i,f]$ is written as a product of the auxiliary (Fermi-liquid type) Green's function $\GH[i,f]$, and {a caparison function} $\widetilde{\mu}[i,f]$ (\disp{product}). The former is a canonical, i.e. unprojected electron type Green's function, while the latter is a dynamical correction, which  arises fundamentally from the removal of double occupancy  from  the Hilbert space. This addresses the difficulty (I) above.
\item{(2)} The introduction of an expansion parameter $\lambda \in (0,1)$, which continuously connects the $\tJ$ model with the free Fermi gas, and enables the formulation of a systematic expansion. {This parameter is related to the extent to which double occupancy is removed, and has a close parallel  to the semiclassical expansion parameter $\frac{1}{2 S}$ arising in the  expansion of  spin $S$ (angular momentum) operators in terms of canonical Bosons\cite{Shastry-AOP}. }
\end{itemize}
In addition the  detailed calculations require certain crucial steps
\begin{itemize} 
\item{(3)} The introduction of a particle-number sum rule for the auxiliary Green's function (\disp{sumrules}), fixing the number of auxiliary fermions to equal the number of physical fermions. This arises from  requiring the charge of the particle to be unaffected by Gutzwiller projection, and  is closely connected to the volume of the Fermi-surface of the physical fermions. In essence it ensures that the theory satisfies the Luttinger-Ward volume theorem\cite{LuttingerWard,Luttinger}.
\item{(4)} The introduction of the second chemical potential $u_0$, which ensures that $\GH[i,f]$ and $\widetilde{\mu}[i,f]$ individually satisfy the shift invariance theorem \cite{Monster}, and together with the original chemical potential $\mu$, facilitates the  {fulfilling } of the two particle-number sum rules.
\end{itemize}
In earlier work these  ingredients are accomplished directly using the Schwinger equation of motion (EOM) for the $\tJ$ model. In particular, the fundamental objects $\GH[i,f]$ and $\widetilde{\mu}[i,f]$ are defined through their respective equations of motion, and the expansion parameter $\lambda$ is inserted directly into the equation of motion. The practical issue of computing objects to various orders in $\lambda$ is also accomplished by iterating the EOM order by order. The technical details are given in \refdisp{ECFL} and \refdisp{Monster}, and are summarized below in section \ref{ECFL}, facilitating a self contained presentation. 

In recent papers, the $O(\lambda^2)$ ECFL has been theoretically benchmarked using Dynamical Mean-Field Theory (DMFT)\cite{ECFLDMFT}, Numerical Renormalization Group (NRG) calculations\cite{ECFLAM}, and high-temperature series \cite{Moments}. In all cases, the low order ECFL calculation compares remarkably well with these well established techniques. On the experimental side, a phenomenological version of ECFL which uses simple Fermi-liquid expressions for the self-energies $\Phi[i,f]$ and $\Psi[i,f]$ (which are simply related $\GH[i,f]$ and $\widetilde{\mu}[i,f]$ respectively) was successful in explaining the anomalous lines shapes of Angle-Resolved Photoemission Spectroscopy (ARPES) experiments \cite{Gweon}.  Encouraged by this,  higher order terms e.g.  $O(\lambda^3)$ are of considerable interest in order to probe
densities closer to the Mott limit than possible with the $O(\lambda^2)$ theories, and in this context the present work is relevant. In this paper, we develop a diagrammatic $\lambda$ expansion. This expansion allows one to calculate the Greens function and related objects to any order in $\lambda$ by drawing diagrams. These diagrams are reminiscent of those in the Feynman series \cite{FW,Nozieres}, although more complicated than the former. This extra complication stems from the non-canonical nature of the $X$-operators and the absence of Wick's theorem. The diagrammatic formulation of the $\lambda$ series has the following advantages:
\begin{itemize}
\item It allows one to calculate the $n^{th}$ order contribution to any object by drawing diagrams directly for that order, without having to iterate the expressions from the previous orders. This not only allows for greater ease of computation of analytical expressions, but is also essential for powerful numerical series summation techniques, such as diagrammatic Monte Carlo \cite{DiagMC}. Ultimately, it will allow the series to be evaluated to high orders in $\lambda$, whereas presently, only a second order calculation has been possible \cite{ECFL2nd}.
\item It allows for the diagrammatic interpretation of the various objects in the theory such as the auxiliary Green's function $\GH[i,f]$ and the caparison factor $\widetilde{\mu}[i,f]$. For example, one can see that the product ansatz (\disp{product}) is a natural consequence of the structure of the $\G[i,f]$ diagrams. In particular, it is necessitated by the extra complexity introduced into the diagrams (over those of the Feynman series) by the projection of the double occupancy.
\item It allows one to visualize the structure of the diagrams to all orders in $\lambda$, therefore facilitating diagrammatic re-summations based on some physical principle.
\end{itemize}
\subsection{Results}

The main result of the paper is the formulation of diagrammatic rules to calculate the Green's function to any order in $\lambda$. More precisely, the rules state how to generate numerical representations (see section \ref{rules}), which are then converted into diagrams. A  subset of these numerical representations (determined by a simple criterion) are in one-to-one correspondence with the standard Feynman diagrams. Therefore, the diagrams given here are a natural generalization of the Feynman diagrams.  In this broader class of diagrams, we obtain a subset of numerical representations which are not in one-to-one correspondence with the resulting non-Feynman diagrams. In particular, two different numerical representations can lead to the same (non-Feynman) diagram. This occurs since in these non-Feynman diagrams, an interaction vertex can have more than two pairs of Green's function lines exiting and entering it (e.g. Fig. (\ref{3rdorderchifromPsi}g)). However, the contributions of both numerical representations must be kept. We also discuss below the relationship between ECFL and a   formalism  using the  high-temperature expansion for the $\tJ$ model due to Zaitsev and Izyumov \cite{Zaitsev,Izyumov}  in section \ref{ZaitsevIzyumov}, and make some connections in the following. 

We find that a certain subset of the $\G[i,f]$ diagrams terminate with a self-energy insertion, rather than a single point, as in the case of the Feynman diagrams. This expresses the diagrammatic  necessity for  the factorization of $\G$ into $\GH$ and $\widetilde{\mu}$. These  are in turn expressed in terms of the two self-energies $\Phi$ and $\Psi$. It is interesting that within the Zaitsev-Izyumov \cite{Zaitsev,Izyumov} formalism, a two self-energy structure for the Green's function is necessary for the exact same reason. The fact that the two self-energy structure comes from three independent approaches, the $\lambda$ expansion, the high-temperature expansion, and the factorization of the Schwinger EOM, shows that it is the correct representation of the Green's function for this model. In addition, as already reported in Ref~(\onlinecite{Shastry-AOP}), the Dyson Maleev approach developed by Harris, Kumar, Halperin and Hohenberg \cite{HKHH} also leads to a similar two self energy scheme in quantum spin systems, where again the algebra of the basic variables is non-canonical.

We derive diagrammatic rules for the constituent objects $\GH$, $\widetilde{\mu}$, $\Phi$, and $\Psi$ from their definitions, starting from the  Schwinger equations of motion. We avoid the use of dressed propagators (leading to skeleton terms), but rather expand various objects in powers of $\lambda$ directly. The fact that these diagrammatic rules are consistent with those of $\G$ and the product ansatz serves as an independent proof of the rules given for $\G$. We find that $\Phi$ consists of two independent pieces. The first can be obtained by adding a single interaction line to the terminal point of a $\Psi$ diagram, while the second one is completely independent of $\Psi$. We denote the second piece by the letter $\chi$, which leads to the relation $\Phi(\vk,i\omega_k) = \epsilon_{\vk} \Psi(\vk,i\omega_k) + \chi(\vk,i\omega_k)$ in momentum space. In a previous work by the same authors \cite{ECFLlarged}, we showed directly from the Schwinger equations of motion, that in the limit of infinite spatial dimensions, $\Phi(\vk,i\omega_k) = \epsilon_{\vk} \Psi(i\omega_k) + \chi(i\omega_k)$. Here, using the diagrammatic $\lambda$ expansion, we show that this relationship continues to make sense in finite dimensions. In going from finite to infinite dimensions, we lose momentum dependence so that  $\Psi(\vk,i\omega_k)\to \Psi(i\omega_k)$ and $\chi(\vk,i\omega_k)\to \chi(i\omega_k)$. We also derive the Schwinger EOM defining the object $\chi$ in finite dimensions.

We derive diagrammatic rules for the three point vertices $\Lambda$ and $\U$, defined as functional derivatives of $\GHI$ and $\widetilde{\mu}$ (\disp{set2}). Diagrammatically, their relationship to $\Phi$ and $\Psi$ is seen to be consistent with the Schwinger equations of motion (\disp{set1}). We also derive a generalized Nozi\`{e}res relation for these vertices, which differs from the standard one for the three-point vertices of the Feynman diagrams. We introduce the concept of a skeleton diagram into our series. This enables us to make the rather subtle connection between our diagrammatic approach for the $\lambda$ expansion, and the iterative one used previously. Finally, we use our diagrammatic approach to derive analytical expressions for the third order skeleton expansion of the objects $\GH$ and $\widetilde{\mu}$, whereas previously only the second order expressions had been derived via iteration of the equations of motion.

\subsection{Outline of the paper}

In section \ref{ECFL}, we begin by reviewing the ECFL formalism from Refs. (\onlinecite{ECFL}) and (\onlinecite{Monster}) in the simplified case of $J=0$. In section \ref{discussion}, we introduce the $\lambda$ expansion diagrams in a heuristic way, drawing an analogy with the standard Feynman diagrams. In section \ref{bare}, we derive the rules for drawing and evaluating the bare diagrams for $\G$ to each order in $\lambda$. We also draw and evaluate the first and second order bare diagrams for $\G$. In section \ref{constituent}, we derive the diagrammatic rules for the constituent objects $\GH$, $\widetilde{\mu}$, $\Phi$, $\Psi$, $\chi$, $\gamma$, $\Lambda$, and $\U$. We also show how to evaluate diagrams in momentum space. We then introduce skeleton diagrams into the series, and complete the full circle by relating our diagrammatic approach to the $\lambda$ expansion to the original iterative one reviewed in section \ref{ECFL}. In section \ref{J}, we review the ECFL formalism \cite{ECFL, Monster} with $J\neq0$, and introduce $J$ into our diagrammatic series. In section \ref{finiteorder}, we compute the skeleton expansion to third order in $\lambda$ for the objects $\GH$ and $\widetilde{\mu}$. We also discuss the high-frequency limit of $\G$ to each order in the bare and skeleton expansions, as well as the ``deviation" of the $\lambda$ series from the Feynman series. Finally, in section \ref{ZaitsevIzyumov}, we discuss the connection between the ECFL and the Zaitsev-Izyumov formalism for the high-temperature expansion of the $\tJ$ model.

\section{ECFL Equations of Motion and the $\lambda$ expansion\label{ECFL}}

 The Greens function is the fundamental object in this theory and is defined as usual by 
\beq
\G_{\si_i,\si_f}[i,f] \equiv - \langle  \langle \X{i}{0 \si_i} \X{f}{\si_f 0} \rangle \rangle= - \frac{1}{Z[\V]}  \tr e^{- \beta H} T_\tau \left(e^{-\cal A } \X{i}{0\si_i}(\tau_i) \X{f}{\si_f 0}(\tau_f)\right) , \label{G-def}
\eeq
where ${\cal A}={\sum_j \int_0^\beta \X{j}{\si \si'}(\tau') \V_j^{\si \si'}(\tau') d\tau'} $, is the additional term in the action due to the Bosonic source
 $\V_i\equiv\V_i(\tau_i)$,  included in the partition functional $Z[\V]= \tr e^{- \beta H} T_\tau e^{-\cal A}$.    The angular brackets represent averages  over the distribution in \disp{G-def}.
The function $\G$ satisfies the Schwinger equation of motion for the \tJ model as derived in Refs. (\onlinecite{ECFL}), (\onlinecite{Monster}), and (\onlinecite{ECQL}).
\barray
&&\{\left[(\partial_{\tau_i}-\mu)\delta[i,\bb{j}]-t[i,\bb{j}]\right]\delta_{\si_1,\si_j}+\V_i^{\si_1,\si_j}\delta[i,\bb{j}]\} \  \G_{\si_j,\si_2}[\bb{j},f] = -\delta[i,f]\delta_{\si_1,\si_2} +\lambda\delta[i,f]\gamma_{\si_1,\si_2}[i] - \lambda t[i,\bb{j}] \gamma_{\si_1,\si_a}[i] \G_{\si_a,\si_2}[\bb{j},f] \nn \\
&&+\lambda t[i,\bb{j}]\si_1\si_a \frac{\delta}{\delta\V_i^{\bb{\sib_1},\bb{\sib_a}}} \G_{\si_a,\si_2}[\bb{j},f] +\frac{\lambda}{2} J[i,\bb{j}] \left( \gamma_{\si_1,\si_a}[\bb{j}] \G_{\si_a,\si_2}[i,f]-\si_1\si_a \frac{\delta}{\delta\V_{\bb{j}}^{\bb{\sib_1},\bb{\sib_a}}} \G_{\si_a,\si_2}[ i,f]\right),
\label{EOMG} 
\earray
where the bold repeated indices are summed over. 
 The functional derivative takes place at time $(\tau_i^+)$, and we have used the notation $\delta[i,m]=\delta_{i,m}\delta(\tau_i-\tau_m)$, $t[i,m]=t_{i,m}\delta(\tau_i-\tau_m)$, and $\gamma_{\si_1,\si_2}[i] = \si_1 \si_2 \G_{\sib_2,\sib_1}[i,i^+]$ .

We next outline how  one  obtains the above equation of motion   from  the definition \disp{G-def}.  We take a time derivative $\partial_{\tau_i}$ of \disp{G-def}, yielding  several terms.  We start with a  simple contribution, namely    the time derivative of the $\theta(\tau_i-\tau_f)$, which involves  the   anticommutator 
\beq
\{\X{i}{0 \si_i}, \X{j}{\si_j 0} \} = \delta_{ij} \left( \delta_{\si_i \si_j} - \lambda \, \si_i \si_j \X{i}{\sib_i \sib_j} \right), \label{ACR}
\eeq
strictly speaking with $\lambda=1$. We  use the  anticommutator, generalized as above  by introducing the parameter $\lambda \in [0,1]$, so that the result interpolates smoothly between the canonical value $\lambda=0$ and the fully Gutzwiller projected value $\lambda=1$.
This process is fundamental to obtaining  the $\lambda$ expansion. From this, we get  $\delta_{i f} \delta(\tau_i-\tau_f)  \left( \delta_{\si_i \si_f} - \lambda \si_i \si_f \langle \X{i}{\sib_i \sib_f}(\tau_i) \rangle \right)$. This is expressed back in terms of the Greens function by writing $\langle \X{i}{\sib_i \sib_f} \rangle  \to \G_{\sib_f \sib_i}[i\tau_i, i \tau_i^+] = \si_i \si_f \gamma_{\si_i \si_f}[i] $, and thus to the first two terms on the right hand side of \disp{EOMG}.

 Another contribution arises from the  $\tau_i$ dependence in the  lower and upper limits of the time integrals in the expression
 \beq
 T_\tau e^{-\cal A_S} \X{i}{0 \si_i}(\tau_i)= e^{- {\sum_j \int_{\tau_i}^\beta \X{j}{\si \si'}(\tau') \V_j^{\si \si'}(\tau') d\tau'} }\X{i}{0 \si_i}(\tau_i) e^{- {\sum_j \int_0^{\tau_i} \X{j}{\si \si'}(\tau') \V_j^{\si \si'}(\tau') d\tau'} },
 \eeq
involving   the equal time commutator $\sum_j \V_j^{\si \si'}(\tau_i) [\X{j}{\si \si'}(\tau_i),\X{i}{0 \si_i}(\tau_i)]= \V_i^{\si_i \si'}(\tau_i) \X{i}{0 \si'}(\tau_i)$. This leads to the third term in the left hand side of \disp{EOMG}.

The non trivial  term  is obtained when the  $\partial_{\tau_i} \X{i}{0 \si}(\tau_i)$  is evaluated from the Heisenberg equation of motion $[H, \X{i}{0 \si}]$ and  the fundamental anticommutator \disp{ACR} yielding
\beq
[\X{i}{0 \si},H]= -\mu \X{i}{0 \si} - t_{ij} \X{j}{0 \si_i} + \lambda \, \sum_{j \si_j} t_{ij} (\si_i \si_j) \X{i}{\sib_i \sib_j} \X{j}{0 \si_j} - \frac{1}{2} \lambda \, \sum_{j \neq i} J_{ij} (\si_i \si_j) \X{j}{\sib_i \sib_j} \X{i}{0 \si_j}.
 \eeq
 Note that the $J$ term has an almost identical structure to the $t$ term, with $i \leftrightarrow j$.
 The term involving $J$  actually does not come with the external $\lambda$, we introduce it so that the $\lambda=0$ limit is the Fermi gas. (This is permissible since we are finally intersted in the limit $\lambda=1$.)
 A higher order Greens function $\langle \langle  \X{i}{\sib_i \sib_j}(\tau_i) \X{j}{0 \si_j}(\tau_i), \X{f}{\si_f 0}(\tau_f) \rangle \rangle$ is generated by the third term and a similar one by the fourth term. These are re-expressed in terms of the Greens function by using the identity due to Schwinger
 \beq
 \langle \langle  \X{i}{\sib_i \sib_j}(\tau_i) \X{j}{0 \si_j}(\tau_i), \X{f}{\si_f 0}(\tau_f) \rangle \rangle =
 \langle \langle  \X{i}{\sib_i \sib_j}(\tau_i)\rangle \rangle \langle \langle  \X{j}{0 \si_j}(\tau_i), \X{f}{\si_f 0}(\tau_f) \rangle \rangle - \frac{\delta}{\delta \V_i^{\sib_i \sib_j}(\tau_i)} \langle \langle  \X{j}{0 \si_j}(\tau_i), \X{f}{\si_f 0}(\tau_f) \rangle \rangle. 
 \eeq
 Using again $\langle \langle  \X{i}{\sib_i \sib_j}(\tau_i)\rangle \rangle= \G_{\sib_j \sib_i}[i\tau_i, i \tau_i^+] = \si_i \si_j \gamma_{\si_i \si_j}[i]$,  we obtain the last four terms on the right hand side of  equation \disp{EOMG}. For ease of presentation we will initially set $J\to0$ and  reinstate it at a later stage.
 
 As demonstrated in \refdisp{ECFL},  the electron Green's function $\G[i,f]$ can be factored via the following product ansatz:
\beq \G[i,f] = \GH[i,\bb{j}] . \widetilde{\mu}[\bb{j},f], \label{product}\eeq
where $\GH[i,f]$ is the auxiliary Green's function, $\widetilde{\mu}[i,f]$ is the caparison factor, all objects have been represented as $2 \times 2$ matrices in spin space, and matrix multiplication has been indicated by a dot. $\GH[i,f]$ and $\widetilde{\mu}[i,f]$ are defined by the their respective Schwinger equations of motion.
\barray
\GHI[i,m]&=&  ( \mu  - \partial_{\tau_i} - \V_i) \ \delta[i,m] + t[i,m] \ ( 1- \lambda\gamma[i]) -\lambda \Phi[i,m].\nn\\
\widetilde{\mu}[i,m]&=& (1- \lambda\gamma[i]) \delta[i,m] +\lambda \Psi[i,m] \nn \\
\Phi[i,m] &=& - t[i, \bb{j}] \ \xi^* . \GH[\bb{j}, \bb{n}]. \Lambda_* [ \bb{n}, m; i]; \;\;\;\;\;\;\; \Psi[i,m] = - t[i, \bb{j}] \ \xi^* . \GH[\bb{j}, \bb{n}]. \U_* [ \bb{n}, m; i].
\label{set1}
\earray
These exact relations give the required objects $\GH$ and $\widetilde{\mu}$  in terms of the vertex functions.
Here we also note that the local (in space and time) Green's function $\gamma[i]$, and the vertices $\Lambda[n,m;i]$ and $\U[n,m;i]$, are defined as
\barray
\gamma[i]= \widetilde{\mu}^{(k)}[\bb{n},i^+] . \GH^{(k)}[i,\bb{n}]; \;\;\;
\Lambda[n,m;i]= - \frac{\delta}{\delta \V_i} \GHI[n,m]; \;\;\; \U[n,m;i]=  \frac{\delta}{\delta \V_i} \widetilde{\mu}[n,m], 
 \label{set2}
\earray
 where we have used the notation $M^{(k)}_{\sigma_1,\sigma_2}= \sigma_1 \sigma_2 M_{\bar{\sigma}_2,\bar{\sigma}_1}$ to denote the time reversed  matrix $M^{(k)}$ of an arbitrary matrix $M$. These exact relations give the vertex functions    in terms of the objects $\GH$ and $\widetilde{\mu}$.
The vertices defined above ($\Lambda$ and $\U$) have four spin indices, those of the object being differentiated and those of the source. For example, $\U^{\sigma_1\sigma_2}_{\sigma_a\sigma_b}[n,m;i]=  \frac{\delta}{\delta \V^{\sigma_a\sigma_b}_i} \widetilde{\mu}_{\sigma_1\sigma_2}[n,m]$. In \disp{set1}, $\xi_{\sigma_a\sigma_b}=\sigma_a\sigma_b$, and the $^*$ indicates that these spin indices should also be carried over (after being flipped) to the bottom indices of the vertex, which is also marked with a $_*$. The top indices of the vertex are given by the usual matrix multiplication. An illustrative example is useful here:  $\left(\xi^* . \GH[j, \bb{n}]. \U_* [ \bb{n}, m; i]\right)_{\sigma_1\sigma_2}= \sigma_1\sigma_{\bb{a}}  \ \GH_{\sigma_{\bb{a}},\sigma_{\bb{b}}}[j, \bb{n}]\frac{\delta}{\delta \V^{\bar{\sigma}_1\bar{\sigma}_{\bb{a}}}_i }\widetilde{\mu}_{\sigma_{\bb{b}},\sigma_2}[\bb{n},m] $.

The $\lambda$ expansion is obtained by expanding \disp{set1} and \disp{set2} iteratively in the continuity parameter $\lambda$. The $\lambda=0$ limit of these equations is the free Fermi gas. Therefore, a direct expansion in $\lambda$ will lead to a series in $\lambda$ in which each term is made up of the hopping $t_{ij}$ and the free Fermi gas Green's function $\GH^{(0)}[i,f]$. As is the case in the Feynman series, this can be reorganized into a skeleton expansion in which only the skeleton graphs are kept and $\GH^{(0)}[i,f] \to \GH[i,f]$. However, one can also obtain the skeleton expansion directly by expanding \disp{set1} and \disp{set2} in $\lambda$, but treating $\GH[i,f]$ as a zeroth order (i.e. unexpanded) object in the expansion. This expansion is carried out to second order in \refdisp{ECFL}. In doing this expansion, one must evaluate the functional derivative $\frac{\delta \GH}{\delta \V}$. This is done with the help of the following useful formula which stems from the product rule for functional derivatives.
\beq  \frac{\delta \GH[i,m]}{\delta \V_r} = \GH[i,\bb{x}].\Lambda[\bb{x},\bb{y},r].\GH[\bb{y},m]. \label{Derivative}\eeq
Within the $\lambda$ expansion, the LHS is evaluated to a certain order in $\lambda$ by taking the vertex $\Lambda$ on the RHS to be of that order in $\lambda$.
\section{Heuristic discussion of $\lambda$ expansion diagrams \label{discussion}}
\subsection{Numerical representations of Feynman diagrams \label{numFeynman}}
Before deriving the precise rules for the $\lambda$ expansion diagrams, it is useful to have a heuristic discussion in which we compare them to the more familiar Feynman diagrams \cite{FW,Nozieres}.  To this end, we introduce numerical representations for the standard Feynman diagrams. These numerical representations will then be generalized to generate the $\lambda$ expansion diagrams.

\begin{figure}[H]
\begin{center}
\includegraphics{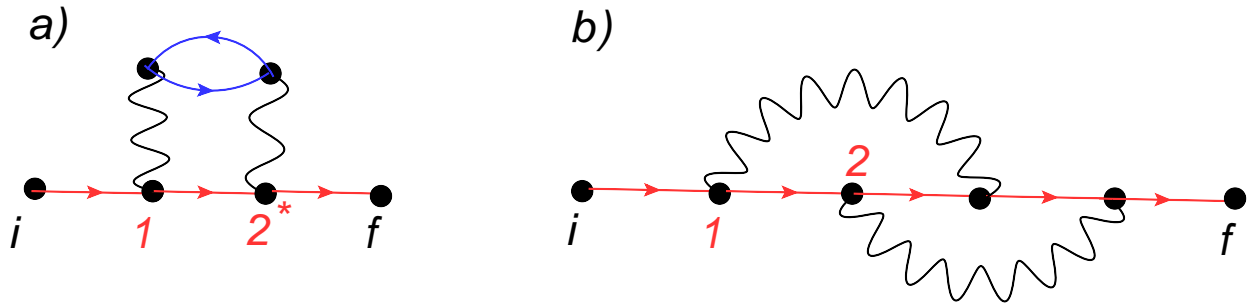}
\caption{Second order Feynman diagrams for $\G[i,f]$. The zeroth Fermi loop, which is the chain running from $i$ to $f$ is colored in red. In panel a), the first Fermi loop is colored in blue. The numerical representation of the diagram in panel a) is 
$ 1 \ 2^* \ f; \  2^{*}: 0 \ _1 \ f$, while that of the diagram in panel b) is $1\ 2 \ _{1 \ 2} \ f$. }
\label{ExFeynman}
\end{center}
\end{figure}

Consider any Feynman diagram for the Green's function $\G[i,f]$ such as those displayed in \figdisp{ExFeynman}. There is a unique path which runs between $i$ and $f$ which uses only Green's function lines, not {counting the }  interaction lines. We denote this as the zeroth Fermi loop. It is drawn in red in \figdisp{ExFeynman}. We number the interaction lines which connect to the zeroth Fermi loop in the order in which they appear in this loop. This list of numbers (along with $f$) is placed in the top row of our numerical representation. In the case of both Fig. (\ref{ExFeynman}a) and Fig. (\ref{ExFeynman}b), it is
\beq
1\ 2 \ f.
\nn
\eeq

If the zeroth Fermi loop does not exhaust all of the Green's function lines in the diagram, such as in Fig. (\ref{ExFeynman}a), we proceed to the first Fermi loop. To identify the first Fermi loop, we find the interaction vertex with the highest number which connects to the zeroth Fermi loop with only one of its two sides. In this case, this is the interaction vertex labeled $2$. The other side has one incoming line and one outgoing line. There is a unique path in the diagram which connects these two lines using only Green's function lines, not interaction lines. This defines the first Fermi loop. It is drawn in blue in Fig. (\ref{ExFeynman}a).

\begin{figure}[H]
\begin{center}
\includegraphics{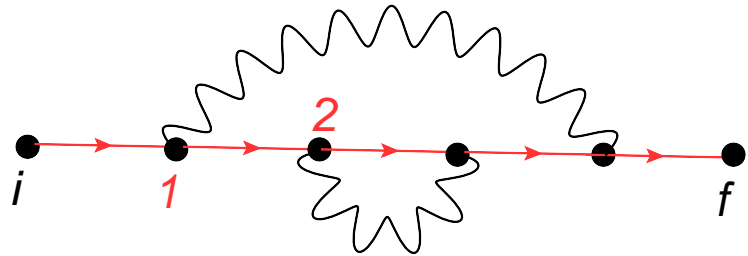}
\caption{This Feynman diagram results from reversing the order of the subscripts in the numerical representation of the Feynman diagram in Fig. (\ref{ExFeynman}b). Therefore, the numerical representation of this diagram is $1\ 2 \ _{2 \ 1} \ f$.}
\label{ExFeynman2}
\end{center}
\end{figure}

Since the interaction vertex $2$ spawned the first Fermi loop, it is starred in the top row of the representation. We also include a lower row for the first Fermi loop. Therefore, the numerical representation of Fig. (\ref{ExFeynman}a) now reads.
\barray
&& 1 \ 2^* \ f \nn\\
2^{*}: && 0 \ f.  \nn  
\earray
The second row, which represents the first Fermi loop, is labeled by $2^*$, since it was spawned by the second interaction vertex in the zeroth Fermi loop. The fact that only $0$ and $f$ are present in the second row tells us that there were no interaction vertices {\bf introduced} in the first Fermi loop. That is to say there are no interaction vertices which connect to the first Fermi loop, but not to the previous ones (in this case the zeroth Fermi loop). Finally, after all of the Fermi loops have been recorded, all nonzero integers which are not starred indicate the position of one side of an interaction vertex in a Fermi loop. We record the position of the other side as a subscript. Therefore, the complete numerical representation of Fig. (\ref{ExFeynman}a) is
\barray
&& 1 \ 2^* \ f \nn\\
2^{*}: && 0 \ _1 \ f.  \nn  
\earray
We can represent this in short as $ 1 \ 2^* \ f; \  2^{*}: 0 \ _1 \ f$, where the semi colon indicates the next line. The complete numerical representation of Fig. (\ref{ExFeynman}b) is
\beq
1\ 2 \ _{1 \ 2} \ f.
\nn
\eeq
Note that the order of appearance of the $1$ and $2$ as subscripts is important. Reversing them would yield the diagram in \figdisp{ExFeynman2}, which has the following numerical representation.
\beq
1\ 2 \ _{2 \ 1} \ f.
\nn
\eeq

\begin{figure}[H]
\begin{center}
\includegraphics{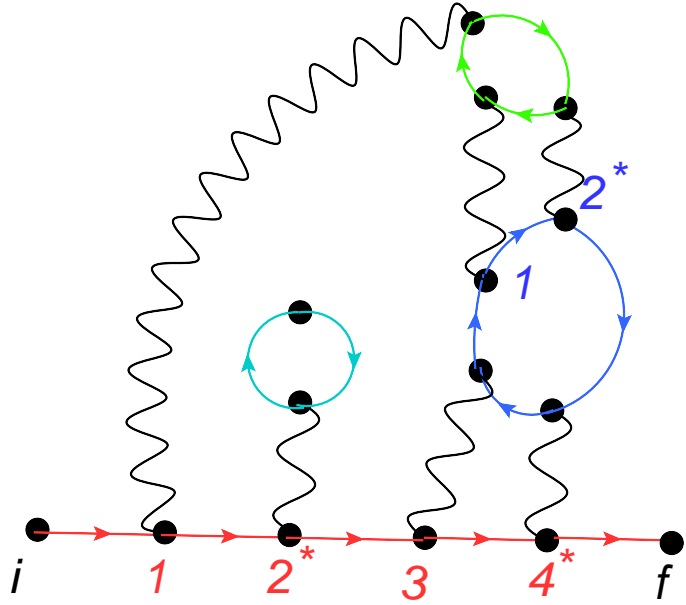}
\caption{Sixth order Feynman diagram. The zeroth, first, second, and third Fermi loops are drawn in red, blue, green, and turquoise respectively. Interaction vertices introduced in a particular Fermi loop are numbered in the same color as that loop. An interaction vertex is starred if it spawns a new Fermi loop. The numerical representation for this diagram is $1 \ 2^* \ 3 \ 4^* \ f; \ 4^{*}: 0 \ _3 \ 1 \ 2^* \ f; \ (4,2)^{*}:  0 \ _{(4,1) \ 1} \ f; \ 2^*: 0 \ f$.}
\label{ExFeynman3}
\end{center}
\end{figure}

We now consider the slightly more complicated diagram in \figdisp{ExFeynman3} to illustrate the scope of this approach. We will now show how the numerical representation of this diagram is derived. We first identify the zeroth Fermi loop, which is drawn in red in \figdisp{ExFeynman3}. The top row now reads
\beq
1\ 2 \ 3 \ 4 \ f.
\nn
\eeq
In this case, the vertex with the highest number which connects to this loop with only one side is $4$. Hence, $4$ spawns another Fermi loop, and gets a star in the top row.
\beq
1\ 2 \ 3 \ 4^* \ f.
\nn
\eeq
We identify this as the first Fermi loop. It is drawn in blue in \figdisp{ExFeynman3}. The numerical representation is modified to read
\barray
&& 1 \ 2 \ 3 \ 4^* \ f \nn\\
4^{*}: && 0 \ 1 \ 2 \ f.  \nn  
\earray
Considering only the interaction vertices introduced in the first Fermi loop, we now search for the one with the highest number which connects to the first Fermi loop with one side, but whose other side is {\bf free}, that is to say that it does not connect to any of the Fermi loops introduced thus far (zeroth and first). This is the interaction vertex 2. Hence, it gets a star, and spawns the second Fermi loop, which is drawn in green. The numerical representation now reads
\barray
&& 1 \ 2 \ 3 \ 4^* \ f \nn\\
4^{*}: && 0 \ 1 \ 2^* \ f  \nn\\
(4,2)^{*}: && 0 \ f.  \nn 
\earray
Here, the ordered pair $(4,2)$ is used to distinguish the $2$ in the first Fermi loop from the $2$ in the zeroth Fermi loop, the latter being denoted simply as $2$. The first number in the pair is $4$ since the fourth interaction vertex in the zeroth Fermi loop spawned the first Fermi loop. Also note that no interaction vertices are introduced in the second Fermi loop, hence its row only has a $0$ and an $f$. Therefore, we have arrived at the end of our first sequence of nested Fermi loops. We now take a step back in this sequence and return to the first Fermi loop. Considering only the interaction vertices introduced in the first loop with number less than $2$, we search for the one with the highest number which connects with one side to the first Fermi loop, but whose other side is free (i.e. does not connect to the zeroth, first, or second Fermi loops). There is no such interaction vertex. Therefore, we take another step back in the sequence, and return to the zeroth Fermi loop. We find that the interaction vertex $2$ connects to this loop with one side, but that the other side is free. Hence, 2 gets a star and spawns the fourth Fermi loop, which is drawn in turquoise. The numerical representation now reads
\barray
&& 1 \ 2^* \ 3 \ 4^* \ f \nn\\
4^{*}: && 0 \ 1 \ 2^* \ f  \nn\\
(4,2)^{*}: && 0 \ f  \nn\\
2^*: && 0 \ f. \nn 
\earray
Since there are no interaction vertices introduced in the fourth Fermi loop, we have arrived at the end of our second sequence of nested Fermi loops. We take a step back to the zeroth Fermi loop and find that there are no  more interaction vertices introduced in this loop which have one side free. Since all of the Fermi loops have been identified, as the final step, we must take the integers which are not starred, and place them in their final locations as subscripts. The complete numerical representation now reads
\barray
&& 1 \ 2^* \ 3 \ 4^* \ f \nn\\
4^{*}: && 0 \ _3 \ 1 \ 2^* \ f  \nn\\
(4,2)^{*}: && 0 \ _{(4,1) \ 1} \ f  \nn\\
2^*: && 0 \ f. \nn 
\earray

If we now wanted to formulate a set of rules for generating the numerical representations obtained from the Feynman diagrams, they would be the following.

\begin{itemize}
\item (1) Write a row of integers $ 1 \ldots m \ f$ where $m\geq1$, e.g. 
\barray
&& 1 \ 2 \ 3 \ 4 \ f. \nn
\earray
\item (2) Assign a star to any of the integers in the row ($f$ does not count as an integer), e.g.
\barray
&& 1 \ 2^* \ 3 \ 4^* \ f. \nn
\earray
\item (3) Every starred integer gives rise to a lower row. The $i^{th}$ lower row also consists of integers $0 \ldots m_i \ f$, where $m_i\geq0$, e.g.
\barray
&& 1 \ 2^* \ 3 \ 4^* \ f \nn\\
4^{*}: && 0 \ 1 \ 2 \ f  \nn\\
2^*: && 0 \ f. \nn 
\earray
\item (4) In the lower rows, assign a star to any of the integers excluding $0$, e.g.
\barray
&& 1 \ 2^* \ 3 \ 4^* \ f \nn\\
4^{*}: && 0 \ 1 \ 2^* \ f  \nn\\
2^*: && 0 \ f. \nn 
\earray
\item (5) The integers starred in step $4$ once again give rise to lower rows, etc. Continue this process until the last rows which you create have no starred integers, e.g.
\barray
&& 1 \ 2^* \ 3 \ 4^* \ f \nn\\
4^{*}: && 0  \ 1 \ 2^* \ f  \nn\\
(4,2)^{*}: && 0 \ f  \nn\\
2^*: && 0 \ f. \nn 
\earray
\item (6) Label each integer with a tuple (an ordered list of numbers) which traces that integer back to the first row through the starred integers. For example, the $0$ in the third row would be labeled $(4,2,0)$.
\item (7) Between any 2 consecutive integers of a row (including $0$'s and $f$'s), one can place as subscripts an ordered list of tuples from the following set: all those corresponding to non-starred integers except 0 whose tuple can be obtained from the tuple of the smaller of the 2 consecutive integers in question, by taking the first $k\leq l$ entries of this tuple (where $l$ is the length of the tuple), and subtracting a non-negative integer from the last entry. For example, suppose that the two consecutive integers in question are the $2$ and $f$ of the second row. Then all tuples (corresponding to non-starred integers) eligible to be used as subscripts between them are: $(4,1)$, $3$, and $1$.  All non-starred integers (except 0's) must be used exactly once in this way, e.g.
\barray
&& 1 \ 2^* \ 3 \ 4^* \ f \nn\\
4^{*}: && 0 \ _3 \ 1 \ 2^* \ f  \nn\\
(4,2)^{*}: && 0 \ _{(4,1) \ 1} \ f  \nn\\
2^*: && 0 \ f. \nn 
\earray
\end{itemize}
If we think back to the order in which we generated Fermi loops (and hence the numerical representation) from a given Feynman diagram, we can see that it complies exactly with rule (7) stated above. Doing things in this way ensures that the mapping between Feynman diagrams and numerical representations is one-to-one.

\subsection {Topologies of $\lambda$ expansion diagrams}

\begin{figure}[H]
\begin{center}
\includegraphics{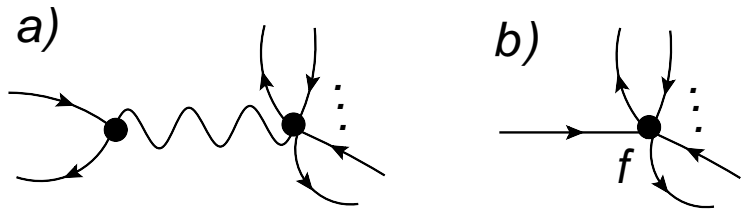}
\caption{The 2 elements used for construction the $\lambda$ expansion diagrams.}
\label{elements}
\end{center}
\end{figure}

\begin{figure}[H]
\begin{center}
\includegraphics{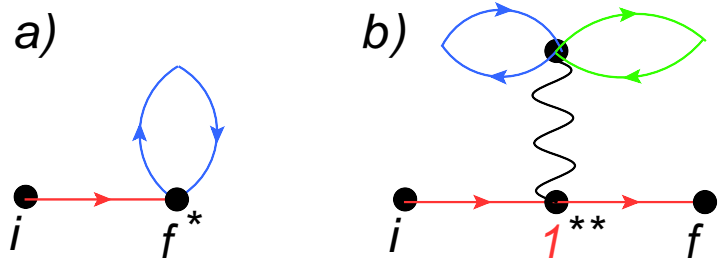}
\caption{Two of the simplest non-Feynman diagrams in the $\lambda$ expansion. A non-Feynman diagram occurs when the terminal point of a Fermi loop spawns another Fermi loop.}
\label{nonFeynman}
\end{center}
\end{figure}

The exact rules for drawing diagrams for the $\lambda$ expansion, as defined in section \ref{ECFL} will be derived in section \ref{bare}. There, it will be shown that the $\lambda$ expansion diagrams are constructed from the $2$ elements displayed in \figdisp{elements}. The one in panel a) is a generalization of the Feynman interaction vertex, in which one of the sides can have any number of pairs of incoming and outgoing lines rather than just one pair. The one in panel b) is a generalization of the terminal point $f$ in a Feynman diagram. In the case of the Feynman diagrams, it is a single point, while in the case of a $\lambda$ expansion diagram, it is a single point along with any number of pairs of incoming and outgoing lines. These extra lines come from the second term on the RHS of \disp{EOMG3}. This term, which itself comes from the anti-commutator of the $X$-operators in \disp{defG}, and which is absent in the EOM of canonical theories, allows a diagram to close in on itself in an iterative expansion of the EOM.
In \figdisp{nonFeynman}, we have drawn two of the simplest non-Feynman diagrams which can be made from these elements. The one in panel a) has the following numerical representation.
\barray
&&   f^* \nn\\
f^{*}: && 0 \ f.  \nn
\earray
The zeroth Fermi loop runs from the site $i$ to the site $f$ and is drawn in red. The site $f$, which is the terminal point of the zeroth Fermi loop spawns the first Fermi loop, drawn in blue. The one in panel b) has the following numerical representation.
\barray
&& 1^{**} \ f \nn\\
1^{**}: && 0 \ f^{*}  \nn\\
(1,f)^{*}: && 0 \ f. \nn
\earray
Here, the interaction vertex $1$, introduced in the zeroth Fermi loop (drawn in red) spawns the first Fermi loop (drawn in blue). Additionally, the terminal point of the first Fermi loop spawns the second Fermi loop (drawn in green).

\begin{figure}[H]
\begin{center}
\includegraphics{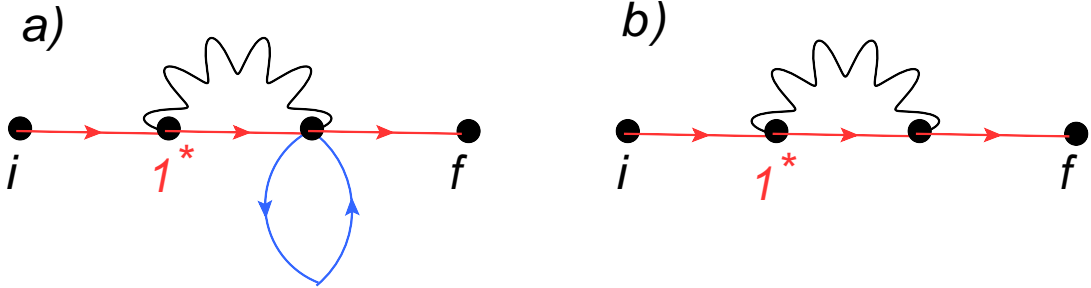}
\caption{The diagram in panel a) is not allowed in the $\lambda$ expansion. This is because first Fermi loop emanates from a point which is represented by a subscript in the Fock diagram displayed in panel b). }
\label{nonlambda}
\end{center}
\end{figure}

\begin{figure}[H]
\begin{center}
\includegraphics{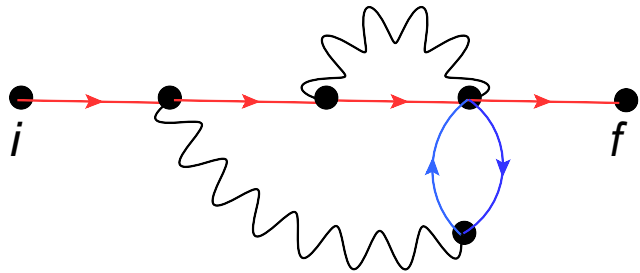}
\caption{A more elaborate version of the diagram in Fig. (\ref{nonlambda}a), which is also not allowed in the $\lambda$ expansion.}
\label{nonlambda2}
\end{center}
\end{figure}

The diagrams drawn in \figdisp{nonFeynman} are both valid $\lambda$ expansion diagrams. However, as will be shown below, the allowed topologies of $\lambda$ expansion diagrams do not include all of the possible ways of combining the two elements in \figdisp{elements}, but rather only a subset of these. To see which subset, consider the plausible diagram displayed in Fig. (\ref{nonlambda}a), which is not an allowed $\lambda$ expansion diagram. This diagram is obtained from the Fock diagram in Fig. (\ref{nonlambda}b) by adding a Fermi loop to the latter. The numerical representation for the diagram in Fig. (\ref{nonlambda}b) is 
\beq
1 \ _ 1 \ f. \nn
\eeq
We see that the point from which the first Fermi loop emanates in Fig. (\ref{nonlambda}a) is represented by a subscript in Fig. (\ref{nonlambda}b). Alternatively, using the terminology introduced in section \ref{numFeynman}, the first Fermi loop is spawned by the interaction vertex 1 of the zeroth Fermi loop. However, the other side of this interaction vertex is not free, but rather connects to the zeroth Fermi loop itself. This is not allowed. In fact, we shall find below that when there are more than one pair of lines connected to a single point of an interaction vertex, each pair must both start and terminate a Fermi loop at that point, as in Fig. (\ref{nonFeynman}b). Another diagram which is not an allowed $\lambda$ expansion diagram is drawn in \figdisp{nonlambda2}.

\begin{figure}[H]
\begin{center}
\includegraphics{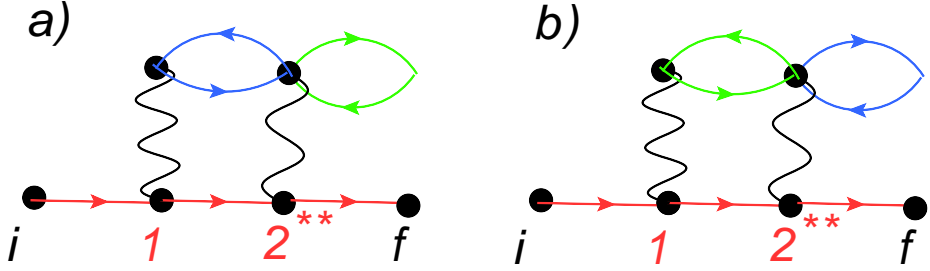}
\caption{A demonstration that unlike Feynman diagrams, $\lambda$ expansion diagrams are not in one-to-one correspondence with their numerical representations. In the diagram in panel a), the interaction vertex 1 of the zeroth Fermi loop also connects to the first Fermi loop. In the diagram in panel b), it connects to the second Fermi loop. The topologies of both diagrams, however, are identical. }
\label{notonetoone}
\end{center}
\end{figure}

Another feature of the $\lambda$ expansion diagrams is that they are not in one-to-one correspondence with their numerical representations. To see this, consider the diagrams drawn in \figdisp{notonetoone}. As usual, the zeroth, first, and second Fermi loops are drawn in red, blue, and green respectively. The diagram in (Fig. \ref{notonetoone}a) has the numerical representation
\barray
&& 1 \ 2^{**} \ f \nn\\
2^{**}: && 0 \ _1 \ f^*  \nn\\
(2,f)^*: && 0 \ f.  \nn
\earray
In words, this says that the interaction vertex $2$ of the zeroth Fermi loop spawns the first Fermi loop. The interaction vertex $1$ of the zeroth Fermi loop connects to the first Fermi loop. Finally, the terminal point of the first Fermi loop spawns the second Fermi loop. On the other hand, the diagram in (Fig. \ref{notonetoone}b) has the numerical representation
\barray
&& 1 \ 2^{**} \ f \nn\\
2^{**}: && 0 \ f^*  \nn\\
(2,f)^*: && 0 \ _1 \ f.  \nn
\earray
In this case, the interaction vertex $1$ of the zeroth Fermi loop connects to the second Fermi loop rather than the first. We see that both of the above numerical representations lead to the same diagram, although they both have a contribution which must be accounted for.

\begin{figure}[H]
\begin{center}
\includegraphics{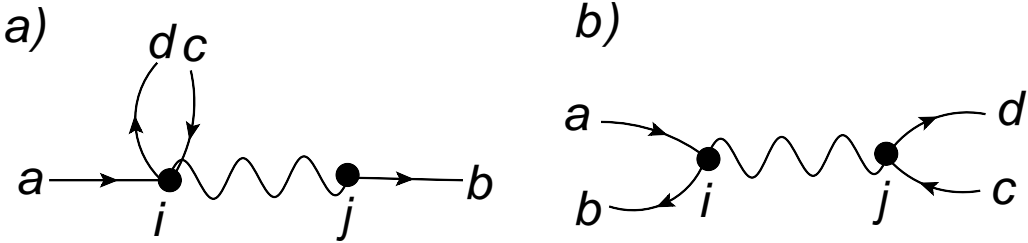}
\caption{$t$-vertices in panel a) versus $J$-vertices in panel b) for the $\lambda$ expansion diagrams which are also Feynman diagrams. The corresponding lines are marked with corresponding letters.}
\label{interactionrealspace}
\end{center}
\end{figure}

\begin{figure}[H]
\begin{center}
\includegraphics{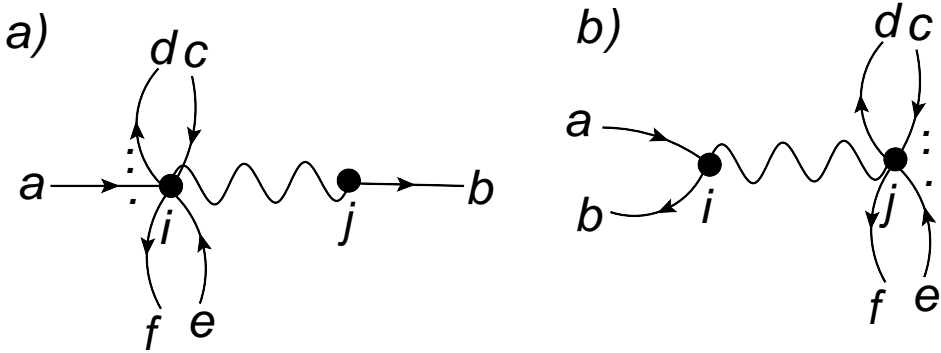}
\caption{$t$-vertices in panel a) versus $J$-vertices in panel b) for the more general $\lambda$ expansion diagrams. The corresponding lines are marked with corresponding letters.}
\label{interactionrealspace2}
\end{center}
\end{figure}

A final point to mention in this discussion of the $\lambda$ expansion diagrams is that when drawing the diagrams in real space, the vertex appropriate for the $t$-interaction differs from the one appropriate for the $J$-interaction. While this is derived rigorously from the EOM below, one can understand it by examining the relevant terms in the Hamiltonian (\disp{tJmodel}). First, we examine the $t$-term. Writing the $X$ operators in terms of canonical creation and destruction operators, we obtain
\beq
 -\sum_{ij\sigma}t_{ij}X_i^{\sigma 0 }X_j^{0\sigma} =  -\sum_{ij\sigma}t_{ij}c_{i\sigma}^\dagger(1-n_{i\sib})c_{j\si} =  -\sum_{ij\sigma}t_{ij}c_{i\sigma}^\dagger c_{j\si} + \sum_{ij\sigma}t_{ij}c_{i\sigma}^\dagger n_{i\sib}c_{j\si}.
\label{tterm}
\eeq
Here, we have used the non-Hermitean mapping described in  \refdisp{Shastry-AOP}
\beq
X_i^{\sigma 0 } \to c_{i\sigma}^\dagger(1-n_{i\sib});\;\;\;\;\; X_j^{0\sigma}\to c_{j\si}.
\eeq 
As discussed in \refdisp{Shastry-AOP}, it is permissible to drop the projection from the destruction operator, since if the system starts in the subspace of no double occupancy, the unprojected destruction operator cannot take it out of this subspace. The second term on the RHS of \disp{tterm} can be represented with the interaction vertex drawn in Fig. (\ref{interactionrealspace}a). Next, we examine the $J$ term. Since a spin flip operator or number operator cannot take the system out of the subspace of no double occupancy, the $X$ operators in the $J$ term can be replaced by their canonical counterparts. Therefore, written in terms of canonical operators, the $J$ term looks like
\beq
\frac{1}{2}\sum_{ij\sigma}J_{ij}n_{i\sigma} + \frac{1}{4}\sum_{ij\sigma_1\sigma_2}\{J_{ij}c^\dagger_{i\si_1}c_{i\si_2}c^\dagger_{j\si_2}c_{j\si_1} -n_{i\si_1}n_{j\si_2}\}.
\eeq
The first term amounts to a shift in the chemical potential $\mu$, while the second one leads to the interaction vertex drawn in Fig. (\ref{interactionrealspace}b). The corresponding lines between Figs. (\ref{interactionrealspace}a) and (\ref{interactionrealspace}b) have been marked with corresponding letters. Throughout the text, we shall sometimes use the term ``Feynman diagrams" to refer to the $\lambda$ expansion diagrams formed solely from the interaction vertices in \figdisp{interactionrealspace}, and sometimes to refer to the usual Feynman diagrams \cite{Nozieres,FW}. It should be clear what we mean from the context. To obtain the more general $\lambda$ expansion diagrams, one must use the vertices drawn in \figdisp{interactionrealspace2}. Once again, the corresponding lines have been marked with corresponding letters between the $t$-vertices in panel a) and the $J$-vertices in panel b).
A $\lambda$ expansion diagram drawn in real space will of course have a mix of $t$-vertices and $J$-vertices. Luckily, when drawing the diagrams in momentum space, we can use only one type of vertex ($t$ or $J$). The details of this procedure are discussed in section \ref{J}. To convert between ``$t$-diagrams" and ``$J$-diagrams", we must rearrange every interaction vertex as indicated in \figdisp{interactionrealspace2}. For example, the Hartree and Fock diagrams, when drawn using $t$-vertices, appear as in Fig. (\ref{G1}b) and Fig. (\ref{G1}c) respectively. In this introductory section, we have used the more familiar $J$-vertices to construct our diagrams, while in the rest of the paper, we shall take the point of view of using the $t$-vertices. The counterparts of the diagrams drawn in Figs. (\ref{ExFeynman}a), (\ref{ExFeynman}b), (\ref{ExFeynman2}), (\ref{nonFeynman}b), and (\ref{notonetoone}), are drawn in Figs. (\ref{G2} g), (\ref{G2} b), (\ref{G2} c), (\ref{G2} n), and (\ref{3rdorderchifromPsi}g) respectively.

To conclude this preliminary discussion, we point out that while it may be possible to define the $\lambda$ diagrams as all inequivalent ways of combining the elements displayed in \figdisp{elements} with some topological constraints, this definition would not have much practical value. It also would not tell us how to evaluate the diagram once we had drawn it. On the other hand, the numerical representations of the $\lambda$ expansion diagrams defined below are both easy to generate in a systematic manner, and easy to evaluate. In fact, one may argue that even for the standard Feynman diagrams, the definition in terms of the numerical representations presented in section \ref{numFeynman} is more useful than the usual one, since it gives a systematic way of generating, and a compact way of representing the diagrams.
\section{Bare Diagrammatic $\lambda$ expansion for $\G[i,f]$.\label{bare}}

\subsection{Integral equation of motion and the first order $\lambda$ expansion.}
As can be seen from \disp{EOMG}, the parameter $\lambda$ adiabatically connects the free Fermi gas at $\lambda=0$ with the fully projected model at $\lambda=1$. Therefore, in the bare $\lambda$ series for $\G$, to each order in $\lambda$, $\G[i,f]$ is expressed as a functional of the free Fermi gas, $\GH^{(0)}[i,f]$ and the hopping $t_{ij}$. In this section, we aim to derive a set of rules for drawing diagrams to compute the $n^{th}$ order contribution to the bare series for $\G[i,f]$. We  do this by rewriting \disp{EOMG} as an integral equation, and then iterating this equation in $\lambda$. An analogous expansion is done for the first couple of orders of the Feynman series in Kadanoff and Baym in \refdisp{KadanoffBaym}. To this end, we rewrite \disp{EOMG} as
\barray
- \GH^{-1(0)}_{\si_1,\si_j}[i,\bb{j}] \G_{\si_j,\si_2}[\bb{j},f] = -\delta[i,f]\delta_{\si_1,\si_2} +\lambda \times\delta[i,f]\gamma_{\si_1,\si_2}[i] - \lambda\times t[i,\bb{j}] \gamma_{\si_1,\si_a}[i] \G_{\si_a,\si_2}[\bb{j},f] +\lambda\times t[i,\bb{j}]\si_1\si_a \frac{\delta}{\delta\V_i^{\bb{\sib_1},\bb{\sib_a}}} \G_{\si_a,\si_2}[\bb{j},f],\nn\\
\label{EOMG2} 
\earray
where $\GH^{-1(0)}[i,f]$, the inverse of the free Fermi gas Green's function is obtained by setting $\lambda=0$ in \disp{set1}.
Rewriting \disp{EOMG2}, we obtain the following integral equation for $\G[i,f]$.
\barray
\G_{\si_1,\si_2}[i,f] &=& \GH^{(0)}_{\si_1,\si_2}[i,f] -\lambda \ \GH^{(0)}_{\si_1,\si_b}[i,f]\si_b\si_2\G_{\sib_2,\sib_b}[f,f^+]\nn\\
&&  -\lambda \times \GH^{(0)}_{\si_1,\si_b}[i,\bb{k}] \left(- t[\bb{k},\bb{j}] \si_b\si_a\G_{\sib_a,\sib_b}[\bb{k},\bb{k}^+]\G_{\si_a,\si_2}[\bb{j},f] + t[\bb{k},\bb{j}]\si_b\si_a \frac{\delta}{\delta\V_\bb{k}^{\bb{\sib_b},\bb{\sib_a}}} \G_{\si_a,\si_2}[\bb{j},f]\right),\nn\\
\label{EOMG3}
\earray
This expression has considerable parallels to a similar expression for the (canonical) Hubbard model, with one exception, the second term on the RHS, (arising from the non-canonical nature of the $X$'s) has no counterpart in the canonical theory. If we drop this term, the series so generated is exactly the Feynman series.

\begin{figure}[H]
\begin{center}
\includegraphics{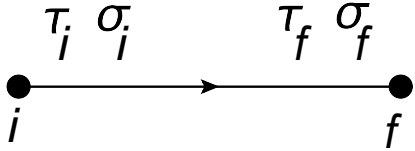}
\caption{The zeroth order contribution to the Green's function: $\GH^{(0)}[i,f]$}
\label{g0}
\end{center}
\end{figure}

We now proceed to draw the diagrams for the zeroth and first order contributions to $\G$. The zeroth order contribution to the Green's function, which is given by the free Fermi gas $\GH^{(0)}[i,f]$, is represented by the diagram in \figdisp{g0}.

\begin{figure}[H]
\begin{center}
\includegraphics[width=.75\columnwidth]{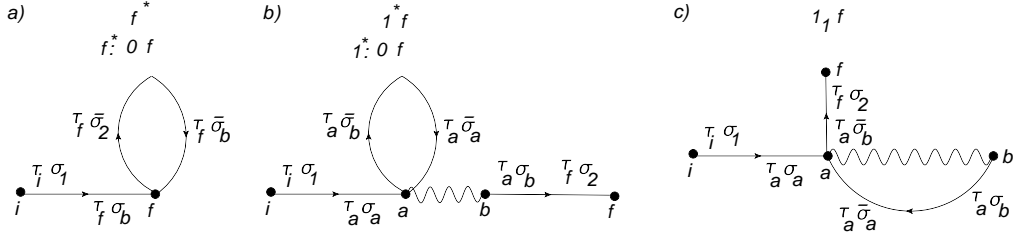}
\caption{The first order contribution to the Green's function: $\G^{(1)}[i,f]$. The diagrams in panels a), b), and c) come from the first, second, and third terms on the RHS of \disp{EOMG3}, respectively.}
\label{G1}
\end{center}
\end{figure}

To obtain the first order contribution to $\G[i,f]$, we plug $\GH^{(0)}[i,f]$ in for $\G[i,f]$ in the RHS of \disp{EOMG3}. This leads to the three diagrams displayed in \figdisp{G1}. The diagrams a), b), and c) in \figdisp{G1} correspond to the three terms in the parenthesis on the RHS of \disp{EOMG3} respectively. They correspond to the analytical expressions a): $- \lambda \si_b\si_2\GH^{(0)}_{\substack{if\\{\sigma_1\sigma_b}}} [\tau_i,\tau_f]\GH^{(0)}_{\substack{ff\\{\sib_2\sib_b}}} [\tau_f,\tau_f^+]$; b): $\lambda \si_a\si_b\GH^{(0)}_{\substack{ia\\{\si_1\si_a}}} [\tau_i,\tau_a] \GH^{(0)}_{\substack{aa\\{\sib_b\sib_a}}} [\tau_a,\tau_a^+]t_{ab}\GH^{(0)}_{\substack{bf\\{\si_b\si_2}}} [\tau_a,\tau_f]$; and c): $- \lambda \si_a \si_b \GH^{(0)}_{\substack{ia\\{\si_1\si_a}}} [\tau_i,\tau_a] t_{ab} \GH^{(0)}_{\substack{ba\\{\si_b\sib_a}}} [\tau_a,\tau_a^+]\GH^{(0)}_{\substack{af\\{\sib_b\si_2}}} [\tau_a,\tau_f]$. In drawing the diagram in Fig. (\ref{G1}c), we have used the Schwinger identity
\beq
\frac{\delta \GH^{(0)}_{\si_a,\si_b}[i,f]}{\delta \V_r^{\si_c\si_d}} = -\GH^{(0)}_{\si_a,\si_x}[i,\bb{x}] \frac{\delta \GH^{-1(0)}_{\si_x,\si_y}[\bb{x},\bb{y}]}{\delta \V_r^{\si_c\si_d}}\GH^{(0)}_{\si_y,\si_b}[\bb{y},f] = \GH^{(0)}_{\si_a,\si_c}[i,r] \GH^{(0)}_{\si_d,\si_b}[r,f]. \label{bypass}
\eeq
In other words, the role of the functional derivative in the \disp{EOMG3} is to pick a line in the diagram for $\G_{\si_a\si_2}[\bb{j},f]$, and to split it into two lines, one entering the point $\bb{k}$, and the other one exiting it. 
The reader would recognize that we  bypassed the  Wicks theorem, by utilizing instead the Schwinger identity \disp{bypass}.

\subsection{Rules for calculating the $n^{th}$ order contribution.\label{rules}}
By plugging in the first and zeroth order diagrams into the RHS of \disp{EOMG3}, we can obtain the second order diagrams. Using this iterative process, we can obtain diagrams for $\G$ to any order in $\lambda$. Moreover, by noticing the pattern in the iterative process, we can derive the rules for obtaining the $n^{th}$ order contribution to $\G$ directly without calculating the lower order contributions. In the case of the Feynman diagrams, this is merely an alternate way of deriving the rules obtained from using Wick's theorem. However, in the present case, in which the standard Wick's theorem is not available, this derivation is essential in going from the EOM definition of the $\lambda$ expansion introduced in \refdisp{ECFL} and the equivalent diagrammatic one developed here. We now present the diagrammatic rules for calculating the $n^{th}$ order contribution to $\G$.

\begin{figure}[H]
\begin{center}
\includegraphics[width=.5\columnwidth]{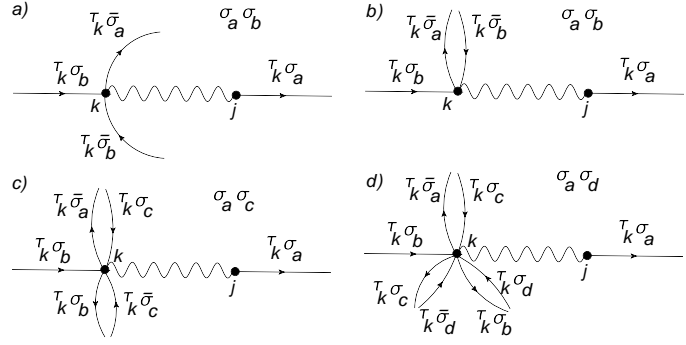}
\caption{Interaction vertices appearing in the diagrams. Panels a), b), c), and d) correspond to 0, 1, 2, and 3 stars on the number representing the interaction vertex, respectively. Note that the lines are broken into pairs based on spin. A pair of two incoming or two outgoing lines share opposite spins, while a pair of one incoming and one outgoing line share the same spin. Moreover, in the case of the former, the spin contributes to the sign of the diagram, while in the case of the latter, it does not. The contribution to the sign is written in the top right of each panel.}
\label{interaction}
\end{center}
\end{figure}

\begin{figure}[H]
\begin{center}
\includegraphics[width=.5\columnwidth]{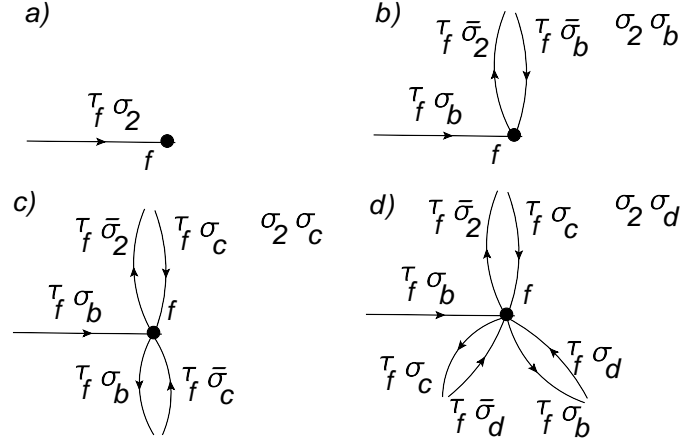}
\caption{Terminal point in the diagram corresponding to the $f$ in the top row. Panels a), b), c), and d) correspond to 0, 1, 2, and 3 stars on the $f$ in the top row, respectively. Same comments regarding spin apply as in \figdisp{interaction}. Note that in the case of one or more stars on the f in the top row, the line labeled by $\sib_2$ is outgoing. This is compensated by the fact there are two more lines entering the point $f$ than exiting it. }
\label{terminalpoint}
\end{center}
\end{figure}

\begin{figure}[H]
\begin{center}
\includegraphics[width=.5\columnwidth]{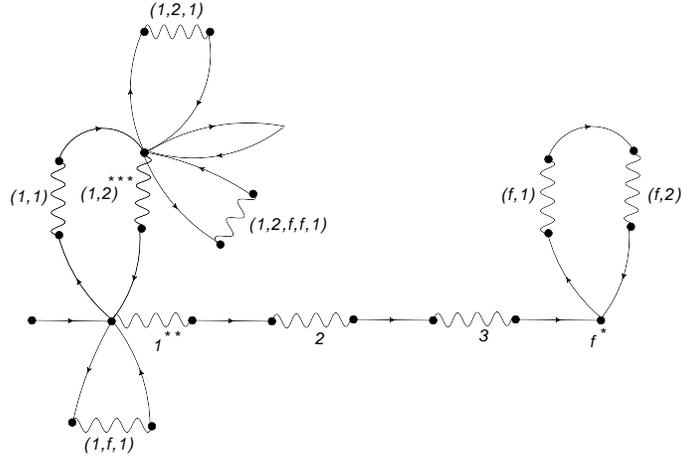}
\caption{Intermediate step in the process of drawing the diagram corresponding to the numerical representation in step 5 of the rules. All of the interaction vertices are drawn in. To complete the diagram, we must split some of the Green's function lines through the unused points in the interaction vertices in a manner indicated by the numerical representation.}
\label{14thorderGexample}
\end{center}
\end{figure}

\begin{itemize}

\item (1) Write a row of consecutive integers followed by the letter $f$, i.e. $ 1 \ldots m \ f$, where $m\geq 0$ (if $m=0$, we simply write $f$), e.g.  
\beq
1 \ 2 \ 3 \ f.
\nn
\eeq
\item(2) Give any number of stars (including no stars) to each these integers (including $f$), e.g.
\beq
1^{**} \ 2 \ 3 \ f^*.\nn
\eeq
\item(3) Each integer (including $f$) with $p$ stars ($p\geq 1$) gives rise to another row of integers which now starts with $0$ (as opposed to $1$), and which ends with an $f$ with $p-1$ stars. Each new row can have any number of integers between the $0$ and the $f$, each of which can have any number of stars, giving rise to further rows. $0$ is not allowed to have any stars, e.g.
\barray
&& 1^{**} \ 2 \ 3 \ f^* \nn\\
1^{**}: && 0 \ 1 \ 2^{***} \ f^* \nn\\
(1,2)^{***}: && 0 \ 1 \ f^{**} \nn\\
(1,2,f)^{**}: && 0 \ f^* \nn\\
(1,2,f,f)^{*}: && 0 \ 1\ f \nn\\
(1,f)^{*}: && 0 \ 1 \  f \nn\\ 
f^*: && 0 \ 1 \  2 \ f.\nn  
\earray
Note that each integer in the above diagram is uniquely specified by a tuple which traces it back to the first row through the starred integers. For example, the number 1 in the fifth row corresponds to the tuple $(1,2,f,f,1)$.

\item(4) Let $\nu$ be the total number of integers without stars excluding $0$'s and $f$'s. Let $s_f$ be the total number of stars on the $f$ in the top row, and let $s$ be the total number of stars excluding those on $f$'s. Then the order $n$ must satisfy the relation $n= \nu + s_f + s$. In the above example, $\nu=8$, $s_f=1$, and $s=5$. Therefore this a $14^{th}$ order diagram.

\item(5) Between any 2 consecutive integers of a row (including $0$'s and $f$'s), one can place as subscripts an ordered list of tuples from the following set: All those corresponding to non-starred integers (except $0$'s and $f$'s) whose tuple can be obtained from the tuple of the smaller of the 2 consecutive integers in question, by taking the first $k\leq l$ entries of this tuple (where $l$ is the length of the tuple), and subtracting a non-negative integer from the last entry. We have taken $f$'s to be integers greater than all other integers in their respective rows. For example, suppose that the two consecutive integers in question are 1 and f in the fifth row of the above diagram. Then all integers eligible to be used as subscripts between them are: $(1,2,f,f,1)$, $(1,2,1)$, and $(1,1)$.  All non-starred integers (except $0$'s and $f$'s) must be used exactly once in this way. e.g.
\barray
&& 1^{**} \ 2 \ 3 \ _{3 \ 2} \ f^* \nn\\
1^{**}: && 0 \ 1 \ 2^{***} \ f^* \nn\\
(1,2)^{***}: && 0 \ 1 \ f^{**} \nn\\
(1,2,f)^{**}: && 0 \ _{(1,2,1)} \ f^* \nn\\
(1,2,f,f)^{*}: && 0 \ 1 \ _{(1,2,f,f,1)}\ f \nn\\
(1,f)^{*}: && 0 \ 1\ _{(1,f,1) \ (1,1) } \  f \nn\\ 
f^*: && 0 \ 1 \  2 \ _{(f,1) \ (f,2)}\ f. \nn  
\earray

\item(6) We use the numerical representation to draw the diagram in the following way. Each integer excluding $0$'s and $f$'s corresponds to an interaction vertex shown in \figdisp{interaction}. The interaction vertices displayed in panels a), b), c), and d) correspond to 0, 1, 2, and 3 stars respectively on the integer in question. On the top right of each panel, we indicate how the spins contribute to the sign of the diagram. Note that when two outgoing or two incoming lines share the same spin label, this spin contributes to the sign of the diagram, while when an outgoing and an incoming line share the same spin label, this spin does not contribute to the sign of the diagram. For example, in panel d) $\si_a$ and $\si_d$ contribute to the sign while $\si_b$ and $\si_c$ do not.

The $f$ in the top row corresponds to a terminal point shown in \figdisp{terminalpoint}. The terminal points displayed in panels a), b), c), and d) correspond to 0, 1, 2, and 3 stars respectively on the $f$ in the top row. On the top right of each panel, we indicate how the spins contribute to the sign of the diagram. Note that the same general rule holds as in the case of the interaction vertices, except that now for the case of 1 or more stars, the spin $\si_2$ also contributes to the sign of the diagram. For the case of one or more stars, one can obtain the terminal points in \figdisp{terminalpoint} from the interaction vertices in \figdisp{interaction} by removing the interaction line and the Green's function line to the right of it, and making the substitution $\si_a\to \si_2$. The interaction vertices displayed in \figdisp{interaction} and the terminal points displayed in \figdisp{terminalpoint} continue to follow the same pattern for greater than three stars.

To actually draw the diagram, let us momentarily ignore the subscripts in our numerical representation, and correspondingly the Green's function lines labeled by $\sib_a$ and $\sib_b$ in panel a) of \figdisp{interaction}, (the case of $0$ stars). Then the top row of the numerical representation corresponds to a chain of interaction lines connected to each other by Green's function lines  running from the point $i$ to the point $f$. The lower rows also correspond to a similar chain running from a single point back to itself. This is the point $k$ (displayed in panels b), c), and d) in \figdisp{interaction}) on the interaction vertex corresponding to the starred integer which gives rise to this lower row. Thus, the number of such chains beginning and ending at a point of a particular interaction vertex is equal to the number of stars on the starred integer which corresponds to this vertex. For the example given above, following this procedure yields the intermediate diagram displayed in \figdisp{14thorderGexample}.

Finally, to put the subscripts back into the diagram, we break each chain at any place where there are subscripts between two consecutive vertices of the chain, and pass the chain through the (non-starred) vertices indicated by the subscripts in the order in which they are written, after which it resumes its original course. This is accomplished with the help of the two Green's function lines labeled by $\sib_a$ and $\sib_b$ on the non-starred vertices, (displayed in panel a) of \figdisp{interaction}),which were ignored in drawing the intermediate diagram in \figdisp{14thorderGexample}. The final diagram is displayed in \figdisp{14thorderGfinal}. 

Note that when drawing a vertex (or terminal point) with multiple stars, such as that displayed in Fig. \ref{interaction}d), the lines $\sib_a$ and $\si_c$ (incoming) correspond to the row with 2 stars on its $f$, the lines $\si_b$ (outgoing) and $\si_d$ correspond to the row with 1 star on its $f$, and the lines $\si_c$ (outgoing) and $\sib_d$ correspond to the row with 0 stars on its $f$. Therefore, in \figdisp{14thorderGfinal}, on the point $k$ corresponding to the vertex $(1,2)^{***}$, the lines $\sib_m$ and $\si_n$ (incoming) are part of the row $(1,2)^{***}$ ($3^{rd}$ row) in the numerical representation, the lines $\si_l$ (outgoing) and $\si_o$ are part of the row $(1,2,f)^{**}$ ($4^{th}$ row) in the numerical representation, and the lines $\si_n$ (outgoing) and $\sib_o$ are part of the row $(1,2,f,f)^*$ ($5^{th}$ row) in the numerical representation.
\item(7) Each solid line in the diagram contributes a non-interacting Green's function, each wavy line contributes a hopping matrix element. An equal-time Green's function is always taken to be $\GH^{(0)}(\tau,\tau^+)$, i.e. the incoming (creation) line is given the greater time.
\item(8) The total sign of the diagram is given by $(-1)^n(-1)^s(-1)^{s_f-1} \times (\text{sign from the spins}) $, where in the case of $s_f=0$, $(-1)^{s_f-1}\equiv1$, and the way in which the spins contribute to the sign is indicated Figs. (\ref{interaction}) and (\ref{terminalpoint}). Therefore, the diagram in \figdisp{14thorderGfinal} has sign $(-1)^{14}(-1)^5(-1)^0(\text{sign from the spins})=-\si_b\si_h\si_c\si_d\si_e\si_g\si_v\si_2\si_w\si_x\si_y\si_z\si_t\si_u\si_j\si_k\si_m\si_o\si_p\si_q\si_r\si_s$.
\item(9) Sum over internal sites and spins, and integrate over internal times.   
\end{itemize}

According to the above rules, the contribution of the diagram drawn in \figdisp{14thorderGfinal} is 
\barray
&&-\si_b\si_h\si_c\si_d\si_e\si_g\si_v\si_2\si_w\si_x\si_y\si_z\si_t\si_u\si_j\si_k\si_m\si_o\si_p\si_q\si_r\si_s\GH^{(0)}_{\substack{ia\\{\si_1\si_a}}} [\tau_i,\tau_a]t_{ab}\GH^{(0)}_{\substack{bc\\{\si_b\si_c}}} [\tau_a,\tau_b]t_{cd}\GH^{(0)}_{\substack{de\\{\si_d\si_e}}} [\tau_b,\tau_c]t_{eg}\GH^{(0)}_{\substack{ge\\{\si_g\sib_e}}} [\tau_c,\tau_c^+]\GH^{(0)}_{\substack{ec\\{\sib_g\sib_c}}} [\tau_c,\tau_b]\nn\\ 
&& \GH^{(0)}_{\substack{cf\\{\sib_d\si_v}}} [\tau_b,\tau_f]\GH^{(0)}_{\substack{ah\\{\sib_b\sib_j}}} [\tau_a,\tau_d]t_{hj}\GH^{(0)}_{\substack{jk\\{\si_k\si_l}}} [\tau_d,\tau_l]t_{kl}\GH^{(0)}_{\substack{la\\{\si_m\si_n}}} [\tau_l,\tau_a]\GH^{(0)}_{\substack{ko\\{\si_e\sib_p}}} [\tau_e,\tau_g]t_{op}\GH^{(0)}_{\substack{pk\\{\si_q\si_n}}} [\tau_g,\tau_e]\GH^{(0)}_{\substack{ko\\{\sib_m\si_p}}} [\tau_e,\tau_g]\GH^{(0)}_{\substack{ok\\{\sib_q\si_o}}} [\tau_g,\tau_e]\GH^{(0)}_{\substack{kq\\{\si_n\si_r}}} [\tau_e,\tau_r]t_{qr}\nn\\
&&\GH^{(0)}_{\substack{rq\\{\si_s\sib_r}}} [\tau_r,\tau_r^+]
\GH^{(0)}_{\substack{qk\\{\sib_s\sib_o}}} [\tau_r,\tau_e]\GH^{(0)}_{\substack{am\\{\si_a\si_t}}} [\tau_a,\tau_s]t_{mn}\GH^{(0)}_{\substack{nm\\{\si_u\sib_t}}} [\tau_s,\tau_s^+]\GH^{(0)}_{\substack{mh\\{\sib_u\sib_j}}} [\tau_s,\tau_d]\GH^{(0)}_{\substack{ha\\{\sib_k\sib_h}}} [\tau_d,\tau_a]\GH^{(0)}_{\substack{fs\\{\sib_2\si_w}}} [\tau_f,\tau_u]t_{st}\GH^{(0)}_{\substack{tu\\{\si_x\si_y}}} [\tau_u,\tau_v]t_{uv}\GH^{(0)}_{\substack{vs\\{\si_z\sib_w}}} [\tau_v,\tau_u]\nn\\
&&\GH^{(0)}_{\substack{su\\{\sib_x\sib_y}}} [\tau_u,\tau_v]\GH^{(0)}_{\substack{uf\\{\sib_z\sib_v}}} [\tau_v,\tau_f].\nn
\earray

Upon turning off the sources, the Green's functions become spin diagonal, i.e. $\GH^{(0)}_{\si_1\si_2}[i,f]=\delta_{\si_1\si_2}\GH^{(0)}_{\uparrow\uparrow}[i,f]=\delta_{\si_1\si_2}\GH^{(0)}_{\downarrow\downarrow}[i,f]\equiv\delta_{\si_1\si_2}\GH^{(0)}[i,f]$. This allows one to evaluate the spin sum and the sign of the above expression. A good way to evaluate the spin sum is to break the diagram into spin loops in the following manner. Recall that at each interaction vertex and at the terminal point, lines are paired according to spin. They share the same spin if one is incoming and the other is outgoing, and they have opposite spins if both lines of the pair are incoming or both are outgoing. Starting with the line exiting $i$, follow the path of Green's function lines created by the spin pairings until you reach the line labeled by $\si_2$ (or $\sib_2$ if $f$ has one or more stars). These spins are all set by the value of $\si_1=\si_2$, and therefore this is the zeroth spin loop. If not all of the lines have been used up by the zeroth loop, find a random line and follow the path created by the spin pairings to reach the line to which it is paired. This is the first spin loop, etc. Continue to do this until you have used up all of the lines in the diagram. Let $F_s$ denote the number of spin loops in the diagram. Then, the spin sum is $2^{F_s}$. We emphasize that unlike the case of the standard Feynman diagrams, the spin loops of the $\lambda$ expansion diagrams do not coincide with the Fermi loops (where each row of the numerical representation can be thought of as a Fermi loop). 

To determine the sign of the diagram, assign values to the spins in a manner consistent with the spin loops (i.e. the value of any one spin in the spin loop determines the values of all of them). Then, plug these values into the analytical expression for the diagram. It is important to note that the reason we can compute the spin sum and the sign independently, is that the choice we make for the values of the spins does not affect the sign of the diagram. To see this note that every spin loop consists of an even number of pairs that have either two incoming lines or two outgoing lines (since it has an equal number of each kind), and an arbitrary number which have one incoming line and one outgoing line. However, only the former contributes to the sign, while the latter does not (see Figs. \ref{interaction} and \ref{terminalpoint}.) Moreover, each pair contributes a distinct spin and appears in exactly one spin loop. Therefore, by flipping all of the spins in a spin loop, we flip an even number of spins, and therefore do not change the sign of the diagram. The only exception to this line of reasoning is the zeroth spin loop, in the case when the terminal point $f$ has $1$ or more stars (see \figdisp{terminalpoint}). In this case, the zeroth spin loop must have one more pair where both lines are incoming than it has pairs where both lines are outgoing. This is due to the fact that in this case both the spins $\si_1$ and $\sib_2$ exit the sites $i$ and $f$ respectively. It is also consistent with the fact that the terminal point $f$ now has one more pair with two incoming lines than two outgoing lines. Therefore, the spin pairs in the zeroth spin loop now contribute an odd number of spins. However, the spin $\si_2$ from the zeroth spin loop now also appears explicitly in the sign. Therefore, flipping all of the spins in the zeroth loop once again does not change the sign of the diagram. In \figdisp{14thorderGfinal}, we find
\barray
&&(\si_1)=(\si_a)=\si_t=\sib_u=\sib_j=\si_b=\si_c=\si_g=\sib_e=\sib_d=\si_v=\si_z=\sib_w=\si_2; \;\;\;\; \si_x= \si_y; \nn\\
&& \si_h=\si_k=(\si_l)=\sib_p=\si_m; \;\;\;\; \si_q=\sib_o=\sib_s=\si_r=(\si_n),
\nn
\earray
where the parenthesis indicates that the spin does not contribute to the sign of the diagram. Therefore, $F_s=3$. The loops contribute $(-1)^{5+0+1+2}=(-1)^8=1$ to the sign. Therefore, the final contribution of the diagram in \figdisp{14thorderGfinal} is
\barray
&&-8 \times \GH^{(0)}_{\substack{ia}} [\tau_i,\tau_a]t_{ab}\GH^{(0)}_{\substack{bc}} [\tau_a,\tau_b]t_{cd}\GH^{(0)}_{\substack{de}} [\tau_b,\tau_c]t_{eg}\GH^{(0)}_{\substack{ge}} [\tau_c,\tau_c^+]\GH^{(0)}_{\substack{ec}} [\tau_c,\tau_b]\GH^{(0)}_{\substack{cf}} [\tau_b,\tau_f]\GH^{(0)}_{\substack{ah}} [\tau_a,\tau_d]t_{hj}\GH^{(0)}_{\substack{jk}} [\tau_d,\tau_l]t_{kl}\nn\\ 
&&\GH^{(0)}_{\substack{la}} [\tau_l,\tau_a]\GH^{(0)}_{\substack{ko}} [\tau_e,\tau_g]t_{op}\GH^{(0)}_{\substack{pk}} [\tau_g,\tau_e]\GH^{(0)}_{\substack{ko}} [\tau_e,\tau_g]\GH^{(0)}_{\substack{ok}} [\tau_g,\tau_e]\GH^{(0)}_{\substack{kq}} [\tau_e,\tau_r]t_{qr}\GH^{(0)}_{\substack{rq}} [\tau_r,\tau_r^+]\GH^{(0)}_{\substack{qk}} [\tau_r,\tau_e]\GH^{(0)}_{\substack{am}} [\tau_a,\tau_s]\nn\\
&&t_{mn}\GH^{(0)}_{\substack{nm}} [\tau_s,\tau_s^+]\GH^{(0)}_{\substack{mh}} [\tau_s,\tau_d]\GH^{(0)}_{\substack{ha}} [\tau_d,\tau_a]\GH^{(0)}_{\substack{fs}} [\tau_f,\tau_u]t_{st}\GH^{(0)}_{\substack{tu}} [\tau_u,\tau_v]t_{uv}\GH^{(0)}_{\substack{vs}} [\tau_v,\tau_u]\GH^{(0)}_{\substack{su}} [\tau_u,\tau_v]\GH^{(0)}_{\substack{uf}} [\tau_v,\tau_f],\nn
\earray
where all sites and times other than $i$ and $f$, and $\tau_i$ and $\tau_f$ are summed/integrated over.

\begin{figure}[H]
\begin{center}
\includegraphics[width=.5\columnwidth]{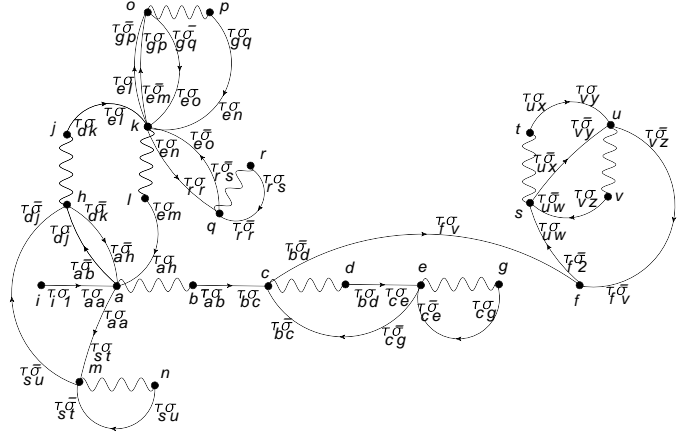}
\caption{Diagram corresponding to the numerical representation in step 5 of the rules.}
\label{14thorderGfinal}
\end{center}
\end{figure}

\subsection{Second order contribution}

\begin{figure}[H]
\begin{center}
\includegraphics[angle=270, scale = 1.4]{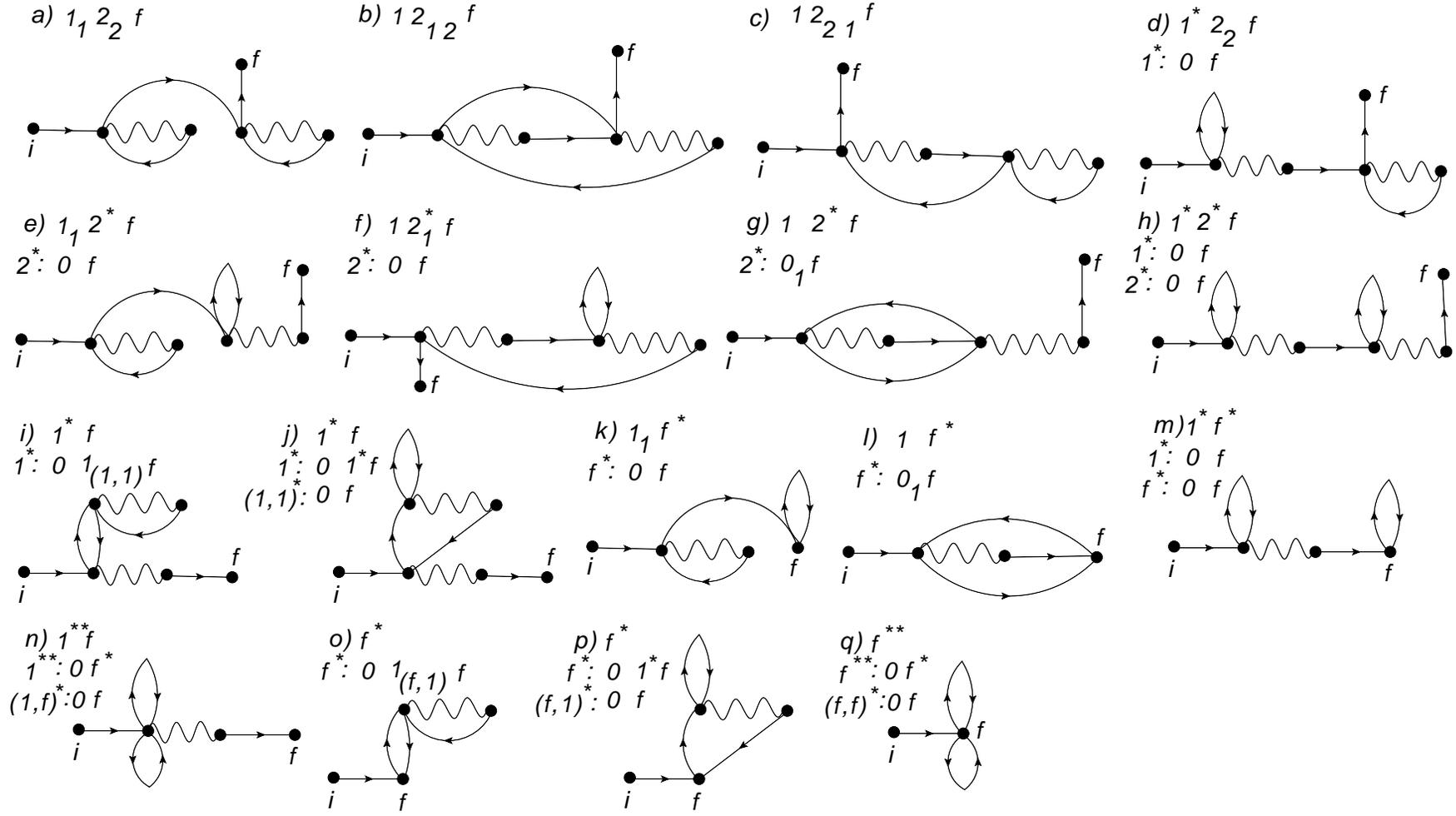}
\caption{The second order diagrams contributing to the Green's function: $\G^{(2)}[i,f]$, and their corresponding numerical representations. Note that the diagrams a) through j) are the standard second order Feynman diagrams. The other diagrams are not.}
\label{G2}
\end{center}
\end{figure}

Using the rules from section \ref{rules}, we draw the diagrams that contribute to $\G[i,f]$ in second order in \figdisp{G2}, and calculate their contributions below.

The contributions of these diagrams are 
\barray
&& a) \ \GH^{(0)}_{ia}[\tau_i,\tau_a]t_{ab}\GH^{(0)}_{ba}[\tau_a,\tau_a^+]\GH^{(0)}_{ac}[\tau_a,\tau_b]t_{cd}\GH^{(0)}_{dc}[\tau_b,\tau_b^+]\GH^{(0)}_{cf}[\tau_b,\tau_f]\nn\\
&& b) \ -2\GH^{(0)}_{ia}[\tau_i,\tau_a]t_{ab}\GH^{(0)}_{bc}[\tau_a,\tau_c]t_{cd}\GH^{(0)}_{da}[\tau_c,\tau_a]\GH^{(0)}_{ac}[\tau_a,\tau_c]\GH^{(0)}_{cf}[\tau_c,\tau_f]\nn\\
&& c) \ \GH^{(0)}_{ia}[\tau_i,\tau_a]t_{ab}\GH^{(0)}_{bc}[\tau_a,\tau_b]t_{cd}\GH^{(0)}_{dc}[\tau_b,\tau_b^+]\GH^{(0)}_{ca}[\tau_b,\tau_a]\GH^{(0)}_{af}[\tau_a,\tau_f]\nn\\
&& d) \ \GH^{(0)}_{ia}[\tau_i,\tau_a]\GH^{(0)}_{aa}[\tau_a,\tau_a^+]t_{ab}\GH^{(0)}_{bc}[\tau_a,\tau_b]t_{cd}\GH^{(0)}_{dc}[\tau_b,\tau_b^+]\GH^{(0)}_{cf}[\tau_b,\tau_f]\nn\\
&& e) \ \GH^{(0)}_{ia}[\tau_i,\tau_a]t_{ab}\GH^{(0)}_{ba}[\tau_a,\tau_a^+]\GH^{(0)}_{ac}[\tau_a,\tau_b]\GH^{(0)}_{cc}[\tau_b,\tau_b^+]t_{cd}\GH^{(0)}_{df}[\tau_b,\tau_f]\nn\\
&& f) \ \GH^{(0)}_{ia}[\tau_i,\tau_a]t_{ab}\GH^{(0)}_{bc}[\tau_b,\tau_b^+]t_{cd}\GH^{(0)}_{da}[\tau_b,\tau_a]\GH^{(0)}_{af}[\tau_a,\tau_f]\nn\\
&& g) \ -2\GH^{(0)}_{ia}[\tau_i,\tau_a] t_{ab} \GH^{(0)}_{bc}[\tau_a,\tau_b]\GH^{(0)}_{ca}[\tau_b,\tau_a]\GH^{(0)}_{ac}[\tau_a,\tau_b]t_{cd}\GH^{(0)}_{df}[\tau_b,\tau_f]\nn\\
&& h) \ \GH^{(0)}_{ia}[\tau_i,\tau_a]\GH^{(0)}_{aa}[\tau_a,\tau_a^+]t_{ab}\GH^{(0)}_{bc}[\tau_a,\tau_b]\GH^{(0)}_{cc}[\tau_b,\tau_b^+]t_{cd}\GH^{(0)}_{df}[\tau_b,\tau_f]\nn\\
&& i) \ \GH^{(0)}_{ia}[\tau_i,\tau_a]t_{ab}\GH^{(0)}_{bf}[\tau_a,\tau_f]\GH^{(0)}_{ac}[\tau_a,\tau_b]t_{cd}\GH^{(0)}_{dc}[\tau_b,\tau_b^+]\GH^{(0)}_{ca}[\tau_b,\tau_a]\nn\\
&& j) \ \GH^{(0)}_{ia}[\tau_i,\tau_a]t_{ab}\GH^{(0)}_{bf}[\tau_a,\tau_f]\GH^{(0)}_{ac}[\tau_a,\tau_b]\GH^{(0)}_{cc}[\tau_b,\tau_b^+]t_{cd}\GH^{(0)}_{da}[\tau_b,\tau_a]\nn\\
&& k) \ -\GH^{(0)}_{ia}[\tau_i,\tau_a]t_{ab}\GH^{(0)}_{ba}[\tau_a,\tau_a^+]\GH^{(0)}_{af}[\tau_a,\tau_f]\GH^{(0)}_{ff}[\tau_f,\tau_f^+]\nn\\
&& l) \ 2\GH^{(0)}_{ia}[\tau_i,\tau_a]t_{ab}\GH^{(0)}_{bf}[\tau_a,\tau_f]\GH^{(0)}_{fa}[\tau_f,\tau_a]\GH^{(0)}_{af}[\tau_a,\tau_f] \nn\\
&& m) \ -\GH^{(0)}_{ia}[\tau_i,\tau_a]\GH^{(0)}_{aa}[\tau_a,\tau_a^+]t_{ab}\GH^{(0)}_{bf}[\tau_a,\tau_f]\GH^{(0)}_{ff}[\tau_f,\tau_f^+]\nn\\
&& n) \ -\GH^{(0)}_{ia}[\tau_i,\tau_a]\GH^{(0)}_{aa}[\tau_a,\tau_a^+]\GH^{(0)}_{aa}[\tau_a,\tau_a^+]t_{ab}\GH^{(0)}_{bf}[\tau_a,\tau_f]\nn\\
&& o) \ - \GH^{(0)}_{if}[\tau_i,\tau_f]\GH^{(0)}_{fa}[\tau_f,\tau_a]t_{ab}\GH^{(0)}_{ba}[\tau_a,\tau_a^+]\GH^{(0)}_{af}[\tau_a,\tau_f]\nn\\
&& p) \ - \GH^{(0)}_{if}[\tau_i,\tau_f]\GH^{(0)}_{fa}[\tau_f,\tau_a]\GH^{(0)}_{aa}[\tau_a,\tau_a^+]t_{ab}\GH^{(0)}_{bf}[\tau_a,\tau_f]\nn\\
&& q) \ \GH^{(0)}_{if}[\tau_i,\tau_f]\GH^{(0)}_{ff}[\tau_f,\tau_f^+]\GH^{(0)}_{ff}[\tau_f,\tau_f^+]. 
\nn
\earray

\section{Diagrammatic $\lambda$ expansion for constituent objects\label{constituent}}
\subsection{Introduction of the two self-energies\label{sigasigb}}
We next consider the auxiliary Green's function $\GH[i,f]$. Using \disp{set1} for $\GHI[i,f]$, we can write the analog of \disp{EOMG3} for $\GH[i,f]$.

\barray
\GH_{\si_1,\si_2}[i,f] &=& \GH^{(0)}_{\si_1,\si_2}[i,f] \nn\\
&&  -\lambda \ \GH^{(0)}_{\si_1,\si_b}[i,\bb{k}] \left(- t[\bb{k},\bb{j}] \si_b\si_a\G_{\sib_a,\sib_b}[\bb{k},\bb{k}^+]\GH_{\si_a,\si_2}[\bb{j},f] + t[\bb{k},\bb{j}]\si_b\si_a \frac{\delta}{\delta\V_\bb{k}^{\bb{\sib_b},\bb{\sib_a}}} \GH_{\si_a,\si_2}[\bb{j},f]\right),\nn\\
\label{EOMg}
\earray
Comparing the iterative expansion of $\G[i,f]$ through \disp{EOMG3} with that of $\GH[i,f]$ through \disp{EOMg}, we see that the terms in the parenthesis are identical in both expansions. However, the second term on the RHS of \disp{EOMG3}, missing in \disp{EOMg}, allows a Green's function diagram to close on itself in the iterative expansion, merging the initial point $i$ and the terminal point $f$. Such a diagram must necessarily have more than one line connected to its terminal point, and therefore at least one star on the $f$ in the top row. Therefore, the diagrams for $\GH[i,f]$ are the subset of the diagrams for $\G[i,f]$ which have no stars on the $f$ in the top row. In \figdisp{G2}, these are diagrams a) through j), and diagram n). 

We see that in the diagrams for $\GH[i,f]$, the terminal point labeled by $f$ is connected to the rest of the diagram only by a single line. Therefore, it will be possible to describe these diagrams in terms of a Dyson equation, with a Dyson self-energy. This is not the case for the other diagrams in \figdisp{G2} (those which do have a star on the $f$ in the top row), and these diagrams require the introduction of a second-self energy. We now proceed to define these two types of self-energies.

We shall denote the Dyson self-energy for $\GH[i,f]$ by $\Sigma_a$. As is the case in the Feynman diagrams, it is obtained from the diagrams for $\GH[i,f]$ by removing the external line coming in from the point $i$, and the one going out to the point $f$. If a diagram for $\Sigma_a$ can be split into two pieces by cutting a single line, then it is reducible. Otherwise, it is irreducible. Denote the irreducible part of $\Sigma_a$ by $\Sigma_a^*$.

Now consider those diagrams which do have a star on the $f$ in the top row. The second self-energy, $\Sigma_b$, is obtained from these diagrams by removing the external line coming from the point $i$. Once again, if a diagram for $\Sigma_b$ can be split into two pieces by cutting a single line, then it is reducible. Otherwise, it is irreducible. Denote the irreducible part of $\Sigma_b$ by $\Sigma_b^*$.

From the diagrammatic structure of the series, it is clear that $\G[i,f] = \GH[i,f] + \GH[i,\bb{j}].\Sigma_b^*[\bb{j},f]$. Comparing with \disp{product}, we see that $\widetilde{\mu}[i,f] = \delta[i,f] + \Sigma_b^*[i,f]$. Also, from Dyson's equation, we know that 
\newline $\GHI[i,f]=\GH^{-1(0)}[i,f] - \Sigma_a^*[i,f]$. We shall give an independent proof of these formulae starting from the equations of motion for $\GHI$ and $\widetilde{\mu}$ (\disp{set1}) in section \ref{ginmu}.

\subsection{$\GHI$ and $\widetilde{\mu}$ \label{ginmu}}
We shall now prove, starting with the equations of motion in \disp{set1}, the observations already made in section \ref{sigasigb}, that
\beq
\widetilde{\mu}[i,f] = \delta[i,f] + \Sigma_b^*[i,f]; \;\;\;\;\;\;\;\; \GHI[i,f]=\GH^{-1(0)}[i,f] - \Sigma_a^*[i,f].
\label{muginsig}
\eeq
This is equivalent to showing that 
\beq
\Sigma_b^*[i,f] = - \lambda \gamma[i] \delta[i,f] + \lambda \Psi[i,f]; \;\;\;\;\;\;\; \Sigma_a^*[i,f] = \lambda \gamma[i] t[i,f] + \lambda \Phi[i,f].
\eeq
We rewrite the EOM for $\widetilde{\mu}[i,f]$ (\disp{set1}) in expanded form.
\barray
\widetilde{\mu}_{\si_1\si_2}[i,f]&=& (\delta_{\si_1\si_2}- \lambda\si_1\si_2\G_{\sib_2\sib_1}[i,i^+]) \delta[i,f] +\lambda \Psi_{\si_1\si_2}[i,f] \nn\\
\Psi_{\si_1\si_2}[i,f] &=& - t[i, \bb{j}] \ \si_1\si_a \GH_{\si_a,\si_b}[\bb{j}, \bb{n}] \frac{\delta}{\delta \V_ i^{\sib_1\sib_a}}\widetilde{\mu}_{\si_b\si_2} [ \bb{n}, f].
\label{mu}
\earray
We now proceed to prove the first of Eqs. (\ref{muginsig}) using induction in $\lambda$. The lowest order contribution to $\Sigma_b^*[i,f]$ comes from diagram a) in \figdisp{G1}. Removing the incoming external line, we obtain $\Sigma_{\substack{b\\{\si_1\si_2}}}^{*(1)}[i,f]=-\lambda\si_1\si_2\delta[i,f]\GH^{(0)}_{\substack{ff\\{\sib_2\sib_1}}} [\tau_f,\tau_f^+]$. Using \disp{mu} to obtain the first order contribution to $\widetilde{\mu}[i,f]$, we get $\widetilde{\mu}_{\si_1\si_2}^{(1)}[i,f] = - \lambda\si_1\si_2\GH^{(0)}_{\sib_2,\sib_1}[i,i^+] \delta[i,f] $. Clearly, these two are equal, and we have that $\Sigma_{\substack{b\\{\si_1\si_2}}}^{*(1)}[i,f]= \widetilde{\mu}_{\si_1\si_2}^{(1)}[i,f]$.

\begin{figure}[H]
\begin{center}
\includegraphics[width=.25\columnwidth]{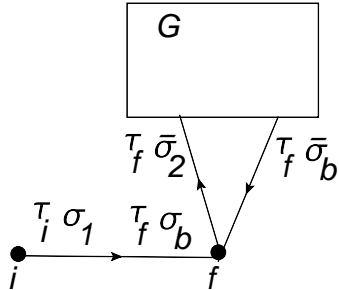}
\caption{Schematic representation for a Green's function diagram with only the number $f^{*\ldots *}$ in the top row. Upon removing the incoming external line, it contributes to $\Sigma_{b1}^*[i,f]$.}
\label{SigbdiagH}
\end{center}
\end{figure}

Now consider the $m^{th}$ order contribution $\Sigma_{\substack{b\\{\si_1\si_2}}}^{*(m)}[i,f]$. This will be obtained from the corresponding $m^{th}$ order $\G[i,f]$ diagram upon dropping the incoming external line. If in the numerical representation for this $\G[i,f]$ diagram, there are no numbers other than $f^{*\ldots *}$ in the top row (e.g. panels o), p), and q) in \figdisp{G2}), then  the contribution of this diagram to $\G[i,f]$ is $\G^{(m)}_{\si_1\si_2}[i,f] = -\lambda  \ \GH^{(0)}_{\si_1\si_b}[i,f] \si_b\si_2 \G^{(m-1)}_{\sib_2\sib_b}[f,f^+]$ (see \figdisp{SigbdiagH}). The resulting contribution to $\Sigma_b^*$, which we shall denote by $\Sigma_{b1}^*$, is 
\beq 
\Sigma_{\substack{b1\\{\si_1\si_2}}}^{*(m)}[i,f]=-\lambda  \ \si_1\si_2 \G^{(m-1)}_{\sib_2\sib_1}[f,f^+]\delta[i,f].
\label{SigbHartree}
\eeq

\begin{figure}[H]
\begin{center}
\includegraphics[width=.5\columnwidth]{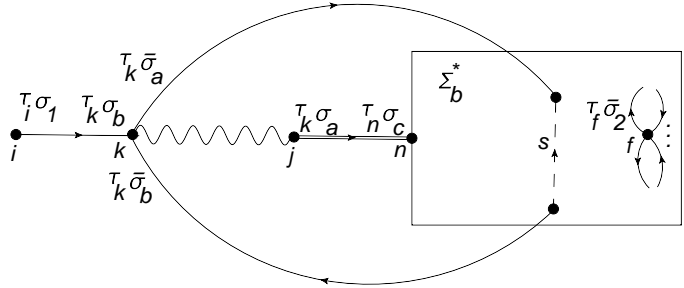}
\caption{Schematic representation for a Green's function diagram whose top row is $1 \ldots f^{*\ldots *}$. Upon removing the incoming external line, it contributes to $\Sigma_{b2}^*[i,f]$.}
\label{Sigb1}
\end{center}
\end{figure}

Alternatively, there are numbers other than $f^{*\ldots *}$ in the top row of the numerical representation of the corresponding $m^{th}$ order $\G[i,f]$ diagram. Then, the top row reads $ 1 \ldots f^{*\ldots *}$  (e.g. panels k) through m) of \figdisp{G2}). In this case, we know that for the resulting $\Sigma_b[i,f]$ diagram to be irreducible, i.e. for it to contribute to $\Sigma_b^*[i,f]$, the number $1$ in the top row should not be starred. Therefore, we can represent the diagram schematically as in \figdisp{Sigb1}. This representation is obtained as follows. If we consider just the part of the diagram between the points $j$ and $f$, we know that a line in this part of the diagram (denoted in \figdisp{Sigb1} by the letter $s$) is split by the point $k$. If we restore $s$ by removing the lines labeled by $\sib_a$ and $\sib_b$ from the point $k$, then the part of the diagram running from $j$ to $f$ is a Green's function diagram which contributes to $\GH[\bb{j},\bb{n}].\Sigma_b^*[\bb{n},f]$ (since the $f$ has at least one star on it). However, the line $s$ can't be contained in the $\GH[\bb{j},\bb{n}]$ part of the diagram (represented in \figdisp{Sigb1} by a double line), since then the resulting $\Sigma_b$ (of the overall diagram) would be reducible. Therefore it must be contained in the $\Sigma_b^*[\bb{n},f]$ part of the diagram. The analytical expression for the diagram in \figdisp{Sigb1} is

\beq
-\lambda \GH^{(0)}_{\si_1\si_b}[i,\bb{k}]t[\bb{k},\bb{j}]\si_b\si_a\GH_{\si_a\si_c}[\bb{j},\bb{n}]\frac{\delta}{\delta \V^{\sib_b\sib_a}_{\bb{k}}}\Sigma^*_{\substack{b\\{\si_c\si_2}}}[\bb{n},f].
\eeq
Removing the incoming external line, and using the inductive hypothesis, we obtain the contribution of these types of diagrams to $\Sigma_b^*[i,f]$, which we shall denote as $\Sigma_{b2}^*[i,f]$.   
\beq
\Sigma^{*(m)}_{\substack{b2\\{\si_1\si_2}}}[i,f]=-\lambda t[i,\bb{j}]\si_1\si_a\GH^{(m_1)}_{\si_a\si_b}[\bb{j},\bb{n}]\frac{\delta}{\delta \V^{\sib_1\sib_a}_{i}}\widetilde{\mu}^{(m_2)}_{\si_b\si_2}[\bb{n},f],
\label{Sigbvertex}
\eeq
where $m=m_1+m_2+1$. Comparing \disp{Sigbvertex} with \disp{mu}, we see that 
\beq \Sigma^*_{\substack{b2\\{\si_1\si_2}}}[i,f] = \lambda \Psi_{\si_1\si_2}[i,f]. \label{PsiSigb2}
\eeq
Combining \disp{SigbHartree} and \disp{PsiSigb2}, we find that
\beq
\Sigma^*_{\substack{b\\{\si_1\si_2}}}[i,f] = \Sigma^*_{\substack{b1\\{\si_1\si_2}}}[i,f]+\Sigma^*_{\substack{b2\\{\si_1\si_2}}}[i,f]=-\lambda  \ \si_1\si_2 \G_{\sib_2\sib_1}[f,f^+]\delta[i,f] +\lambda\Psi_{\si_1\si_2}[i,f].
\label{Sigbstar}
\eeq
Therefore, comparing \disp{Sigbstar} with \disp{mu}, we have shown the first of Eqs. (\ref{muginsig}) to be true.

\begin{figure}[H]
\begin{center}
\includegraphics[width=.25\columnwidth]{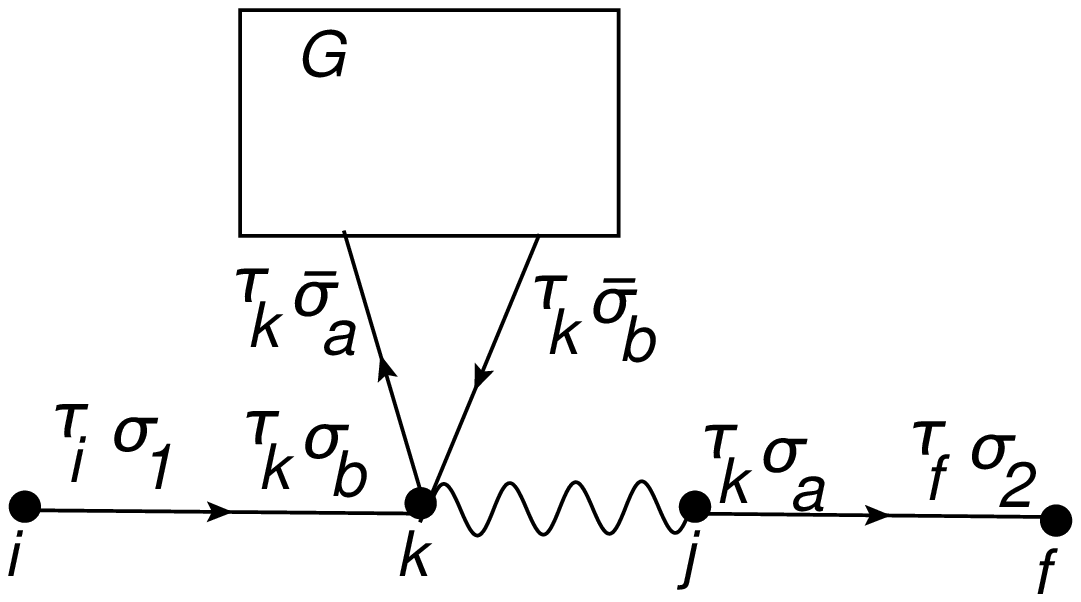}
\caption{Schematic representation for Green's function diagram whose top row is $ 1^{*\ldots*} \ f$. Upon removing the incoming and outgoing external lines, it contributes to $\Sigma_{a1}^*[i,f]$. One can obtain $\Sigma_{b1}^*[i,f]$ displayed in \figdisp{SigbdiagH} by also removing the interaction line exiting the point $k$.}
\label{SigadiagH}
\end{center}
\end{figure}

\begin{figure}[H]
\begin{center}
\includegraphics[width=.5\columnwidth]{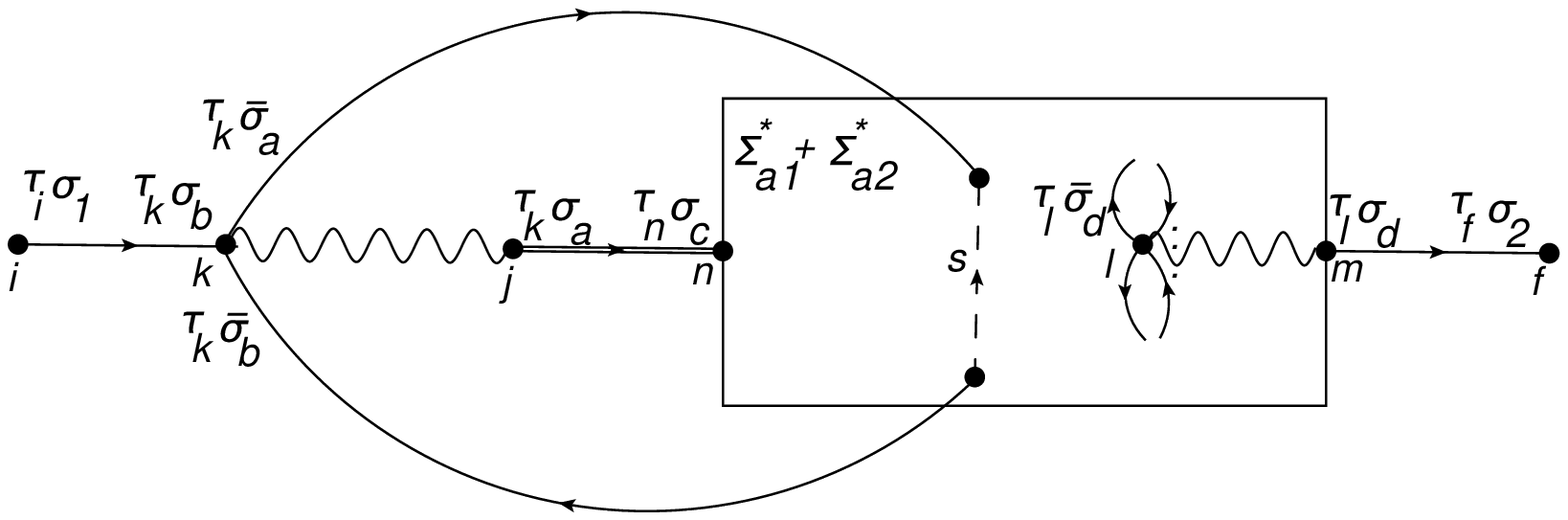}
\caption{Schematic representation for Green's function diagram whose top row is $ 1 \ldots c^{*\ldots*} \ f$. Upon removing the outgoing and incoming external lines, it contributes to $\Sigma_{a2}^*[i,f]$. One can obtain $\Sigma_{b2}^*[i,f]$ displayed in \figdisp{Sigb1} by also removing the interaction line exiting the point $l$.}
\label{Siga1}
\end{center}
\end{figure}

Now consider the EOM for $\GHI[i,f]$ (\disp{set1}) in expanded form.
\barray
\GHI_{\si_1\si_2}[i,f]&=& \GH_{\si_1\si_2}^{-1(0)}[i,f]- \lambda \ t[i,f]\si_1\si_2\G_{\sib_2\sib_1}[i,i^+] -\lambda \Phi_{\si_1\si_2}[i,f] \nn\\
\Phi_{\si_1\si_2}[i,f] &=& t[i, \bb{j}] \ \si_1\si_a \GH_{\si_a,\si_b}[\bb{j}, \bb{n}] \frac{\delta}{\delta \V_ i^{\sib_1\sib_a}}\GHI_{\si_b\si_2} [ \bb{n}, f].
\label{gin}
\earray
Our goal is to prove the second of Eqs. (\ref{muginsig}) using \disp{gin}. To this end, we note that diagrams for $\Sigma_a^*[i,f]$ can be split into four groups. Recall that a diagram for $\Sigma_a[i,f]$ is obtained from a $\GH[i,f]$ diagram (or equivalently from a $\G[i,f]$ diagram with no stars on the $f$ in the top row) by removing the incoming and outgoing external lines. Consider a $\GH[i,f]$ diagrams whose numerical representation has the following property. There are no subscripts between the number immediately to the left of $f$ in the top row (which we shall denote by $c$) and $f$ (e.g. panels e), g) through j), and n) of \figdisp{G2}). This implies that $c$ has at least one star, as otherwise $c$ must be a subscript between $c$ and $f$. Therefore, the top row looks like $ 1 \ldots c^{*\ldots*} \ f$. In the case that $c=1$ (e.g. panels i), j), and n) of \figdisp{G2}), these diagrams can be represented schematically as in \figdisp{SigadiagH}. We denote the corresponding contribution to $\Sigma_a^*$ by $\Sigma_{a1}^*$. If $c>1$ (e.g. panels e), g), and h) of \figdisp{G2}), then the diagrams can be represented schematically as in \figdisp{Siga1}. We denote the corresponding contribution to $\Sigma_a^*$ by $\Sigma_{a2}^*$. Comparing \figdisp{SigbdiagH} with \figdisp{SigadiagH} and \figdisp{Sigb1} with \figdisp{Siga1}, and removing the external lines, we find that 
\beq
\Sigma^*_{\substack{a1\\{\si_1\si_2}}}[i,f] = -\Sigma^*_{\substack{b1\\{\si_1\si_2}}}[i,\bb{j}] t[\bb{j},f]; \;\;\;\;\;\;\;\; \Sigma^*_{\substack{a2\\{\si_1\si_2}}}[i,f] = -\Sigma^*_{\substack{b2\\{\si_1\si_2}}}[i,\bb{j}] t[\bb{j},f].
\eeq
Here, the minus comes from rule (8) of section \ref{rules}, where there is a minus sign discrepancy between the factors $(-1)^{s_f-1}$ (applicable to $f^{*\ldots*}$ in $\Sigma^*_b$) and $(-1)^s$ (applicable to $c^{*\ldots*}$ in $\Sigma^*_a$). Using \disp{SigbHartree} and \disp{PsiSigb2}, we find that
\beq
-\Sigma^*_{\substack{a1\\{\si_1\si_2}}}[i,f]= -\lambda  \ \si_1\si_2 \G_{\sib_2\sib_1}[i,i^+]t[i,f] ;\;\;\;\;\;\;\;\; -\Sigma^*_{\substack{a2\\{\si_1\si_2}}}[i,f]=\lambda\Psi_{\si_1\si_2}[i,\bb{j}]t[\bb{j},f].
\label{Siga1star}
\eeq

Motivated by this observation, we define a new object $\chi_{\si_1\si_2}[i,f]$ defined by the formula
\beq \Phi_{\si_1\si_2}[i,f] = -\Psi_{\si_1\si_2}[i,\bb{j}]t[\bb{j},f] + \chi_{\si_1\si_2}[i,f]
\label{chi}
\eeq
Plugging this formula into \disp{gin}, we obtain
\beq
\GHI_{\si_1\si_2}[i,f]= \GH_{\si_1\si_2}^{-1(0)}[i,f]- \lambda \ t[i,f]\si_1\si_2\G_{\sib_2\sib_1}[i,i^+] +\lambda \Psi_{\si_1\si_2}[i,\bb{j}]t[\bb{j},f] - \lambda \chi_{\si_1\si_2}[i,f]
\label{gin2}
\eeq
Plugging \disp{gin2} into the equation for $\Phi$ (\disp{gin}), we obtain
\beq
\Phi_{\si_1\si_2}[i,f] = -t[i,\bb{j}]\si_1\sib_2 \GH_{\sib_2\sib_1}[\bb{j},i]\delta[i,f]-\Psi_{\si_1\si_2}[i,\bb{j}]t[\bb{j},f] -\lambda t[i,\bb{j}]\si_1\si_a\GH_{\si_a\si_b}[\bb{j},\bb{n}]\frac{\delta}{\delta\V_i^{\sib_1\sib_a}}\chi_{\si_b\si_2}[\bb{n},f],
\label{Phi2}
\eeq
where we have used \disp{mu} to handle the second and third terms on the RHS of \disp{gin2}.
Comparing \disp{Phi2} with \disp{chi}, we obtain the following EOM for $\chi_{\si_1\si_2}[i,f]$.
\beq
\chi_{\si_1\si_2}[i,f] = -t[i,\bb{j}]\si_1\sib_2 \GH_{\sib_2\sib_1}[\bb{j},i]\delta[i,f]-\lambda t[i,\bb{j}]\si_1\si_a\GH_{\si_a\si_b}[\bb{j},\bb{n}]\frac{\delta}{\delta\V_i^{\sib_1\sib_a}}\chi_{\si_b\si_2}[\bb{n},f].
\label{chi2}
\eeq

Comparing \disp{gin2} and \disp{Siga1star}, we see that $-\Sigma^*_{\substack{a1\\{\si_1\si_2}}}[i,f]$ and $-\Sigma^*_{\substack{a2\\{\si_1\si_2}}}[i,f]$ account for the second and third terms on the RHS of \disp{gin2} respectively. Therefore, we now show that the remainder of the $-\Sigma_a^*[i,f]$ diagrams account for the fourth term. To this end, we consider all $\GH[i,f]$ diagrams which do have a subscript between the number $c$ (the number immediately to the left of $f$ in the top row) and $f$ in the top row. The top row now looks like $1 \ldots c _{\ldots} f$ (e.g. panels a) through d) and f) of \figdisp{G2}). Note that for the resulting $\Sigma_a[i,f]$ diagram to be irreducible, the number $1$ cannot have any stars. We further subdivide this group of  $\GH[i,f]$ diagrams into 2 groups. In the first group, whose contribution to $\Sigma_a^*[i,f]$ shall be denoted by $\Sigma_{a3}^*[i,f]$, the subscript immediately preceding $f$ in the top row is $1$. The top row for these diagrams looks like $1 \ldots c _{\ldots 1} f$ (e.g. panels c) and f) of \figdisp{G2}). In the second group, whose contribution to $\Sigma_a^*[i,f]$ shall be denoted by $\Sigma_{a4}^*[i,f]$, the subscript immediately preceding $f$ in the top row is not $1$. The top row for these diagrams looks like $1 \ldots c _{\ldots d} f$, where $d\neq1$ (e.g. panels a), b), and d) of \figdisp{G2}). Our goal is to show that $\lambda \chi_{\si_1\si_2}[i,f] = \Sigma^*_{\substack{a3\\{\si_1\si_2}}}[i,f] + \Sigma^*_{\substack{a4\\{\si_1\si_2}}}[i,f]$. 

\begin{figure}[H]
\begin{center}
\includegraphics[width=.25\columnwidth]{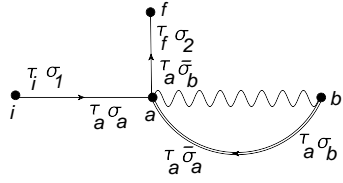}
\caption{Schematic representation for Green's function diagram whose top row is  $1 \ldots c _{\ldots 1} f$. Upon removing the incoming and outgoing external lines, it contributes to $\Sigma_{a3}^*[i,f]$. This is the analog of the Fock diagram in the standard Feynman skeleton expansion.}
\label{SigaFock}
\end{center}
\end{figure}

We do this by induction. The $\GH[i,f]$ diagrams contributing to $\Sigma_{a3}^*[i,f]$ are shown in \figdisp{SigaFock}. The contribution of this diagram becomes
\beq 
- \lambda \si_a \si_b \GH^{(0)}_{\substack{i\bb{a}\\{\si_1\si_a}}} [\tau_i,\tau_\bb{a}] t_{\bb{a}\bb{b}} \GH_{\substack{\bb{b
}\bb{a}\\{\si_b\sib_a}}} [\tau_\bb{a},\tau_\bb{a}^+]\GH^{(0)}_{\substack{\bb{a}f\\{\sib_b\si_2}}} [\tau_\bb{a},\tau_f].
\eeq  
After removing the two external lines, we find that
\beq
\Sigma^*_{\substack{a3\\{\si_1\si_2}}}[i,f] = - \lambda \si_1 \sib_2 t_{i\bb{b}} \GH_{\substack{\bb{b}i\\{\sib_2\sib_1}}} [\tau_i,\tau_i^+]\delta[i,f].
\label{Siga2star}
\eeq

\begin{figure}[H]
\begin{center}
\includegraphics[width=.5\columnwidth]{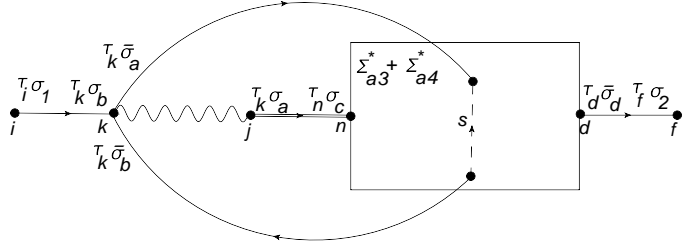}
\caption{Schematic representation for Green's function diagram whose top row is $1 \ldots c _{\ldots d} f$, where $d\neq1$. Upon removing the incoming and outgoing external lines, it contributes to $\Sigma_{a4}^*[i,f]$.}
\label{Siga3}
\end{center}
\end{figure}

Thus, $\Sigma^*_{\substack{a3\\{\si_1\si_2}}}[i,f]$ is equal to the first term on the RHS of \disp{chi2}. Note that since this term contains the lowest order contribution to $\chi[i,f]$, this covers the base case of the induction. We now want to show that $\Sigma^*_{\substack{a4\\{\si_1\si_2}}}[i,f]$ equals the second term on the RHS of \disp{chi2}. The $\GH[i,f]$ diagrams contributing to $\Sigma^*_{\substack{a4\\{\si_1\si_2}}}[i,f]$ can be represented schematically as in \figdisp{Siga3}. Here, the reasoning is similar to that which led to \figdisp{Sigb1}. If the line $s$ were contained in $\GH_{\substack{\bb{j}\bb{n}\\{\si_a\si_c}}} [\tau_\bb{k},\tau_\bb{n}]$, then the resulting $\Sigma_a$ (of the overall diagram) would be reducible, while if $s$ was the bare line $\GH^{(0)}_{\substack{\bb{d}f\\{\sib_d\si_2}}} [\tau_\bb{d},\tau_f]$, the diagram would contribute to $\Sigma^*_{a3}$ (see \figdisp{SigaFock}). The box can be either a $\Sigma^*_{a3}$ insertion or a $\Sigma^*_{a4}$ insertion, but can't be a $\Sigma^*_{a1}$ insertion or a $\Sigma^*_{a2}$ insertion, since in this case the diagram would contribute to $\Sigma^*_{a2}$ (see \figdisp{Siga1}). The analytical contribution of \figdisp{Siga3} is
\beq
-\lambda\GH^{(0)}_{\si_1\si_b}[i,\bb{k}]t[\bb{k},\bb{j}]\si_b\si_a\GH_{\si_a\si_c}[\bb{j},\bb{n}]\frac{\delta}{\delta\V^{\sib_b\sib_a}_{\bb{k}}}\left(\Sigma^*_{\substack{a3\\{\si_c\sib_d}}}[\bb{n},\bb{d}] +\Sigma^*_{\substack{a4\\{\si_c\sib_d}}}[\bb{n},\bb{d}] \right)\GH^{(0)}_{\sib_d\si_2}[\bb{d},f].
\eeq
Dropping the external lines, and using the inductive hypothesis, we obtain
\beq
\Sigma^*_{\substack{a4\\{\si_1\si_2}}}[i,f] = -\lambda^2 t[i,\bb{j}]\si_1\si_a\GH_{\si_a\si_b}[\bb{j},\bb{n}]\frac{\delta}{\delta\V^{\sib_1\sib_a}_{i}} \chi_{\si_b\si_2}[\bb{n},f] .
\label{Siga3star}
\eeq
Combining this with \disp{Siga2star} and comparing with \disp{chi2}, we find that
\beq
\Sigma^*_{\substack{a3\\{\si_1\si_2}}}[i,f] + \Sigma^*_{\substack{a4\\{\si_1\si_2}}}[i,f] = \lambda \chi_{\si_1\si_2}[i,f].
\label{chi3}
\eeq
Using \disp{chi3}, \disp{Siga1star}, and \disp{gin2}, we have proven the second of Eqs. (\ref{muginsig}).

\subsection{Diagrams in momentum space}

\begin{figure}[H]
\begin{center}
\includegraphics[width=.25\columnwidth]{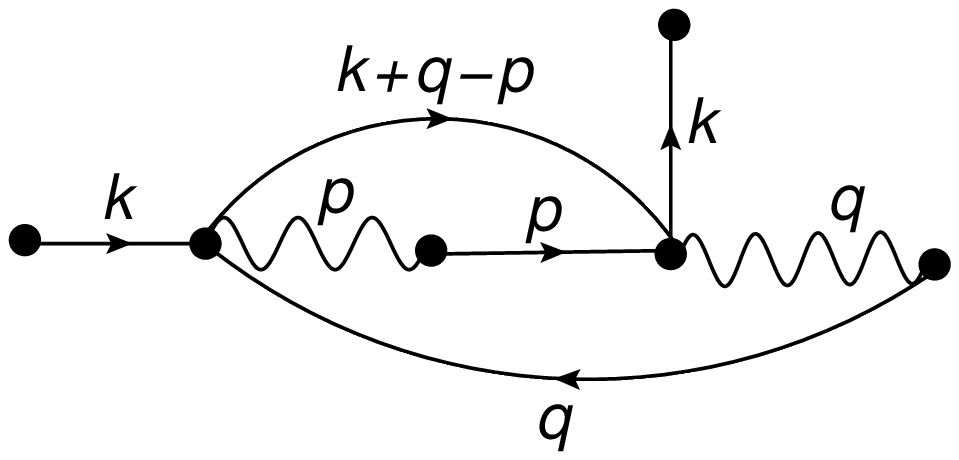}
\caption{Momentum space representation of diagram for $\GH(k)$ from Fig. (\ref{G2}b). Upon removing the incoming and outgoing external lines, it contributes to $\chi(k)$.}
\label{G2bmom}
\end{center}
\end{figure}

Upon turning off the sources, all objects become translationally-invariant in both space and time. We define the Fourier transform of all objects with two external points (e.g. $\G[i,f]$), denoted below by the generic symbol $Q[i,f]$, as
\beq
Q[i,f] = \frac{1}{N_s\beta}\sum_k e^{i k (i-f)} Q(k),
\eeq
where $N_s$ is the number of sites on the lattice, $\beta$ is the inverse temperature, $k\equiv (\vk,i\omega_k)$, and 
\newline $k(i-f) \equiv \vk\cdot(\vec{R}_i - \vec{R}_f)- \omega_k (\tau_i-\tau_f)$. For the rest of the paper, we shall not write the explicit factor $\frac{1}{N_s\beta}$ that goes along with each momentum sum. To obtain the momentum space contribution of a given $\GH(k)$ diagram, we assign momentum $k$ to the outgoing and incoming external lines, and sum over the momenta of the internal lines, in such a way that momentum is conserved at each point in the diagram. We also associate with each Green's function line the factor $\GH^{(0)}(q)$, where q is the momentum label of that line, and with each interaction line the factor $-\epsilon_q$, where $q$ is the momentum label of that interaction line, and $t[i,f] \equiv -\sum_q e^{i q (i-f)} \epsilon_q$.  The other rules are the same as in the coordinate space evaluation. For example, consider the diagram in panel b) of \figdisp{G2}, whose momentum space labels are displayed in \figdisp{G2bmom}. The momentum space contribution of this diagram is
\beq
-2 \ \GH^{(0)}(k)\epsilon_p\GH^{(0)}(p)\epsilon_q\GH^{(0)}(q)\GH^{(0)}(k+q-p)\GH^{(0)}(k)
\eeq
where a sum over the internal momenta $p$ and $q$ is implied.
Upon removing the external lines, we obtain the following contribution to $\Sigma_a^*(k)$, or equivalently to $\chi(k)$:

\beq
-2 \ \epsilon_p\GH^{(0)}(p)\epsilon_q\GH^{(0)}(q)\GH^{(0)}(k+q-p).
\eeq

\begin{figure}[H]
\begin{center}
\includegraphics[width=.25\columnwidth]{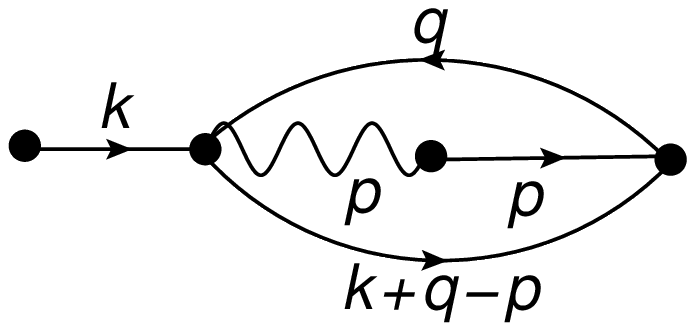}
\caption{Momentum space representation of diagram for $\G(k)$ from Fig. (\ref{G2}l). Upon removing the incoming external line, it contributes to $\Psi(k)$.}
\label{G2lmom}
\end{center}
\end{figure}

Additionally, consider the diagram for $\G(k)$ displayed in panel l) of \figdisp{G2}, whose momentum space labels are displayed in \figdisp{G2lmom}. The incoming external line carries momentum $k$ into the diagram, while the terminal point absorbs this momentum without transferring it to an outgoing external line. The momentum space contribution of this diagram is

\beq
-2 \ \GH^{(0)}(k)\epsilon_p\GH^{(0)}(p)\GH^{(0)}(q)\GH^{(0)}(k+q-p)
\eeq

Upon removing the incoming external line, we obtain the following contribution to $\Sigma_b^*(k)$, or equivalently to $\Psi(k)$:

\beq
-2 \ \epsilon_p\GH^{(0)}(p)\GH^{(0)}(q)\GH^{(0)}(k+q-p)
\eeq

\subsection{ The vertices $\Lambda$ and $\U$}

In section \ref{ginmu}, we showed that our diagrammatic series is consistent with the ECFL EOM, \disp{set1} and \disp{set2}. We rewrite them here for convenience.

\barray
\GHI[i,m]&=&  ( \mu  - \partial_{\tau_i} - \V_i) \ \delta[i,m] + t[i,m] \ ( 1- \lambda\gamma[i]) -\lambda \Phi[i,m].\nn\\
\widetilde{\mu}[i,m]&=& (1- \lambda\gamma[i]) \delta[i,m] +\lambda \Psi[i,m] \nn \\
\Phi[i,m] &=& - t[i, \bb{j}] \ \xi^* . \GH[\bb{j}, \bb{n}]. \Lambda_* [ \bb{n}, m; i]; \;\;\;\;\;\;\; \Psi[i,m] = - t[i, \bb{j}] \ \xi^* . \GH[\bb{j}, \bb{n}]. \U_* [ \bb{n}, m; i].
 \nn\\
\label{set3}
\earray
\barray
\gamma[i]= \widetilde{\mu}^{(k)}[\bb{n},i^+] . \GH^{(k)}[i,\bb{n}]; \;\;\;
\Lambda[n,m;i]= - \frac{\delta}{\delta \V_i} \GHI[n,m]; \;\;\; \U[n,m;i]=  \frac{\delta}{\delta \V_i} \widetilde{\mu}[n,m]. 
 \label{set4}
\earray

\begin{figure}[H]
\begin{center}
\includegraphics[width=.5\columnwidth]{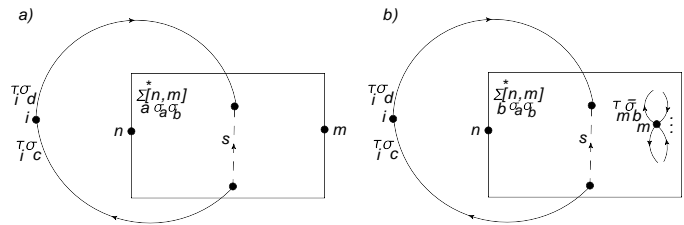}
\caption{Schematic diagram for the vertices. $\Lambda^{\si_a\si_b}_{\si_c\si_d}[n,m;i]$ is displayed in panel a) while $\U^{\si_a\si_b}_{\si_c\si_d}[n,m;i]$ is displayed in panel b).}
\label{VerticesSig}
\end{center}
\end{figure}

We now examine the vertices $\Lambda^{\si_a\si_b}_{\si_c\si_d}[n,m;i]\equiv - \frac{\delta}{\delta \V^{\si_c\si_d}_i} \GHI_{\si_a\si_b}[n,m]$ and $\U^{\si_a\si_b}_{\si_c\si_d}[n,m;i]\equiv  \frac{\delta}{\delta \V^{\si_c\si_d}_i} \widetilde{\mu}_{\si_a\si_b}[n,m]$ in more detail. The zeroth order vertices, also called the bare vertices, are given by
\beq
\Lambda^{(0)\si_a\si_b}_{\;\;\;\;\si_c\si_d}[n,m;i] = \delta[n,m]\delta[n,i]\delta_{\si_a\si_c}\delta_{\si_b\si_d}; \;\;\;\;\;\;\; \U^{(0)\si_a\si_b}_{\;\;\;\;\si_c\si_d}[n,m;i] = 0.
\label{barevertices}
\eeq
The higher order terms  contributing to $\Lambda^{\si_a\si_b}_{\si_c\si_d}[n,m;i]$ arise from splitting a line in $\Sigma^*_{\substack{a\\{\si_a\si_b}}}[n,m]$ through the point $i$. The higher order terms contributing to $\U^{\si_a\si_b}_{\si_c\si_d}[n,m;i]$ arise from splitting a line in $\Sigma^*_{\substack{b\\{\si_a\si_b}}}[n,m]$ through the point $i$. These terms can be represented schematically as in \figdisp{VerticesSig}. 

\begin{figure}[H]
\begin{center}
\includegraphics[width=.5\columnwidth]{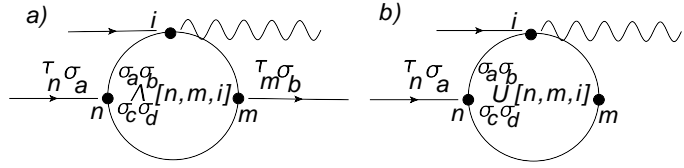}
\caption{Schematic diagram for the vertices. $\Lambda^{\si_a\si_b}_{\si_c\si_d}[n,m;i]$ is displayed in panel a) while $\U^{\si_a\si_b}_{\si_c\si_d}[n,m;i]$ is displayed in panel b).  }
\label{Vertices}
\end{center}
\end{figure}

From Fig. (\ref{VerticesSig}a), we see that in $\Lambda^{\si_a\si_b}_{\si_c\si_d}[n,m;i]$, the external points $n$ and $m$ accommodate an incoming and outgoing external Green's function line, respectively, while the external point $i$ accommodates an incoming external Green's function line and an external interaction line (Compare with Fig. (\ref{interaction}a)). In Fig. (\ref{VerticesSig}b), we see that in $\U^{\si_a\si_b}_{\si_c\si_d}[n,m;i]$, the external point $n$ accommodates an incoming external Green's function line, while the external point $i$ accommodates an incoming external Green's function line and an external interaction line. However, the external point $m$ is the terminal point and does not accommodate any external lines. Therefore, the vertices are represented schematically as in \figdisp{Vertices}. In the case of the bare vertex $\Lambda^{(0)}[n,m,i]$, the diagram in Fig. (\ref{Vertices}a) collapses onto a single point, which corresponds to the point $k$ in Fig. (\ref{interaction}a).

\begin{figure}[H]
\begin{center}
\includegraphics[width=.5\columnwidth]{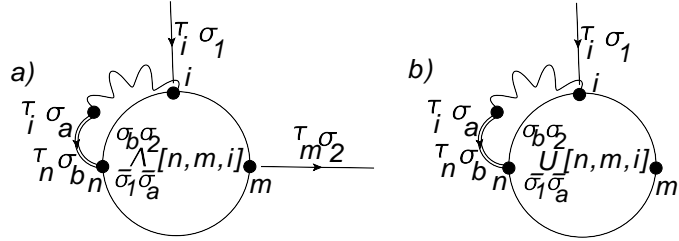}
\caption{Schematic diagram for the self-energies in terms of the vertices. In panel a) $\Phi_{\si_1\si_2}[i,m]$ is expressed in terms of $\Lambda^{\si_a\si_b}_{\si_c\si_d}[n,m;i]$  and in panel b) $\Psi_{\si_1\si_2}[i,m]$ is expressed in terms of $\U^{\si_a\si_b}_{\si_c\si_d}[n,m;i]$.}
\label{SigVertices}
\end{center}
\end{figure}

In \disp{set3}, the self-energies $\Phi$ and $\Psi$ are expressed in terms of the vertices $\Lambda$ and $\U$ respectively. These relationships can be expressed diagrammatically as in \figdisp{SigVertices}.

\begin{figure}[H]
\begin{center}
\includegraphics[width=.5\columnwidth]{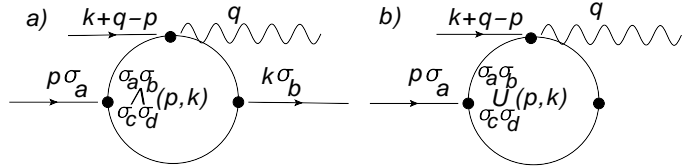}
\caption{Schematic diagram for the vertices in momentum space. $\Lambda^{\si_a\si_b}_{\si_c\si_d}(p,k)$ is displayed in panel a) while $\U^{\si_a\si_b}_{\si_c\si_d}(p,k)$ is displayed in panel b).}
\label{Verticesmom}
\end{center}
\end{figure}

We now turn the sources off, so that we can represent the vertices in  momentum space, as in \figdisp{Verticesmom}. In the case of $\Lambda(p,k)$, the external lines carry a total of zero momentum out of the vertex. In the case of $\U(p,k)$, the terminal point (the one with no external lines coming in or out) absorbs momentum $k$, and therefore the remainder of the external lines have to bring momentum $k$ into the vertex. Therefore, comparing \figdisp{Verticesmom} with \figdisp{Vertices}, the Fourier transform of the three point vertices, denoted below by the generic symbol $Q[n,m,i]$, is:
\beq
Q[n,m,i] = \sum_{kp} e^{i p n} e^{-i k m}e^{i (k-p) i} Q(p,k)=\sum_{kp} e^{i p (n-i)} e^{i k (i-m)} Q(p,k).
\eeq

Furthermore, there are only four non-zero spin configurations contributing to the vertex. These are $Q^{(1)}\equiv Q^{\si\si}_{\si\si}$, $Q^{(2)}\equiv Q^{\si\si}_{\sib\sib}$, $Q^{(3)}\equiv Q^{\si\sib}_{\si\sib}$, and $Q^{(4)}\equiv Q^{\si\sib}_{\sib\si}$. These four spin configurations are related by the equation
\beq
Q^{(1)} - Q^{(2)}= Q^{(3)}+Q^{(4)}.
\label{Nozieresrel}
\eeq
We shall now state the rules for computing the $Q^{(i)}$ and derive \disp{Nozieresrel}. Recall that to obtain a diagram for $\Lambda$ ($\U$), we must split a line in the self-energy $\Sigma_a^*$ ($\Sigma_b^*)$. This will give us an extra Green's function line in the diagram, and we must assign momenta to the external lines as indicated in \figdisp{Verticesmom}, at the same time summing over the momenta of the internal lines in such a way as to conserve momentum at each point of the diagram. Also recall from section \ref{rules} that the Green's function lines in the diagrams for $\Sigma_a^*$ ($\Sigma_b^*$) are partitioned into anywhere between 0 and $F_s$ spin loops, where the zeroth loop contains the lines with the labels $\si_1$ and $\si_2$. The spins carried by the Green's function lines in a single loop are allowed to alternate. However, the spin carried by each Green's function line in the loop is determined by that of any one of them (in the case of the zeroth loop it is the fixed spin $\si_1$).

Now, in the case that the line split in going from $\Sigma^*\to Q$ is from a loop which is not the zeroth loop, the resulting vertex diagram contributes only to $Q^{(1)}$ and $Q^{(2)}$ with a factor of $\frac{1}{2}$ relative to the contribution of the original diagram to $\Sigma^*$. In the case that the line split in going from $\Sigma^*\to Q$ is from the zeroth loop, the line split could either carry spin $\si_1$ in the original $\Sigma^*$ diagram or spin $\sib_1$. In the case of the former, the resulting vertex diagram contributes to both $Q^{(1)}$ and $Q^{(3)}$ with a factor of $1$ relative to the contribution of the original diagram to $\Sigma^*$. In the case of the latter, the resulting vertex diagram contributes to $Q^{(2)}$ with a factor of $1$, and to $Q^{(4)}$ with a factor of $(-1)$, relative to the contribution of the original diagram to $\Sigma^*$. \disp{Nozieresrel} immediately follows. Note that in the Feynman diagrams, we have the simpler situation in which all of the Green's function lines in a single spin loop (also referred to as Fermi loop), carry the same spin \cite{FW}. Then, the very last case described above becomes impossible, $Q^{(4)} \to 0$, and \disp{Nozieresrel} reduces to the standard Nozi\`{e}res relation $Q^{(1)} - Q^{(2)}= Q^{(3)}$ \cite{Nozieres}.

\begin{figure}[H]
\begin{center}
\includegraphics[width=.5\columnwidth]{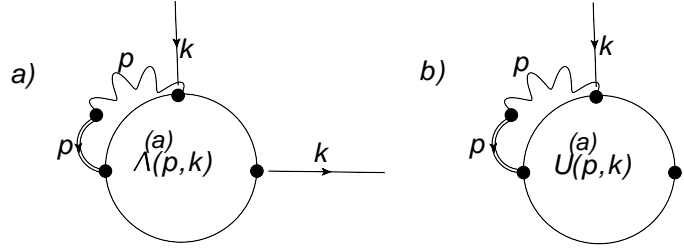}
\caption{Schematic diagram for the self-energies in terms of the vertices. In panel a) $\Phi(k)$ is expressed in terms of $\Lambda^{(a)}(p,k)$  and in panel b) $\Psi(k)$ is expressed in terms of $\U^{(a)}(p,k)$.}
\label{SigVerticesmom}
\end{center}
\end{figure}

Following \refdisp{ECFL}, we define $Q^{(a)}\equiv Q^{(2)}-Q^{(3)}$. Fourier transforming \disp{set3}, we obtain:

\beq
\Phi(k) = \sum_p\epsilon_p\GH(p)\Lambda^{(a)}(p,k);\;\;\;\;\;\;\;\; \Psi(k) = \sum_p\epsilon_p\GH(p)\U^{(a)}(p,k)
\eeq
These relations are represented diagrammatically in \figdisp{SigVerticesmom}

\subsection{Skeleton diagrams}

Consider the diagrammatic expansion for the irreducible self-energies that we have been using thus far, in which each diagram  is composed of bare Green's function lines $\GH^{(0)}[i,f]$, and hopping matrix elements $t_{if}$. We aim to reorganize this expansion in such a way that we only keep a subset of these diagrams, in which we replace each bare Green's function line $\GH^{(0)}[i,f]$, by the full auxiliary Green's function $\GH[i,f]$, thereby accounting for the diagrams which we discarded. We shall now define this subset of diagrams, which is referred to as the skeleton diagrams.

\begin{figure}[H]
\begin{center}
\includegraphics[width=.5\columnwidth]{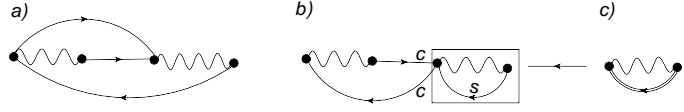}
\caption{Examples of skeleton and non-skeleton diagrams for the irreducible self-energy $\Sigma_a^*$. The diagram in panel a) is a skeleton diagram. The diagram in panel b) is not, since we can isolate a self-energy insertion by cutting the two lines labeled by $c$. The non-skeleton diagram in panel b) can be obtained from the skeleton diagram in panel c) by inserting the self-energy insertion enclosed by the box into the Green's function line. However, by splitting the line labeled by $s$ in the diagram in panel b), through another external point, we obtain a skeleton diagram for the vertex $\Lambda$.}
\label{skeleton}
\end{center}
\end{figure}

The skeleton diagrams are those diagrams in which one can't separate a self-energy insertion $\Sigma_a$ from the rest of the diagram by cutting two Green's function lines. For example, consider the $\Sigma_a^*$ diagrams in \figdisp{skeleton} (the same considerations will apply to $\Sigma_b^*$ diagrams). From left to right, these are the irreducible self-energies corresponding to the $\GH$ diagrams in Fig. (\ref{G2}b), Fig. (\ref{G2}c), and Fig. (\ref{G1}c). We see that the $\Sigma_a^*$ diagram in panel b) of \figdisp{skeleton} is a non-skeleton diagram, since by cutting the two Green's function lines labeled by the letter $c$, we isolate the $\Sigma_a$ self-energy insertion enclosed in the box. In contrast, the $\Sigma_a^*$ diagram in panel a) of \figdisp{skeleton} is a skeleton diagram, since it is impossible to isolate a $\Sigma_a$ insertion by cutting two Green's function lines. Finally, the diagram in panel c) of \figdisp{skeleton} is also a skeleton diagram. Furthermore, we see that by placing the self-energy insertion enclosed in the box into the Green's function line of the diagram in Fig. (\ref{skeleton}c), we reproduce the diagram in Fig. (\ref{skeleton}b). Since a full auxiliary Green's function line consists of an arbitrary self-energy insertion $\Sigma_a$ surrounded by two bare Green's function lines $\GH^{(0)}$, we see that the whole series is reproduced by keeping only the skeleton diagrams and making the substitution $\GH^{(0)}[i,f]\to\GH[i,f]$. 

Now, consider the vertices $\Lambda[n,m;i]$ and $\U[n,m;i]$. Recall from \figdisp{VerticesSig}, that these correspond to splitting a Green's function line through the point $i$ in $\Sigma_a^*[n,m]$ and $\Sigma_b^*[n,m]$, respectively. How do we obtain the skeleton diagrams for the vertices? A naive guess would be that we do so by splitting a Green's function line in the skeleton diagrams for the irreducible self-energies. However, this is only partially correct. To see this, consider again the non-skeleton $\Sigma_a^*$ diagram in panel b) of \figdisp{skeleton}. If we choose to split either of the two lines labeled by $c$, then we leave the self-energy insertion surrounded by the box intact, and the resulting diagram for $\Lambda$ is a non-skeleton diagram. However, if we split the Green's function line labeled by $s$, this breaks up this self-energy insertion, and leads to a skeleton diagram for $\Lambda$.

\begin{figure}[H]
\begin{center}
\includegraphics[width=.5\columnwidth]{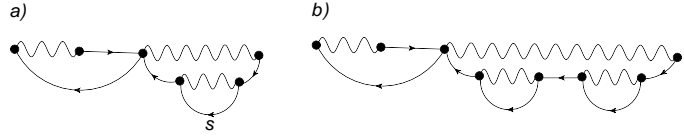}
\caption{In this case both $\Sigma_a^*$ diagrams displayed in panels a) and b) are non-skeleton diagrams. However, the diagram in panel a) contains only irreducible self-energy insertions, while the one in panel b) contains a reducible self-energy insertion. One can obtain a skeleton diagram for the vertex $\Lambda$ only by splitting the line labelled by $s$ in the diagram in panel a). It is impossible to obtain a skeleton diagram for the vertex $\Lambda$ from the diagram in panel b) regardless of which line we split. }
\label{Lambdaskel}
\end{center}
\end{figure}

Taking this reasoning a step further, consider the diagram for $\Sigma_a^*$ in panel a) of \figdisp{Lambdaskel}. This diagram can be obtained from the diagram in Fig. (\ref{skeleton}b) by inserting the self-energy insertion enclosed by the box into the line labeled by $s$ in Fig. (\ref{skeleton}b). Once again, if we split any line other than the one labeled by $s$ in Fig. (\ref{Lambdaskel}a), the resulting diagram for $\Lambda$ will be a non-skeleton diagram, while if we split the line labeled by $s$, the resulting diagram for $\Lambda$ will be a skeleton diagram. Meanwhile, for the $\Sigma_a^*$ diagram in Fig. (\ref{Lambdaskel}b), obtained from the diagram in Fig. (\ref{skeleton}b) by putting a reducible self-energy insertion into the line labeled by $s$ in Fig. (\ref{skeleton}b), it is not possible to split any line in such a way that the resulting diagram for $\Lambda$ will be a skeleton diagram. 

\begin{figure}[H]
\begin{center}
\includegraphics[width=.5\columnwidth]{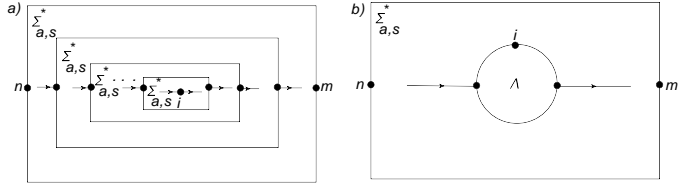}
\caption{Panel a) demonstrates the general procedure for obtaining a skeleton diagram for the vertex $\Lambda$ from a $\Sigma_a^*$ diagram consisting of a sequence of $\Sigma_a^*$ skeleton diagrams. The original $\Sigma_a^*$ diagram is itself a skeleton diagram only if there is only one skeleton diagram in the sequence, i.e. the $\Sigma_a^*$ diagram in question. If we remove the outermost box in panel a), we are still left with a general skeleton diagram for the vertex $\Lambda$. Therefore, to obtain a skeleton diagram for $\Lambda$, one must insert the full vertex $\Lambda$ into a green's function line of a skeleton diagram for $\Sigma_a^*$. This is displayed in panel b). In the case that the original $\Sigma_a^*$ diagram in panel a) is itself a skeleton diagram (i.e. there is only one skeleton diagram in the sequence), the $\Lambda$ vertex in panel b) is a bare vertex.}
\label{Lambdaskelgeneral}
\end{center}
\end{figure}

Therefore, we see that to construct the skeleton diagrams for $\Lambda[n,m;i]$ ($\U[n,m;i]$), we have to use the following procedure. Take a skeleton diagram for $\Sigma_a^*[n,m]$ ($\Sigma_b^*[n,m]$), and insert into at most one line of this diagram, a skeleton diagram for $\Sigma_a^*$. Then, insert into at most line of that diagram, a skeleton diagram for $\Sigma_a^*$, and so on. This produces a sequence of skeleton diagrams for the irreducible self-energies. Then, in the last skeleton diagram of the sequence, split a single Green's function line through the point $i$. This procedure is represented schematically (for the case of $\Lambda[n,m;i]$) in Fig. (\ref{Lambdaskelgeneral}a).

Now consider the part of Fig. (\ref{Lambdaskelgeneral}a) enclosed by the second box (counting from the very outer box). This is itself a skeleton diagram for the vertex $\Lambda[w,v;i]$, where $w$ and $v$ are internal variables. Therefore, we see that one can obtain the skeleton expansion for $\Lambda[n,m;i]$ ($\U[n,m;i]$) from the skeleton expansion for $\Sigma_a^*[n,m]$ ($\Sigma_b^*[n,m]$) by replacing in each skeleton diagram for $\Sigma_a^*[n,m]$ ($\Sigma_b^*[n,m]$), a single Green's function line $\GH[x,y]$, with $\GH[x,\bb{w}].\Lambda[\bb{w},\bb{v};i].\GH[\bb{v},y]$, where $\Lambda[\bb{w},\bb{v};i]$ is the full vertex. This is represented schematically in Fig. (\ref{Lambdaskelgeneral}b). The case in which there is only one box in Fig. (\ref{Lambdaskelgeneral}a) corresponds to plugging in the bare vertex into Fig. (\ref{Lambdaskelgeneral}b). 

We now have three skeleton expansions. The first is the original skeleton expansion of the self-energies in terms of the auxiliary Green's function.
\beq
\GHI\equiv\GHI[\GH];\;\;\;\;\;\;\;\widetilde{\mu}\equiv\widetilde{\mu}[\GH],
\label{skeleton1}
\eeq
The second is the skeleton expansion for the vertices in terms of the auxiliary Green's function. This is the skeleton expansion represented in Fig. (\ref{Lambdaskelgeneral}a).
\beq \Lambda\equiv\Lambda[\GH];\;\;\;\;\;\;\U\equiv\U[\GH].
\label{skeleton2}
\eeq
The third is the skeleton expansion for the vertices in terms of the auxiliary Green's function and the full vertex $\Lambda$. This is the skeleton expansion represented in Fig. (\ref{Lambdaskelgeneral}b).
\beq \Lambda\equiv\Lambda[\GH,\Lambda];\;\;\;\;\;\;\U\equiv\U[\GH,\Lambda].
\label{skeleton3}
\eeq

Using the diagrammatic rules developed here, we have access to all three of these skeleton expansions at any order. However, in the absence of these rules, we could derive the terms in these skeleton expansions by using Eqs. (\ref{skeleton1}), (\ref{skeleton2}), (\ref{skeleton3}), and (\ref{set3}) in the following manner. Suppose that we have the skeleton expansions in Eqs. (\ref{skeleton1}) - (\ref{skeleton3}) through $m^{th}$ order in $\lambda$. Then, plugging the $m^{th}$ order term of the skeleton expansion from \disp{skeleton2} into \disp{set3} yields the $m+1^{st}$ order term of the skeleton expansion in \disp{skeleton1}. Then, applying the rule $\GH\to\GH\Lambda\GH$ to the $m+1^{st}$ order term of the skeleton expansion in \disp{skeleton1}, yields the $m+1^{st}$ order contribution to the skeleton expansion in \disp{skeleton3}. Finally, plugging the $k^{th}$ order term of the skeleton expansion from \disp{skeleton2} ($0\leq k\leq m$) into the $m+1-k^{th}$ term of the skeleton expansion from \disp{skeleton3} yields the $m+1^{st}$ order term of the skeleton expansion from \disp{skeleton2}, after which we can iterate the process again. This process starts at zeroth order by plugging the bare vertex into 
\disp{set3} and calculating the first order contribution to the skeleton expansion in \disp{skeleton1}, and so on. This is the approach used in the original ECFL papers \cite{ECFL,Monster}, and reviewed in section \ref{ECFL}. It reveals the power of the Schwinger approach in that it enables one to bypass the bare series and work directly with the skeleton expansion. However, the utility of the diagrams developed here is that they enable one to obtain the contribution of a given order directly, without iteration, and also to visualize all the higher order terms diagrammatically, therefore facilitating diagrammatic re-summations.

\section{Putting $J$ back into the equations \label{J}}

Let us rewrite \disp{EOMG} in the form of an integral equation as in \disp{EOMG3}, but this time keeping $J$.
\barray
\G_{\si_1,\si_2}[i,f] &=& \GH^{(0)}_{\si_1,\si_2}[i,f] -\lambda \ \GH^{(0)}_{\si_1,\si_b}[i,f]\si_b\si_2\G_{\sib_2,\sib_b}[f,f^+] \nn\\
&&  -\lambda \ \GH^{(0)}_{\si_1,\si_b}[i,\bb{k}] \left(- t[\bb{k},\bb{j}] \si_b\si_a\G_{\sib_a,\sib_b}[\bb{k},\bb{k}^+]\G_{\si_a,\si_2}[\bb{j},f] + t[\bb{k},\bb{j}]\si_b\si_a \frac{\delta}{\delta\V_\bb{k}^{\sib_b,\sib_a}} \G_{\si_a,\si_2}[\bb{j},f]\right),\nn\\
&&  -\lambda \ \GH^{(0)}_{\si_1,\si_b}[i,\bb{k}] \left( \frac{1}{2} J[\bb{k},\bb{j}] \si_b\si_a\G_{\sib_a,\sib_b}[\bb{j},\bb{j}^+]\G_{\si_a,\si_2}[\bb{k},f] -\frac{1}{2} J[\bb{k},\bb{j}]\si_b\si_a \frac{\delta}{\delta\V_\bb{j}^{\sib_b,\sib_a}} \G_{\si_a,\si_2}[\bb{k},f]\right).
\label{EOMGJ2}
\earray

\begin{figure}[H]
\begin{center}
\includegraphics[width=.5\columnwidth]{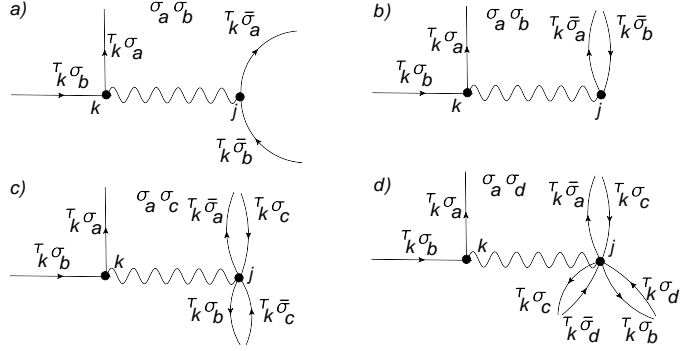}
\caption{The $J$-vertices in the diagrams of the $\lambda$ expansion. They are more reminiscent of the Feynman diagram vertices than the $t$-vertices displayed in \figdisp{interaction}. The two types of vertices can be obtained from each other by interchanging lines between the two points of the vertex.}
\label{interactionlinesJ}
\end{center}
\end{figure}

The $\lambda$-expansion of \disp{EOMGJ2} is given by the same set of rules as in section \ref{rules}, with the only difference being that now each vertex can be either a $t$-vertex or a $J$-vertex. Comparing the second and third lines on the RHS of \disp{EOMGJ2}, we see that the $J$-vertices can be obtained from the $t$-vertices in \figdisp{interaction} by moving the line labeled by $\si_a$ from the point $j$ to the point $k$, and moving all lines but the one labeled by $\si_b$ from the point $k$ to the point $j$. The $J$-vertices are displayed in \figdisp{interactionlinesJ}. They are more reminiscent of the standard Feynman diagram vertices.

\begin{figure}[H]
\begin{center}
\includegraphics[width=.5\columnwidth]{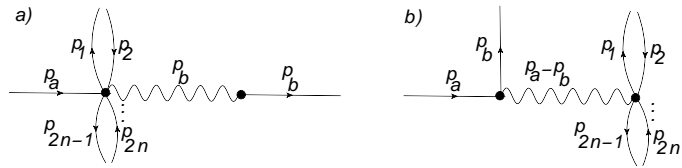}
\caption{The $t$ and $J$ vertices are displayed in momentum space in panels a) and b) respectively. For each interaction vertex of a diagram, we can choose to use either one as long as we associate with it the factor $\epsilon_{p_b}+\frac{1}{2}J_{p_a-p_b}$.}
\label{tJvertices}
\end{center}
\end{figure}

\begin{figure}[H]
\begin{center}
\includegraphics[width=.5\columnwidth]{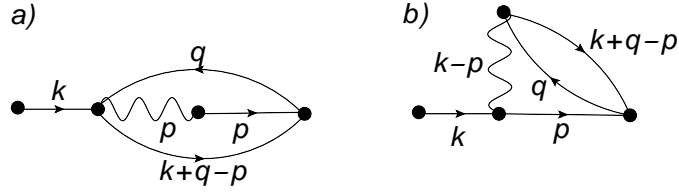}
\caption{The $\G(k)$ diagrams drawn in panel a) and panel b) correspond to the same diagram. The one in panel a) is drawn using a $t$-vertex, while the one in panel b) is drawn using a $J$-vertex. In both cases, we associate the factor $\epsilon_p + \frac{1}{2} J_{k-p}$ with the interaction vertex.}
\label{G2lmomtJ}
\end{center}
\end{figure}

Now, let us compare an arbitrary $t$-vertex and an arbitrary $J$-vertex in momentum space. The $t$-vertex is shown in panel a) of \figdisp{tJvertices}, while the $J$-vertex is shown in panel b). Conserving momentum at each point of the $t$ and $J$ vertices yields the relation
\beq
p_a + \sum_{m=1}^np_{2m} = p_b + \sum_{m=1}^np_{2m-1}.
\label{consmom}
\eeq
In Fig. (\ref{tJvertices}a), the interaction line contributes a factor of $\epsilon_{p_b}$, while in Fig. (\ref{tJvertices}b), the interaction line contributes a factor of $\frac{1}{2}J_{p_a-p_b}$. At first, it seems as though for each diagram with $i$ interaction vertices, we must now draw $2^i$ separate diagrams, since for each vertex we must decide whether it will be a $t$-vertex or a $J$-vertex. For example, consider the diagram in \figdisp{G2lmom}, also displayed in Fig. (\ref{G2lmomtJ}a), in which the interaction vertex is a $t$-vertex. In Fig. (\ref{G2lmomtJ}b), it is drawn with a $J$-vertex. However, we see that the Green's function lines in both diagrams have the same momentum labels, and the only difference is the momentum label of the interaction line. This is because the two things which determine the  momentum labels of the Green's function lines are
\begin{itemize}
\item(1) the interconnections (via Green's function lines) between the interaction vertices (irrespective of where on these vertices these lines appear),
\item(2) \disp{consmom}.
\end{itemize} 
Since the $J$-vertex simply reshuffles the lines on the $t$-vertex, and \disp{consmom} applies equally well to both types of vertices, both (1) and (2) are unaffected by the choice of $t$ vertex vs. $J$ vertex. Therefore, we can choose to use either the diagram in Fig. (\ref{G2lmomtJ}a) or the diagram in Fig. (\ref{G2lmomtJ}b) if we associate with the interaction line in each diagram the factor $\epsilon_p + \frac{1}{2} J_{k-p}$. In general, we can construct diagrams either from the vertices in Fig. (\ref{tJvertices}a) (as we have already been doing), or from the vertices in Fig. (\ref{tJvertices}b) (which would be more reminiscent of the Feynman diagrams), as long as we associate with each interaction vertex the factor $\epsilon_{p_b}+\frac{1}{2}J_{p_a-p_b}$. 

The ECFL equations with $J$ included are given as follows \cite{ECFL}.

\barray
\GHI[i,m]&=&  ( \mu  - \partial_{\tau_i} - \V_i) \ \delta[i,m] + t[i,m] \ ( 1- \lambda\gamma[i]) +\frac{\lambda}{2}J[i,\bb{j}]\gamma[\bb{j}]\delta[i,m]-\lambda \Phi[i,m],\nn\\
\widetilde{\mu}[i,m]&=& (1- \lambda\gamma[i]) \delta[i,m] +\lambda \Psi[i,m], \nn \\
\Phi[i,m] &=& \bb{L}[i,\bb{n}].\GHI[\bb{n},m];\;\;\;\;\;\;\;\;\;\;\;\;\;\;\;\;\; \Psi[i,m] = -\bb{L}[i,\bb{n}].\widetilde{\mu}[\bb{n},m],
 \nn\\
\label{set1J}
\earray
where the operator $\bb{L}$ is given by:
\beq
\bb{L}_{\si_1\si_2}[i,m] = t[i,\bb{j}]\si_1\si_a\GH_{\si_a\si_2}[\bb{j},m]\frac{\delta}{\delta\V_i^{\sib_1\sib_a}} -\frac{1}{2}J[i,\bb{j}]\si_1\si_a\GH_{\si_a\si_2}[i,m]\frac{\delta}{\delta\V_{\bb{j}}^{\sib_1\sib_a}}.
\eeq
Using the same decomposition as in \disp{chi}, i.e. 
\beq
\Phi[i,m] = -\Psi[i,\bb{j}] t[\bb{j},m] + \chi[i,m],
\label{chiJ}
\eeq
we find that
\beq
\chi[i,m] = \bb{L}[i,\bb{n}].\GH^{-1(0)}[\bb{n},m] + \frac{\lambda}{2} J[m,\bb{k}]\bb{L}[i,m]\gamma[\bb{k}] - \lambda \bb{L}[i,\bb{n}]\chi[\bb{n},m],
\eeq
where
\beq
(\bb{L}[i,\bb{n}].\GH^{-1(0)}[\bb{n},m])_{\si_1\si_2} = t[i,\bb{j}]\si_1\si_2\GH_{\sib_2\sib_1}[\bb{j},i]\delta[i,m] -\frac{1}{2}J[i,m]\si_1\si_2\GH_{\sib_2\sib_1}[i,m].
\eeq

Finally, we note that the equations associated with \figdisp{SigVerticesmom} now become
\beq
\Phi(k) = \sum_p(\epsilon_p+\frac{1}{2}J_{k-p})\GH(p)\Lambda^{(a)}(p,k);\;\;\;\;\;\;\;\; \Psi(k) = \sum_p(\epsilon_p+\frac{1}{2}J_{k-p})\GH(p)\U^{(a)}(p,k).
\eeq

\section{Finite order calculations\label{finiteorder}}
\subsection{Zeroth through third order calculation}

In this section, we compute the skeleton expansion for the objects $\gamma$, $\Psi$, and $\chi$ through second order in $\lambda$ in momentum space. As can be seen from \disp{set1mom} below, this yields the skeleton expansion for $\GHI$ and $\widetilde{\mu}$ through third order in $\lambda$. Before proceeding with this computation, we follow \refdisp{Monster} in introducing a second chemical potential $u_0$ into the theory. As explained in \refdisp{Monster}, there is a so-called shift identity of the $\tJ$ model, which states that adding an onsite term to the hopping affects $\G$ only through a shift of the chemical potential $\mu$. However, the same is not true of the constituent factors $\GH$ and $\widetilde{\mu}$, which will be affected by such a shift. To remedy this, in \refdisp{Monster}, the second chemical potential $u_0$ is introduced directly into the definitions of $\GHI$ and $\widetilde{\mu}$ (\disp{set1J}) through the formula $t[i,j]\to t[i,j] + \frac{u_0}{2}\delta[i,j]$ in every term but the $t[i,f]$ term in the equation for $\GHI[i,f]$. Now, an onsite shift in the hopping affects $\GH$ and $\widetilde{\mu}$ only through a shift in the second chemical potential $u_0$. Moreover, the fact that $\G$ will not be affected for any value of $u_0$ (other than through a shift of the original chemical potential $\mu$) is a consequence of the shift identity. Furthermore, the two chemical potentials $\mu$ and $u_0$ can now be used to satisfy the two sum rules
\beq
\sum_k \G(k) = \frac{n}{2};\;\;\;\;\;\;\;\;\;\sum_k \GH(k) = \frac{n}{2}.
\label{sumrules}
\eeq  
The first of these ensures the correct particle sum-rule for the physical electrons. The second one states that the auxiliary fermions must satisfy the same particle sum-rule as the physical ones. We can think of the Hubbard operator $X_i^{0\si} = c_{i\si}(1-n_{i\sib})$ as representing the physical fermions, and the canonical operator $c_{i\si}$ as representing the auxiliary fermions. Since, the number operator is a charge neutral object, charge conservation implies that the physical and auxiliary fermions must satisfy the same particle sum-rule. As a consequence of this, the physical electrons have a Fermi-surface which complies with the Luttinger-Ward volume theorem (see \refdisp{ECFL} where these sum rules were originally introduced and their implications discussed). 

We now proceed to present the diagrams and analytical expressions for $\GHI$ and $\widetilde{\mu}$ through third order in $\lambda$.
Taking the Fourier transform of \disp{set1J} and \disp{product}, and using \disp{chiJ}, we obtain

\barray
\GHI(k) &=& i\omega_k + \mu' - (\epsilon_k-\frac{u_0}{2})\widetilde{\mu}(k) - \lambda \chi(k), \nn\\
\mu' &=& \mu -\frac{u_0}{2}+ \frac{\lambda}{2}\gamma J_0 \nn\\
\widetilde{\mu}(k) &=& (1-\lambda \gamma) + \lambda \Psi(k), \nn\\
\G(k) &=& \GH(k) \widetilde{\mu}(k),
\label{set1mom}
\earray  
where $J_0$ is the zero-momentum component of the Fourier transform of $J_{ij}$.  Our strategy is to compute the skeleton expansion for $\gamma$, $\Psi$, and $\chi$ through second order in $\lambda$ (i.e. $\gamma = \gamma^{(0)} +\gamma^{(1)}+\gamma^{(2)}$, etc.) After plugging in the expressions from this skeleton expansion into \disp{set1mom}, we must set $\lambda=1$, and solve the resulting integral equations. The two Lagrange multipliers $\mu$ and $u_0$ are then determined by the sum rules in \disp{sumrules}.

\begin{figure}[H]
\begin{center}
\includegraphics[width=.5\columnwidth]{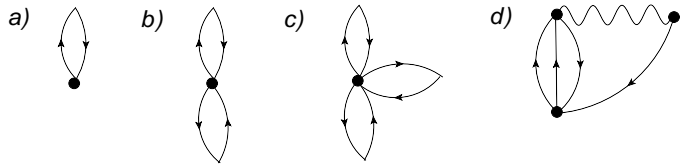}
\caption{Second order skeleton expansion for $\gamma$. Only the diagram in panel a) is a standard Feynman diagram. $\gamma^{(0)}$ is given by the diagram in panel a), $\gamma^{(1)}$ is given by the diagram in panel b), and $\gamma^{(2)}$ is given by diagrams in panels c) and d). We conserve momentum at each interaction vertex as indicated in Figs. (\ref{tJvertices}) and (\ref{G2lmomtJ}).}
\label{3rdordergamma}
\end{center}
\end{figure}

In \figdisp{3rdordergamma}, we have drawn the skeleton diagrams for $\gamma$ (which is just a constant when the sources are off) through second order in $\lambda$. Therefore, $\gamma$ is the sum of the following terms 
\barray
&& a) \ \frac{n}{2} \nn\\
&& b) \ -\lambda\left(\frac{n}{2}\right)^2 \nn\\
&& c) \ \lambda^2\left(\frac{n}{2}\right)^3 \nn\\
&& d) \ -2\lambda^2 \sum_{plq}\GH(p)\GH(l)\GH(q)\GH(p+l-q)(\epsilon_q-\frac{u_0}{2}+\frac{1}{2}J_{p-q}). \nn\\
\earray

\begin{figure}[H]
\begin{center}
\includegraphics[width=.5\columnwidth]{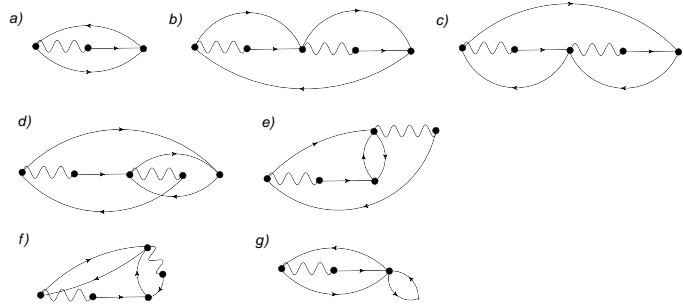}
\caption{Second order skeleton expansion for $\Psi(k)$. All diagrams but the one in panel g) are standard Feynman diagrams (with one interaction line set to unity). $\Psi^{(1)}$ is given by the diagram in panel a), and $\Psi^{(2)}$ is given by the diagrams in panels b) through g). We conserve momentum at each interaction vertex as indicated in Figs. (\ref{tJvertices}) and (\ref{G2lmomtJ}).}
\label{3rdorderPsi}
\end{center}
\end{figure}

In \figdisp{3rdorderPsi}, we have done the same for $\Psi(k)$. Therefore, $\Psi(k)$ is the sum of the following terms

\barray
&& a) \ -2\lambda \sum_{pq}\GH(p)\GH(q)\GH(k+q-p)(\epsilon_p-\frac{u_0}{2}+\frac{1}{2}J_{k-p})\nn\\
&& b) \ -4 \lambda^2 \sum_{pql}\GH(p)\GH(l)\GH(q)\GH(k+q-p)\GH(k+q-l)(\epsilon_p-\frac{u_0}{2}+\frac{1}{2}J_{k-p})(\epsilon_l-\frac{u_0}{2}+\frac{1}{2}J_{p-l})\nn\\
&& c) \ - \lambda^2 \sum_{pql}\GH(p)\GH(l)\GH(q)\GH(k+q-p)\GH(q+l-p)(\epsilon_p-\frac{u_0}{2}+\frac{1}{2}J_{k-p})(\epsilon_l-\frac{u_0}{2}+\frac{1}{2}J_{p-l})\nn\\
&& d) \ - \lambda^2 \sum_{pql}\GH(p)\GH(l)\GH(q)\GH(k+q-p)\GH(p+l-q)(\epsilon_p-\frac{u_0}{2}+\frac{1}{2}J_{k-p})(\epsilon_q-\frac{u_0}{2}+\frac{1}{2}J_{p-q})\nn\\
&& e) \ - \lambda^2 \sum_{pql}\GH(p)\GH(l)\GH(q)\GH(k+q-p)\GH(k+l-p)(\epsilon_p-\frac{u_0}{2}+\frac{1}{2}J_{k-p})(\epsilon_q-\frac{u_0}{2}+\frac{1}{2}J_{l-q})\nn\\
&& f) \ - \lambda^2 \sum_{pql}\GH(p)\GH(l)\GH(q)\GH(k+q-p)\GH(l+p-k)(\epsilon_p-\frac{u_0}{2}+\frac{1}{2}J_{k-p})(\epsilon_l-\frac{u_0}{2}+\frac{1}{2}J_{p-k})\nn\\
&& g)  \ \lambda^2 \frac{n}{2}\sum_{pq}\GH(p)\GH(q)\GH(k+q-p)(\epsilon_p-\frac{u_0}{2}+\frac{1}{2}J_{k-p})\nn\\
\earray

\begin{figure}[H]
\begin{center}
\includegraphics[width=.5\columnwidth]{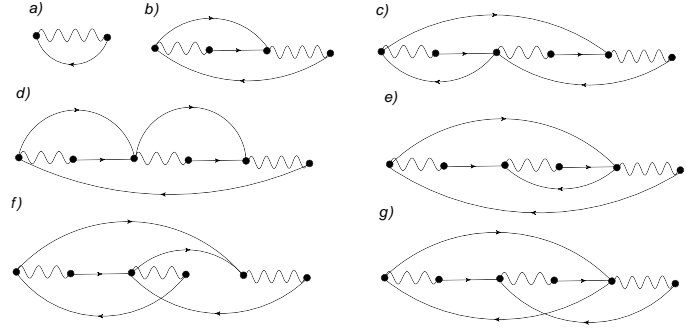}
\caption{Second order skeleton expansion for $\chi_A(k)$. These diagrams are independent of those for $\Psi(k)$. All diagrams are standard Feynman diagrams. The diagram in panel a) contributes to $\chi^{(0)}$, the diagram in panel b) contributes to $\chi^{(1)}$, and the diagrams in panels c) through g) contribute to $\chi^{(2)}$. We conserve momentum at each interaction vertex as indicated in Figs. (\ref{tJvertices}) and (\ref{G2lmomtJ}).}
\label{3rdorderchi}
\end{center}
\end{figure}

\begin{figure}[H]
\begin{center}
\includegraphics[width=.5\columnwidth]{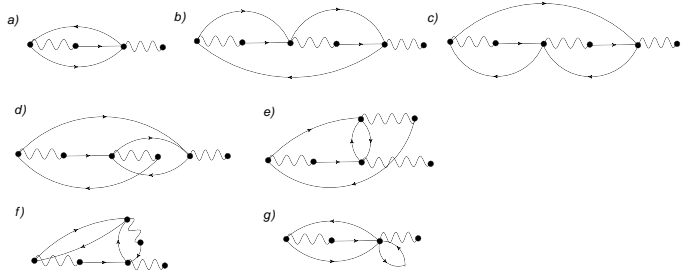}
\caption{Second order skeleton expansion for $\chi_B(k)$, which vanishes when $J=0$. These diagrams can be obtained from those for $\Psi(k)$ in \figdisp{3rdorderPsi} by adding an interaction line to the terminal point of those diagrams. However, this interaction line contributes only a factor of $J$, and not a factor of $\epsilon$. All diagrams but the one in panel g) are standard Feynman diagrams. The diagram in panel a) contributes to $\chi^{(1)}$, and the diagrams in panels b) through g) contribute to $\chi^{(2)}$. We conserve momentum at each interaction vertex as indicated in Figs. (\ref{tJvertices}) and (\ref{G2lmomtJ}).}
\label{3rdorderchifromPsi}
\end{center}
\end{figure}

The skeleton diagrams for $\chi(k)$ have been split into two groups. Those drawn in \figdisp{3rdorderchifromPsi}, whose contribution will be denoted by $\chi_B(k)$, can be obtained from the $\Psi(k)$ diagrams in \figdisp{3rdorderPsi} by attaching an interaction line to the terminal point of those diagrams. Due to the decomposition \disp{chiJ}, this interaction line will contribute only a $J$ term, but no $\epsilon$ term, to the expression for $\chi_B(k)$. The rest of the $\chi(k)$ diagrams, whose contribution will be denoted by $\chi_A(k)$, are drawn in \figdisp{3rdorderchi}. Then, $\chi(k) = \chi_A(k) + \chi_B(k)$, where $\chi_A(k)$ is the sum of the terms in \disp{chiA} and $\chi_B(k)$ is the sum of the terms in \disp{chiB}. 
\barray
&& a) \ -\sum_{p}\GH(p)(\epsilon_p-\frac{u_0}{2}+\frac{1}{2}J_{k-p})\nn\\
&& b) \ -2 \lambda \sum_{pq}\GH(p)\GH(q)\GH(k+q-p)(\epsilon_p-\frac{u_0}{2}+\frac{1}{2}J_{k-p})(\epsilon_q-\frac{u_0}{2}+\frac{1}{2}J_{p-q})\nn\\
&& c) \ - \lambda^2 \sum_{pql}\GH(p)\GH(l)\GH(q)\GH(k+q-p)\GH(l+q-p)(\epsilon_p-\frac{u_0}{2}+\frac{1}{2}J_{k-p})(\epsilon_l-\frac{u_0}{2}+\frac{1}{2}J_{p-l})(\epsilon_{l+q-p}-\frac{u_0}{2}+\frac{1}{2}J_{p-q})\nn\\
&& d) \ -4 \lambda^2 \sum_{pql}\GH(p)\GH(l)\GH(q)\GH(k+q-p)\GH(k+q-l)(\epsilon_p-\frac{u_0}{2}+\frac{1}{2}J_{k-p})(\epsilon_l-\frac{u_0}{2}+\frac{1}{2}J_{p-l})(\epsilon_q-\frac{u_0}{2}+\frac{1}{2}J_{l-q})\nn\\
&& e) \ - \lambda^2 \sum_{pql}\GH(p)\GH(l)\GH(q)\GH(k+q-p)\GH(k+l-p)(\epsilon_p-\frac{u_0}{2}+\frac{1}{2}J_{k-p})(\epsilon_l-\frac{u_0}{2}+\frac{1}{2}J_{p-l})(\epsilon_q-\frac{u_0}{2}+\frac{1}{2}J_{l-q})\nn\\
&& f) \ - \lambda^2 \sum_{pql}\GH(p)\GH(l)\GH(q)\GH(k+q-p)\GH(p+l-q)(\epsilon_p-\frac{u_0}{2}+\frac{1}{2}J_{k-p})(\epsilon_q-\frac{u_0}{2}+\frac{1}{2}J_{p-q})(\epsilon_l-\frac{u_0}{2}+\frac{1}{2}J_{k+q-p-l})\nn\\
&& g) \ - \lambda^2 \sum_{pql}\GH(p)\GH(l)\GH(q)\GH(k+q-p)\GH(k+l-p)(\epsilon_p-\frac{u_0}{2}+\frac{1}{2}J_{k-p})(\epsilon_l-\frac{u_0}{2}+\frac{1}{2}J_{p-l})(\epsilon_{k+l-p}-\frac{u_0}{2}+\frac{1}{2}J_{p-k})\nn\\
\label{chiA}
\earray

\barray
&& a) \ - \lambda \sum_{pq}\GH(p)\GH(q)\GH(k+q-p)(\epsilon_p-\frac{u_0}{2}+\frac{1}{2}J_{k-p})J_{p-k}\nn\\
&& b) \ -2 \lambda^2 \sum_{pql}\GH(p)\GH(l)\GH(q)\GH(k+q-p)\GH(k+q-l)(\epsilon_p-\frac{u_0}{2}+\frac{1}{2}J_{k-p})(\epsilon_l-\frac{u_0}{2}+\frac{1}{2}J_{p-l})J_{l-k}\nn\\
&& c) \ - \frac{\lambda^2 }{2} \sum_{pql}\GH(p)\GH(l)\GH(q)\GH(k+q-p)\GH(q+l-p)(\epsilon_p-\frac{u_0}{2}+\frac{1}{2}J_{k-p})(\epsilon_l-\frac{u_0}{2}+\frac{1}{2}J_{p-l})J_{l-k}\nn\\
&& d) \ -\frac{\lambda^2 }{2}  \sum_{pql}\GH(p)\GH(l)\GH(q)\GH(k+q-p)\GH(p+l-q)(\epsilon_p-\frac{u_0}{2}+\frac{1}{2}J_{k-p})(\epsilon_q-\frac{u_0}{2}+\frac{1}{2}J_{p-q})J_{q-p}\nn\\
&& e) \ -\frac{\lambda^2 }{2}  \sum_{pql}\GH(p)\GH(l)\GH(q)\GH(k+q-p)\GH(k+l-p)(\epsilon_p-\frac{u_0}{2}+\frac{1}{2}J_{k-p})(\epsilon_q-\frac{u_0}{2}+\frac{1}{2}J_{l-q})J_{p-k}\nn\\
&& f) \ - \frac{\lambda^2 }{2}  \sum_{pql}\GH(p)\GH(l)\GH(q)\GH(k+q-p)\GH(l+p-k)(\epsilon_p-\frac{u_0}{2}+\frac{1}{2}J_{k-p})(\epsilon_l-\frac{u_0}{2}+\frac{1}{2}J_{p-k})J_{p-k}\nn\\
&& g)  \ \lambda^2 \frac{n}{4} \sum_{pq}\GH(p)\GH(q)\GH(k+q-p)(\epsilon_p-\frac{u_0}{2}+\frac{1}{2}J_{k-p})J_{p-k}\nn\\
\label{chiB}
\earray

\subsection{High frequency limit}
We know from the anti-commutation relations for the Hubbard $X$ operators, that the high frequency limit of the Green's function is $\Lim{i\omega_k\to\infty}\G(k)=\frac{1-\frac{n}{2}}{i\omega_k}$. From \disp{set1mom}, we see that the high frequency limit of the Green's function can also be expressed as $\Lim{i\omega_k\to\infty}\G(k)=\frac{1-\lambda \gamma}{i\omega_k}$. Since $\gamma = \sum_k \G(k)=\frac{n}{2}$ in the exact theory, after setting $\lambda=1$ the two expressions for the high frequency limit are equivalent.

From \disp{set1mom}, we see that to obtain $\GHI(k)$ and $\widetilde{\mu}(k)$ to $m^{th}$ order in $\lambda$, we must calculate $\gamma$, $\Psi(k)$, and $\chi(k)$ to order $m-1$. If we are doing this using the bare expansion, then in order to satisfy the sum rules in \disp{sumrules} order by order, we must also expand the two chemical potentials $\mu$ and $u_0$ in $\lambda$ \cite{KohnLuttinger}.
\beq 
\mu = \mu^{(0)} + \mu^{(1)} + \ldots \;\;\;\;\;\;\;\;\;\;\; u_0 = u_0^{(0)} + u_0^{(1)} + \ldots \ ,
\label{chemexp}
\eeq
where $\mu^{(0)}$ is zeroth order in $\lambda$, $\mu^{(1)}$ is first order in $\lambda$, etc. Denoting $\GH$, $\widetilde{\mu}$, $\gamma$, $\Psi$, and $\chi$ by the generic symbol $Q$, and plugging the expansions from \disp{chemexp} into the bare expansion for $Q^{(m)} = Q^{(m)}(\mu,u_0)$, the latter is rearranged with the various orders being mixed due to the expansion of the chemical potentials. Then, we can solve for the various quantities $\mu^{(0)}$, $\mu^{(1)}$, etc such that in the rearranged series for $\gamma^{(m)} = \gamma^{(m)}(n)$ and $\GH^{(m)} = \GH^{(m)}(n)$,
\beq
\gamma^{(m)} = \delta_{m,0}\frac{n}{2}; \;\;\;\;\;\;\;\;\;\; \sum_k\GH^{(m)}(k)=\delta_{m,0}\frac{n}{2}.
\eeq
Then, substituting the expression for $\gamma^{(m)}$ back into \disp{set1mom}, we see that only $\G^{(0)}(k)$ and $\G^{(1)}(k)$ contribute to the high-frequency limit of the Green's function, and that  $\Lim{i\omega_k\to\infty}\G(k)=\frac{1-\lambda \gamma}{i\omega_k}=\frac{1-\lambda\frac{n}{2}}{i\omega_k}$.

In the skeleton expansion, the situation is different. In this case, after we set $\lambda=1$, the diagrams from all orders in the skeleton expansion are mixed together on equal footing to generate one integral equation which together with the sum rules in \disp{sumrules} determines $\GH$, $\mu$, and $u_0$. The other objects are then obtained from these. In this case, if the skeleton expansions for $\gamma$, $\Psi(k)$, and $\chi(k)$ have been carried out to $m-1^{st}$ order before being plugged into \disp{set1mom}, then the sum rule \disp{sumrules} implies that (after setting $\lambda=1$) $\sum_{l=0}^m \gamma^{(l)} = \frac{n}{2}$. However, from \disp{set1mom}, the high frequency limit is given by $\Lim{i\omega_k\to\infty}\G(k)=\frac{1-\sum_{l=0}^{m-1} \gamma^{(l)}}{i\omega_k}$. Therefore, the error in the high frequency limit is equal to $\gamma^{(m)}$, and we have that
\beq
\Lim{i\omega_k\to\infty}\G(k)=\frac{1-\frac{n}{2}+ \gamma^{(m)}}{i\omega_k}.
\eeq
This error vanishes as $m\to\infty$.

\subsection{Analysis of the $\lambda$ expansion: Feynman type diagrams and non-Feynman diagrams.}

The $\lambda$ series for $\G$ differs from the Feynman series for $\G$ in two fundamental ways. The first is the presence of the term $-\lambda \gamma+\lambda \Psi(k)$ in the numerator of $\G(k)$. In the Feynman series, this term is absent. To discuss the second one, let us identify $\lambda\gamma$ with the Hartree term in the Feynman diagrams, and $\lambda\Phi$ with all self-energy diagrams other than the Hartree term. $\Psi$ forms a subset of $\Phi$ (except for a missing interaction line which is not important for the present discussion), and hence all considerations which apply to $\Phi$ will apply equally well to $\Psi$. Hence, the second important difference is that there are diagrams which contribute to $\lambda\gamma$ which do not contribute the Hartree term of the Feynman series, and there are diagrams that contribute to $\lambda\Phi$ which do not contribute to the other self-energy diagrams of the Feynman series. 

From \figdisp{3rdordergamma}, we can see that the first order $\lambda \gamma$ diagram is exactly the Hartree term of the Feynman series, while the others are all diagrams which do not contribute to the Hartree term of the Feynman series. However, from \figdisp{3rdorderchi} and \figdisp{3rdorderchifromPsi}, we can see that the only diagram in the 3rd order skeleton expansion for $\lambda \Phi$ which is not a Feynman diagram, is diagram g) in \figdisp{3rdorderchifromPsi} (Feynman diagrams are the same order in $\lambda$ as they are in the interaction, while non-Feynman diagrams are not). Therefore, the deviation of $\lambda\Phi$ and $\lambda\Psi$ from the Feynman series grows rather slowly as compared with the growth of the series itself. Moreover, if we consider the fact that the infinite series for $\gamma$ must sum to $\frac{n}{2}$, we see that to ``leading order", the only difference between the $\lambda$ series and the Feynman series is the presence of the term $-\lambda \gamma+\lambda \Psi(k)$ in the numerator of $\G(k)$. This leads us to the point of view taken in the phenomenological ECFL \cite{ECFL,Anatomy,Gweon,ECFLDMFT}, in which $\gamma\to\frac{n}{2}$, and the self-energies $\Psi(k)$ and $\Phi(k)$ are given simple Fermi-liquid forms. Then, the main correction to Fermi-liquid behavior is not seen as coming from the self-energies themselves, but from the interplay between the numerator and denominator of the single-particle Green's function.

\section{Connection with Zaitsev-Izyumov formalism \label{ZaitsevIzyumov}}

The Zaitsev-Izyumov formalism\cite{Izyumov,Zaitsev} is a technique for doing an expansion in $t$ and $J$ around the atomic limit of the $\tJ$ model (given by $t\to0$ and $J\to0$ in \disp{tJmodel}). This can also be viewed as a high-temperature expansion since each factor of $t$ and $J$ must necessarily appear with a factor of $\beta$. The diagrams of this series give rise to the same two self-energy structure for the single-particle Green's function as found in ECFL. In particular, Eq. (3.6) of \refdisp{Izyumov} reads
\beq
\G_\si = \frac{\langle F^{\si0}\rangle + \Delta_\si }{(G^0_\si)^{-1}-\Sigma_\si}.
\eeq
We can make the identifications
\beq
\langle F^{\si0}\rangle \to 1-\gamma; \;\;\;\;\;\;\;\;\Delta_\si\to\Psi(k);\;\;\;\;\;\;\;\;(G^0_\si)^{-1}\to\GH^{-1(0)};\;\;\;\;\;\;\;\; \Sigma_\si \to -\epsilon_k\gamma +\Phi(k).
\eeq
As is the case in the $\lambda$ series, the fundamental object in the Zaitsev-Izyumov high-temperature series is the auxiliary Green's function $\GH$.

\begin{figure}[H]
\begin{center}
\includegraphics{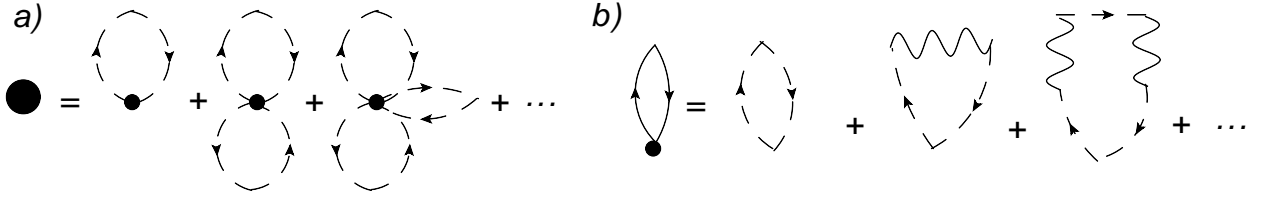}
\caption{In panel a), the zeroth order $\gamma$ diagram from the high-temperature series (the big dot) is expanded as an infinite series in $\lambda$. The dashed lines indicate auxiliary Green's functions in the atomic limit $\GH_{t\to0,J\to0}(i\omega_k)$. In panel b), the zeroth order $\gamma$ diagram from the $\lambda$ series is expanded as an infinite series in the hopping. The solid line indicates a bare propagating auxiliary Green's function $\GH^{(0)}(k)$, while the wavy line indicates the hopping $\epsilon_k$. }
\label{IzyumovECFL}
\end{center}
\end{figure}

The {main} difference between the two series is the {dimensionless} expansion parameter. In the case of ECFL, it is the continuity parameter $\lambda$. In the case of the high-temperature series, it is {$\beta t$ and $\beta J$}. To see this more explicitly, consider the simplest diagram in both series, which is the zeroth order diagram for $\gamma$. In ECFL, this is the diagram in Fig. (\ref{3rdordergamma}a). In \refdisp{Izyumov}, it is represented by a dot. The relationship between the two is shown in \figdisp{IzyumovECFL}. In this figure, the dashed line indicates an atomic limit auxiliary Green's function $\GH_{t\to0,J\to0}(i\omega_k) = \frac{1}{i\omega_k+\mu}$. The big dot indicates the atomic limit value of $\gamma$, i.e. $\gamma_{t\to0,J\to0}=\frac{\rho}{2}$, where $\rho= \frac{2 e^{\beta \mu}}{1 + 2 e^{\beta \mu}}$ is the atomic limit density. The wiggly line indicates a hoping $\epsilon_k$. Finally, the solid line indicates the bare auxiliary Green's function $\GH^{(0)}(k)=\frac{1}{i\omega_k+\mu-\epsilon_k}$. In panel a), the zeroth order $\gamma$ from the high-temperature series is expanded as an infinite series in $\lambda$. Here, each loop corresponds to $\sum_{i\omega_k} \GH_{t\to0,J\to0}(i\omega_k) = \frac{\frac{\rho}{2}}{1-\frac{\rho}{2}}$, and there is a minus sign between the successive terms of the series. Summing the geometric series, we find that $\frac{\frac{\rho}{2}}{1-\frac{\rho}{2}} \cdot \frac{1}{1+\frac{\frac{\rho}{2}}{1-\frac{\rho}{2}}}=\frac{\rho}{2}$. In panel b), the zeroth order $\gamma$ from the $\lambda$ series is expanded as an infinite series in the hopping $\epsilon_k$. This gives the geometric series $\sum_k \GH^{(0)}(k) = \sum_k\sum_{n=0}^\infty \frac{\epsilon_k^n}{(i\omega_k+\mu)^{n+1}}$. We see that to get from the high-temperature series to the $\lambda$ series, one would have to break up all atomic limit objects into an infinite series in terms of $\lambda$, and replace every atomic limit auxiliary Green's function with a bare propagating one. 

We can summarize the fundamental difference between the two approaches as follows. In the case of zero magnetic field, the high-temperature series is an expansion around {the atomic limit}, i.e. {an exponentially}   degenerate manifold of states,  without giving preference to any one of them. In doing so, it is difficult to recover {the adiabatic continuity aspect of} physics relating to the Fermi-surface and the Luttinger-Ward volume theorem \cite{Nozieres}. In contrast, ECFL builds the Fermi-surface into the $\lambda$ expansion at zeroth order, by expanding around the free Fermi gas and {by maintaining continuity in $\lambda$. Finally,} by enforcing that the number of auxiliary fermions equals the number of physical ones through the second chemical potential $u_0$, ECFL is able to satisfy the Luttinger-Ward volume theorem. 

\section{Conclusion}

In conclusion, starting with the $\lambda$ expansion as defined through iteration of the Schwinger EOM around the free Fermi gas \cite{ECFL,Monster}, we derived a set of diagrammatic rules to calculate the $n^{th}$ order contribution to the physical Green's function $\G$ in the \tJ model. The resulting diagrams suggested the need for two self-energies, which we denoted by $\Sigma_a$ and $\Sigma_b$.  Using the Schwinger equations of motion defining the ECFL objects, $\GH$, $\widetilde{\mu}$, $\gamma$, $\Phi$, and $\Psi$, we derived diagrammatic rules for calculating these objects and found that they could be related simply to $\Sigma_a^*$ and $\Sigma_b^*$, the irreducible parts of $\Sigma_a$ and $\Sigma_b$. We also discovered diagrammatically that $\Psi$ diagrams are simply a subset of the $\Phi$ diagrams, with an interaction line missing. Denoting the remainder of the $\Phi$ diagrams by the symbol $\chi$, this implied the expression $\Phi(k) = \epsilon_k\Psi(k) + \chi(k)$.  We had already found this to be the case in the limit of infinite spatial dimensions with $\chi(k)\to\chi(i\omega_k)$ and $\Psi(k)\to\Psi(i\omega_k)$ in \refdisp{ECFLlarged}, and here we generalized it to finite dimensions.  We also derived the Schwinger EOM for the object $\chi$. We derived diagrammatic rules for the three point vertices $\Lambda$ and $\U$, defined as the functional derivatives of $\GHI$ and $\widetilde{\mu}$ respectively, with respect to the source. We derived a generalized Nozi\`{e}res relation for these vertices, which differs from the standard one for the Feynman diagrams. We then introduced skeleton diagrams into our series, thereby allowing us to make the connection with the iterative expansion of the Schwinger equations of motion (as done in Refs. (\onlinecite{ECFL}) and (\onlinecite{Monster})), which deals exclusively with skeleton diagrams. 

We then derived the third order skeleton expansion for $\GH$ and $\widetilde{\mu}$. Previously, this had been done only up to second order. We then discussed the error in the high-frequency limit incurred in the skeleton expansion carried out to any order in $\lambda$. We also discussed the ``deviation" of the $\lambda$ series from the Feynman series, thereby justifying on a qualitative level, the phenomenological ECFL\cite{ECFL, Anatomy}, which has already been successful in explaining lines shapes found both from ARPES experiments\cite{Gweon}, and from DMFT calculations\cite{ECFLDMFT}. Finally, we discussed the connection between ECFL and the Zaitsev-Izyumov high-temperature series. We found that while both formalisms dealt with the projection of double occupancy by introducing two self-energies, they had fundamentally different approaches to dealing with the problem of the Fermi-surface. While the high-temperature series is an expansion around a completely degenerate manifold of states, ECFL makes an adiabatic connection with the Fermi-surface and preserves the Luttinger-Ward volume theorem.

Our {main motivation} in deriving these diagrammatic rules is that they will allow the $\lambda$ expansion to be evaluated to high orders using powerful numerical techniques such as diagrammatic Monte Carlo, and also that the intuition gained from the diagrams themselves {could} facilitate infinite re-summations guided by some physical principles.

\section{Acknowledgements}
This work was supported by DOE under Grant No. FG02-06ER46319. The authors  thank Professor Antoine Georges, CPHT-Ecole Polytechnique and Coll\`ege de France, supported in part by the DARP/MURI-OLE program, for their warm hospitality, where this work was completed.


\begin{thebibliography}{235}
\bibitem{Anderson}P. W. Anderson, Science {\bf235}, 1196 (1987); The Theory of Superconductivity , Princeton University Press, Princeton, NJ,1997;  arXiv:0709.0656 (unpublished).
\bibitem{ECQL}  B. S. Shastry, Phys. Rev. {\bf B 81}, 045121 (2010).  
\bibitem{Wang} Z. Wang, Y. Bang, and G. Kotliar, Phys. Rev. Lett. {\bf 67}, 2733-2736 (1991).
\bibitem{Greco} A. Greco and R. Zeyher, Phys. Rev. {\bf B 63}, 064520 (2001).
\bibitem{Foussats} A. Foussats and A. Greco, Phys. Rev. {\bf B 70}, 205123 (2004).
\bibitem{Zeyher} R. Zeyher and A. Greco, Phys. Rev. {\bf B 80}, 064519 (2009).
\bibitem{Zaitsev} R. O. Zaitsev, Sov. Phys - JETP {\bf 43}, 574 (1976).
\bibitem{Izyumov} Yu. A. Izyumov and B. M. Letfulov, J. Phys: Condens. Matter  {\bf 2}, 8905-8923 (1990).
\bibitem{ECFL} B. S. Shastry, Phys. Rev. Letts. {\bf107}, 056403 (2011).
\bibitem{Monster} B. S. Shastry, Phys. Rev. {\bf B 87}, 125124 (2013).
\bibitem{Shastry-AOP} B. S. Shastry, Ann. Phys. {\bf 343}, 164 (2014).

\bibitem{LuttingerWard} J. M. Luttinger and J. C. Ward, Phys. Rev. {\bf 118}, 1417 (1960).
\bibitem{Luttinger} J. M. Luttinger, Phys. Rev. {\bf 119}, 1153 (1960).
\bibitem{ECFLDMFT}  R. \v{Z}itko, D. Hansen, E. Perepelitsky, J. Mravlje, A. Georges, and B. S. Shastry, Phys. Rev. {\bf B 88}, 235132 (2013).
\bibitem{ECFLAM} B. S. Shastry , E. Perepelitsky, and A.C. Hewson, Phys. Rev. B {\bf 88}, 205108 (2013).
\bibitem{Moments} E. Khatami, D. Hansen, E. Perepelitsky, M. Rigol, and B. S. Shastry, Phys. Rev. {\bf B 87}, 161120 (2013).
\bibitem{Gweon} G. H. Gweon, B. S. Shastry, and G. D. Gu, Phys. Rev. Lett. {\bf107}, 056404 (2011); K. Matsuyama, G. H. Gweon,  Phys. Rev. Lett. {\bf 111}, 246401 (2013).
\bibitem{FW} A. L. Fetter and J. D. Walecka, Quantum Theory of Many-Particle Systems, Dover Publications, INC. Mineola, NY, 2003.
\bibitem{Nozieres} P. Nozi\`{e}res, Theory of Interacting Fermi Systems, W. A. Benjamin, Amsterdam, 1964.
\bibitem{DiagMC} N. V. Prokof'Ev, B. V. Svistunov, and I. S. Tupitsyn, 
Journal of Experimental and Theoretical Physics {\bf 87}, Issue 2, 310-321 (1998); K. V. Houcke, E. Kozik, N. Prokof'ev, and B. Svistunov, Physics Procedia {\bf 6}, 95-105 (2010).
\bibitem{ECFL2nd} D. Hansen and B. S. Shastry, Phys. Rev. {\bf B 87}, 245101 (2013).

\bibitem{HKHH} A. B. Harris. D. Kumar, B. I. Halperin and P. C. Hohenberg, Phys. Rev. {\bf B 3}, 961 (1971).
\bibitem{ECFLlarged} E. Perepelitsky and B. S. Shastry, Annals of Physics {\bf 338},  283 (2013).
\bibitem{KadanoffBaym} L. P. Kadanoff and G. Baym, Quantum Statistical Mechanics: Green's Function Methods in Equilibrium and Nonequilibrium Problems, Benjamin, NY, 1962.
\bibitem{KohnLuttinger} W. Kohn and J. M. Luttinger, Phys. Rev. {\bf 118}, 41 (1960). 
\bibitem{Anatomy} B. S. Shastry, Phys. Rev. {\bf B 84}, 165112 (2011).
\end{thebibliography}
\end{document}